# Resolving oxidation states and Sn–halide interactions of perovskites through Auger parameter analysis in XPS

*Alexander Wieczorek[1], Huagui Lai[2], Johnpaul Pious[2], Fan Fu*[2], Sebastian Siol*[1]*

[1]Laboratory for Surface Science and Coating Technologies, Empa − Swiss Federal Laboratories for Materials Science and Technology, Switzerland.

[2]Laboratory for Thin Films and Photovoltaics, Empa − Swiss Federal Laboratories for Materials Science and Technology, Switzerland.

Corresponding authors: fan.fu@empa.ch, sebastian.siol@empa.ch

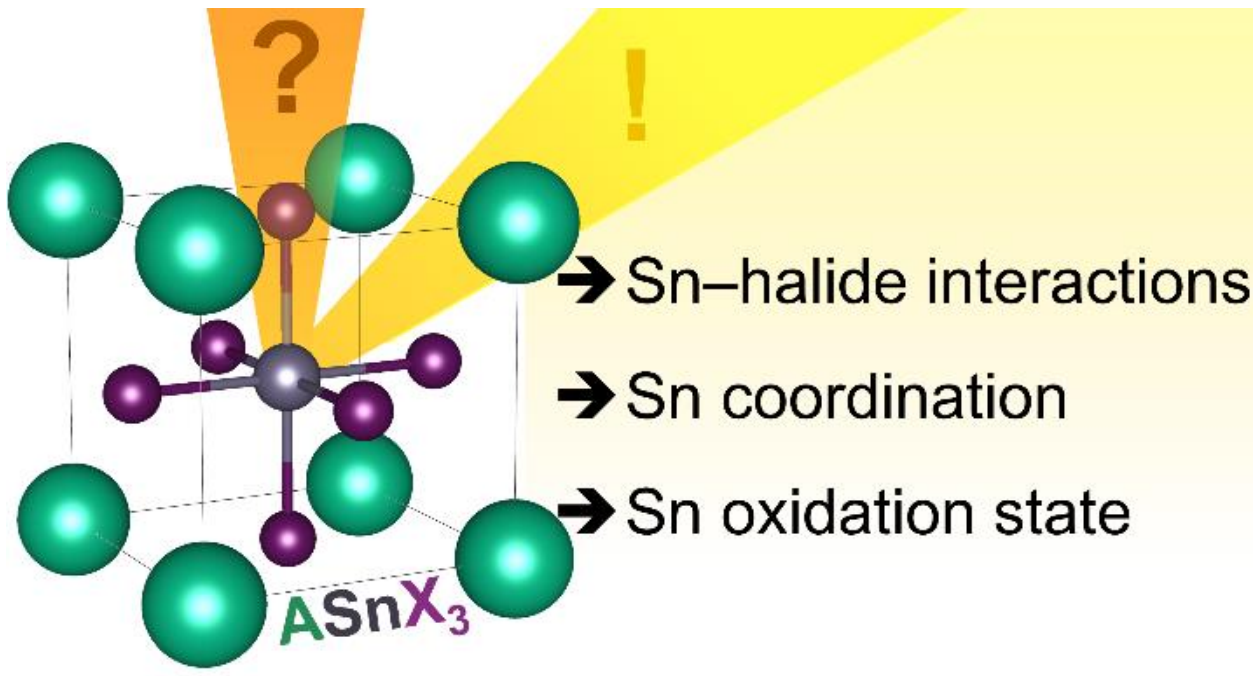

Abstract

Reliable chemical state analysis of Sn semiconductors by XPS is hindered by the marginal observed shift in the Sn 3d region. For hybrid Sn-based perovskites especially, errors associated with charge referencing can easily exceed chemistry-related shifts. Studies based on the modified Auger parameter $\alpha'$ provide a suitable alternative and have been used previously to resolve different chemical states in Sn alloys and oxides. However, the meaningful interpretation of Auger parameter variations on Sn-based perovskite semiconductors requires fundamental studies. In this work, we perform a comprehensive Auger parameter study through systematic compositional variations of Sn halide perovskites. We find that in addition to the oxidation state, $\alpha'$ is highly sensitivity to the composition of the halide-site, inducing shifts of up to $\Delta\alpha'$ = 2 eV between $ASnI_3$ and $ASnBr_3$ type perovskites. The reported dependencies of $\alpha'$ on the Sn oxidation state, coordination and local chemistry provide a framework that enables reliable tracking of degradation as well as X-site interaction for Sn-based perovskites and related compounds. The higher robustness and sensitivity of such studies not only enables more in-depth surface analysis of Sn-based perovskites than previously performed, but also increases reproducibility across laboratories.

## 1. Introduction

Across a wide range of applications, semiconductors based on Sn are being investigated. Besides kesterites (e.g. $Cu_2ZnSnS_4$),[1] chalcogenides (e.g. SnSe)[2] and group IV alloys (e.g. GeSn),[3] Sn halide perovskites have peaked interest for optoelectronic applications.[4] Their composition follows the formula $ASnX_3$ where typically A = $Cs^+$, methylammonium ($MA^+$), formamidinium ($FA^+$) and X = $I^-$, $Br^-$, $Cl^-$. Furthermore, they are classified by their unique perovskite crystal structure (**Figure 1a**). The lower toxicity compared to their more established $APbX_3$ counterparts and relatively narrower yet tunable bandgap make them attractive candidates for a range of optoelectronic applications.[5] However, the facile oxidation of the Sn(II) species to Sn(IV) compounds severely limits their lifetime, especially under ambient conditions.[6]

The development of Sn-based semiconductors necessitates a reliable determination of the chemical state of the constituent elements. X-ray photoelectron spectroscopy (XPS) is an important tool in this regard as resulting insights are routinely used in the development of novel energy materials and devices.[7,8]

Typically, these studies are based on the observation of the core level photoelectron emission features and their position on the binding energy ($E_b$) scale. To counteract charging effects, referencing based on adventitious carbon is often performed. However, this results in a typical inaccuracy of ± 0.2 eV, whereas even errors of more than 1 eV have been observed.[9]

For many Sn compounds with different Sn oxidation states and compositions, the main Sn 3d core level emission exhibits only minimal $E_b$ shifts, which lie below this error range.[10] Consequently, probing degradation as well as assigning the effects of compositional changes and additives to the surface chemistry of Sn-based perovskites with conventional XPS can be challenging.

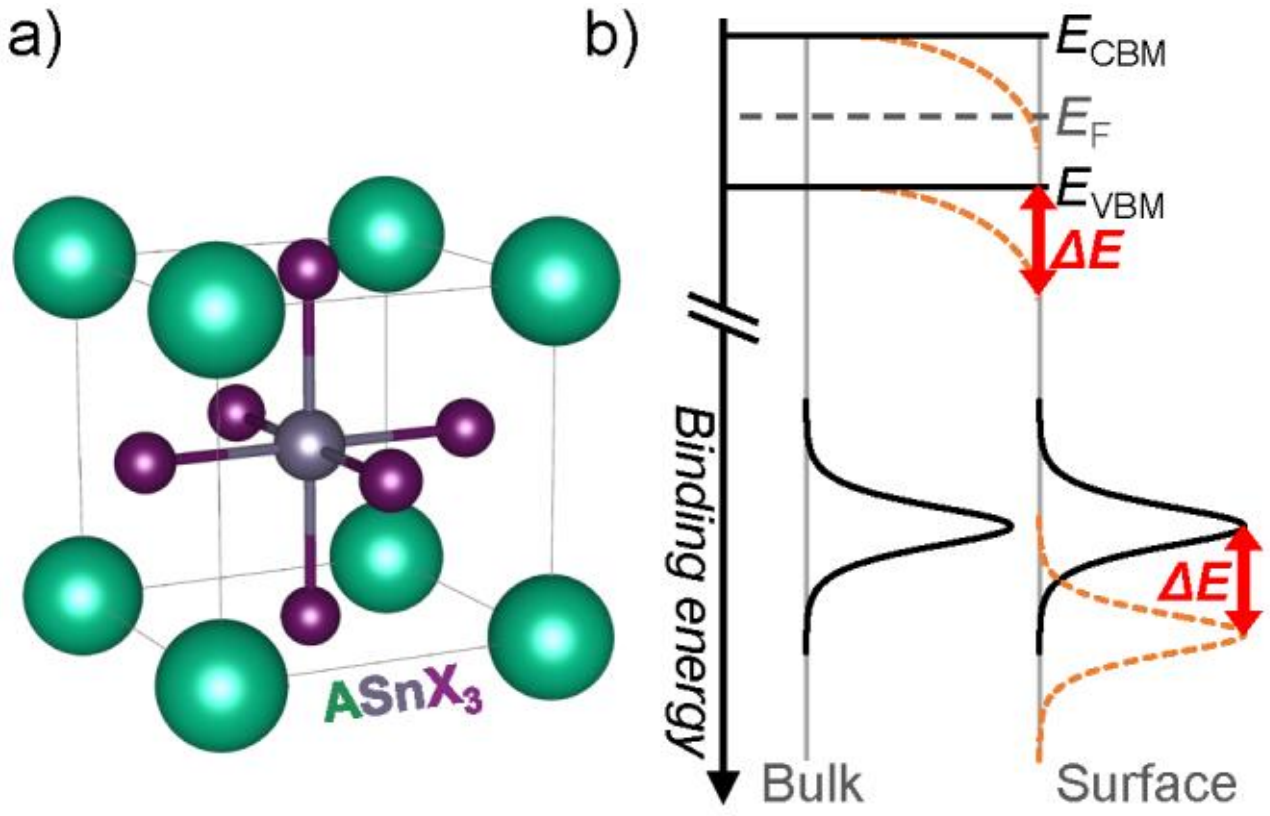

**Figure 1.** a) Perovskite crystal structure. b) Schematic of the band positions and core-level features observed using XPS with (orange dashed) or without (black solid) surface band bending effects. Shown are the valence band maximum $E_{VBM}$, conduction band minimum $E_{CBM}$ and Fermi level position $E_F$. A change of the surface potential $\Delta E$ (i.e. the Fermi level position) due to differential charging or surface photo voltages results in surface band bending and consequently in a shift of the observed XPS features, depending on the probing depth.

In addition, shifts on the binding energy scale not originating from chemical interactions are especially pronounced for semiconductors (**Figure 1b**). Similar to effects observed in UPS,[11] excitation during measurement with an X-Ray source may induce photovoltages,[12] resulting in surface bend bending. Surface charges e.g. induced by ionic additives may lead to similar effects which originate from the resulting space charge and not the bonding to the additive itself.[13] Consequently, a probing depth-dependent shift is induced, which may further inhibit reliable charge referencing.

For chemical state analysis of Sn, XPS analysis based on the modified Auger parameter (AP) concept offers an effective alternative to distinguish different chemical states.[14] This approach relies on the observation of both the photoelectrons (i.e. Sn 3d) and Auger electrons (i.e. Sn MNN). Since the AP is only determined from the relative positions of these features to another, the AP is insensitive to charging effects and erroneous binding energy calibration. Furthermore, the Auger emission is more sensitive to changes in the local chemical environment than the photoelectron emission, due double core-hole final state in the Auger process.[15] Using this additional feature may therefore increase the shifts observed due to altered chemical interactions. This holds especially true for the Sn MNN Auger emission feature resulting from valence shell transitions which is accessible using standard XPS excitation sources (i.e. Al kα).

While first attempts have been made to use the AP approach for Sn compounds,[16] there is a lack of systematic studies on the effects of oxidation state and X-site interactions for Sn-based perovskites specifically. Since a large amount of factors such as oxidations state or nearest neighbor interactions govern the resulting AP value,[17] empirical studies on reference samples may help to unravel each contributing factor and understand more complex samples.[18]

In this work we perform AP analysis on a comprehensive set of reference compounds typically found in literature. The perovskite structure for each sample was confirmed by XRD, while the transfer into the XPS measurement system was performed in inert-gas atmosphere. Based on this systematic study, we identified a high sensitivity of the AP on Sn–X interactions and construct a framework which links the AP to the perovskite crystal structure and Sn chemistry. This enables more in-depth and robust chemical state analysis via XPS than studies which are based on the observation of the photoelectron emission alone. In addition, the results from this work can be used by other researchers and groups to track and resolve the Sn chemical state in related compounds. Finally, the methods presented here are applicable for interface studies on perovskites devices providing additional and valuable insights for interface engineering.[19]

## 2. Results and Discussion

To obtain high-quality perovskite reference samples representing the typical compositional space for optoelectronics applications, we prepared $ASnX_3$ type perovskites with A = $Cs^+$, $FA^+$, $MA^+$ and X = $Br^-$, $I^-$. For additional variation of the B-site, $FA(Sn_{1-x}Pb_x)I_3$ films with x = 0.25, 0.5, 0.75 were prepared. Samples were transferred to the XPS chamber via inert-gas transfer. Short-term measurements of the Sn 3d core level were performed before and after each measurement to ensure the observed features remained stable during these measurements (**Figure S 1**). This ensured that the reported APs reflect the chemical states of pristine samples instead of aged samples with surface oxides or X-ray induced damage. To confirm the successful conversion of the precursors to the perovskite phase, each film was analyzed using X-ray diffraction (XRD) after the XPS analysis.

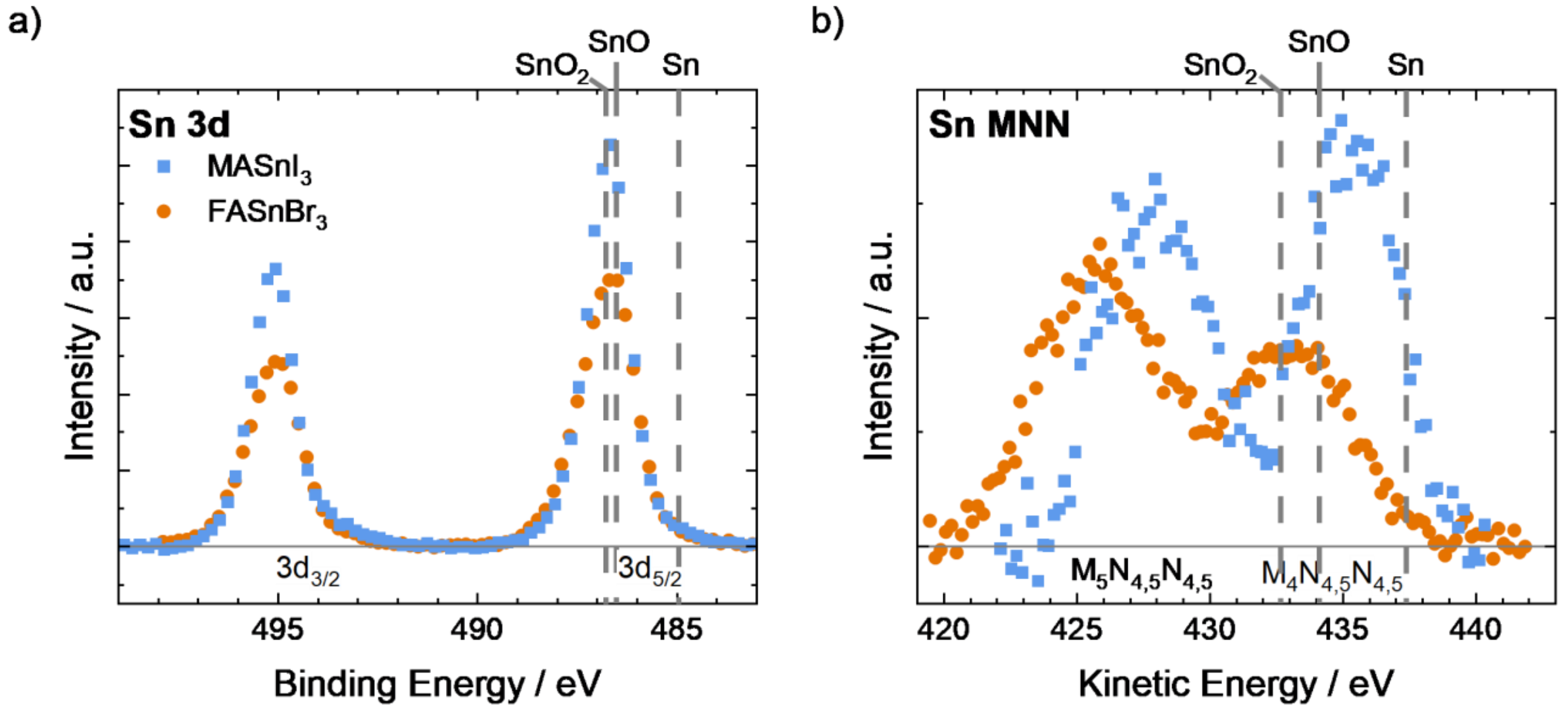

**Figure 2.** Representative spectra of the a) Sn 3d core level and b) Sn MNN Auger electron emission of $MASnI_3$ and $FASnBr_3$ measured by XPS. In the Sn 3d region, both compounds exhibit similar emission features with no visible shift in binding energy. For the Sn MNN region, a clear difference in peak shape and position was found. This highlights the sensitivity of the Auger emission to the chemical environment. The results are in alignment with reference data obtained

from the NIST database. For Sn, SnO and $SnO_2$ the relative shifts of the electron energy is pronounced to a greater extend for the Sn MNN emission.[20]

As evident from the XPS measurements of the Sn 3d and Sn MNN regions for $MASnI_3$ and $FASnBr_3$, no clear shift can be observed for the Sn $3d_{5/2}$ feature between these compounds (**Figure 2a**). These results are well in alignment for the minor reported shifts of < 0.2 eV even between SnO (i.e. Sn(II)) and $SnO_2$ (i.e. Sn(IV)) from literature.[21,22] In contrast, shifts of > 2.5 eV can be observed for the maxima of the $M_4N_{4,5}N_{4,5}$ emissions (**Figure 2b**) in alignment with the increased sensitivity of the Auger emission to the local chemical environment of Sn. While the determination of the Sn chemical state on the Auger emission alone may thus appear attractive, the absolute values on the kinetic energy ($E_{kin}$) scale remain dependent on external charge referencing as well as potential surface band bending. In contrast, the AP is largely independent from the Fermi level position at the surface.

Deconvolution of the Auger emission is often impractical, as well as ambiguous, due to the complexity of the underlying physical process, resulting in a large amount of overlapping Auger electron features.[23] Consequently, as often applied for Sn alloys and oxides,[24–26] we performed the AP analysis based on the maximum of the Sn $M_4N_{4,5}N_{4,5}$ feature. The same applies for the Sn 3d core level. Especially, when binding energy shifts between different compounds are minute, peak fitting with multiple components can add ambiguity to the results and hinder reproducibility of the results between different groups. Here we used the binding energy of the Sn $3d_{5/2}$ feature maximum.

The modified Auger parameter $\alpha'$ can be calculated according to

$$\alpha' = E_{\mathrm{kin}}\left(\mathrm{Sn\ M_4N_{4,5}N_{4,5}}\right) + E_{\mathrm{b}}\left(\mathrm{Sn\ 3d_{5/2}}\right) \quad (1)$$

, where $E_{kin}$(Sn $M_4N_{4,5}N_{4,5}$) denotes the kinetic energy of the Sn $M_4N_{4,5}N_{4,5}$ feature maximum and $E_b$(Sn $3d_{5/2}$) denotes the binding energy of the Sn $3d_{5/2}$ feature maximum.[27]

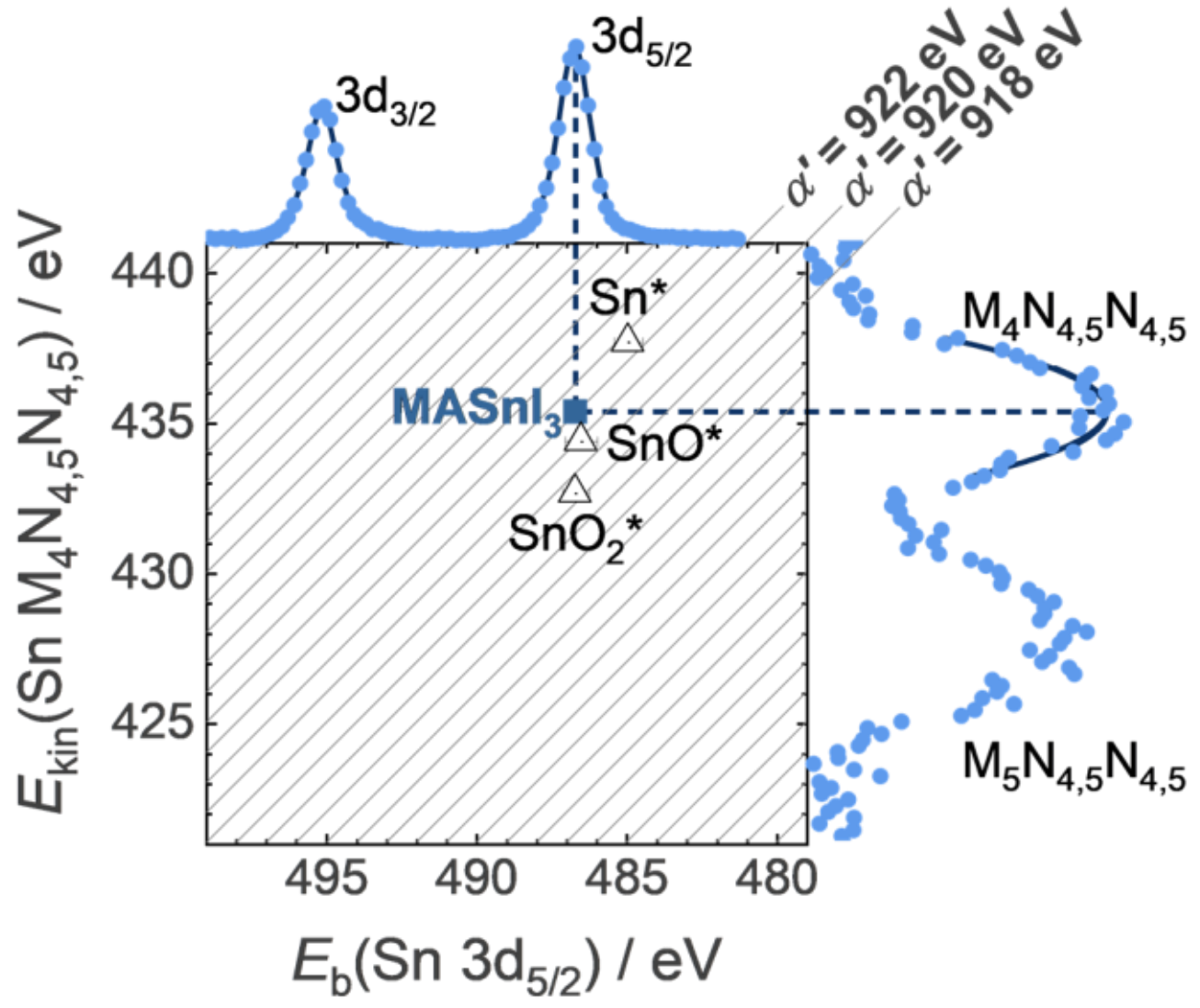

**Figure 3.** Peak identification of the Sn 3d and Sn MNN features to obtain the modified Auger parameter *α'*. Peak determination for the Sn $3d_{5/2}$ and Sn $M_4N_{4,5}N_{4,5}$ features constructs a point on the Wagner plot for a single measurement. Reference values obtained from the NIST database are marked with an asterisk and related error bars are calculated based on the standard deviation between these values.

This relationship can be visualized by a Wagner plot (**Figure 3**), where $E_b$(Sn $3d_{5/2}$) and $E_{kin}$(Sn $M_4N_{4,5}N_{4,5}$) are plotted against each other.[28] As a result, $\alpha'$ as the crucial value appears as a diagonal line in this graph. Charging effects may only shift the data points along these diagonal lines. Furthermore, as evident from the reference values of SnO and $SnO_2$ in the graph, shifts in the AP are mostly originating from the shifts of their $E_{kin}$(Sn $M_4N_{4,5}N_{4,5}$) values. Since this approach relies on the comparison of features across wide energy ranges, a careful calibration of the linearity of the kinetic energy scale of the instrument is critical.

We performed detailed XPS analysis on a large number of reference compounds and calculated $\alpha'$ of Sn (eq 1). The resulting values for all reference Sn-based perovskites are displayed in a single Wagner plot (**Figure 4**).

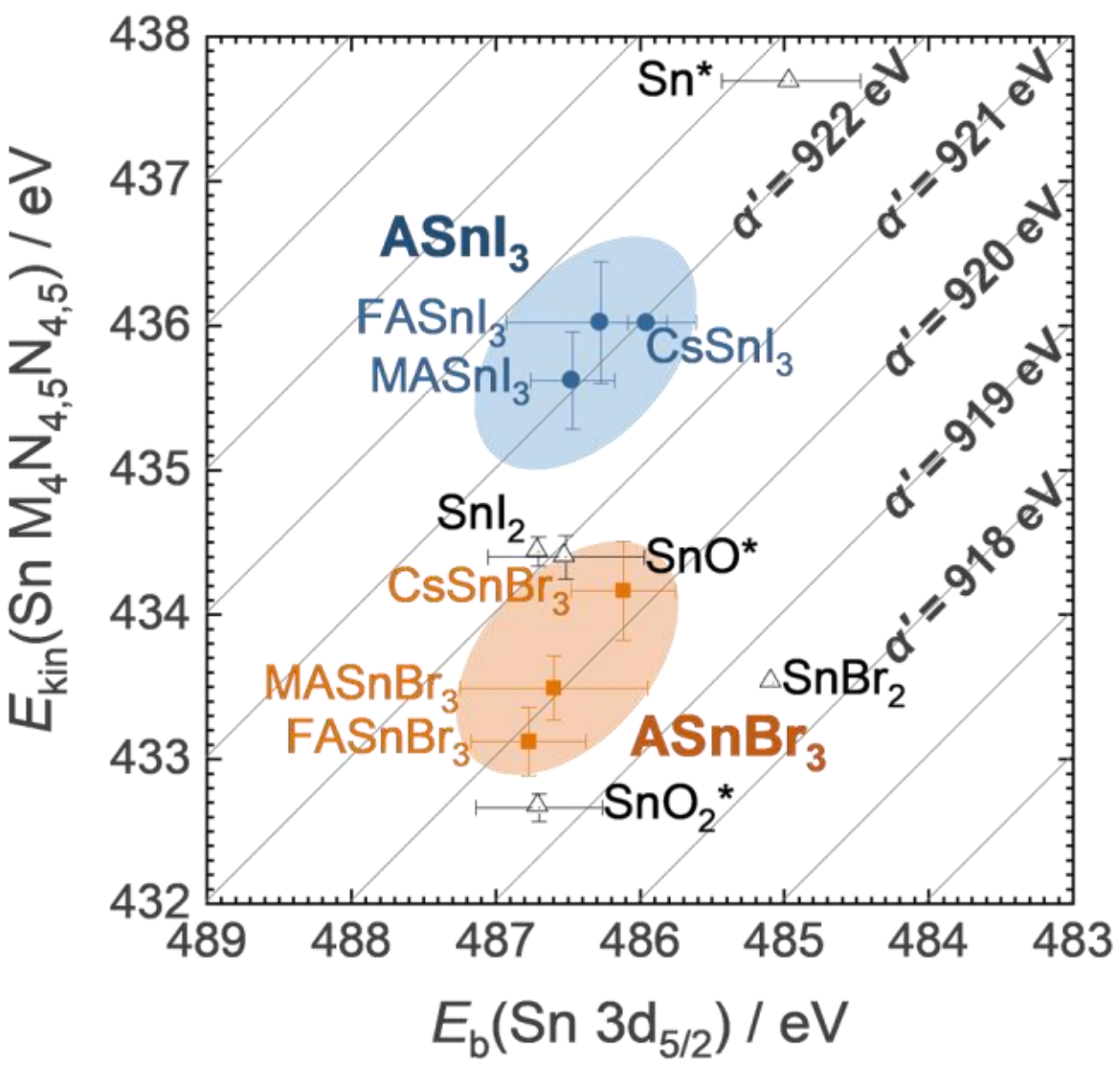

**Figure 4.** Wagner plot depicting all reference $ASnX_3$ perovskites of this study. Most strikingly, clear regions for $ASnI_3$ and $ASnBr_3$ perovskites could be identified on the Wagner plot. The difference of the AP by $\Delta\alpha' \approx 2$ eV highlights its sensitivity for Sn–X interactions. $E_{kin}$ error bars were derived from the estimated error for each peak fit as well as the standard deviation of maxima across measurements. $E_b$ error bars were derived from the standard deviation of the maxima across measurements alone. Reference values obtained from the NIST database are marked with an asterisk and related error bars are calculated based on the standard deviation between these values.

Most notably a clear clustering is observed based on the X-site chemistry of the materials, allowing to resolve trends beyond those observed based on the Sn 3d feature alone. For all $ASnI_3$ type perovskites, the AP was clustered around $\alpha' = 922$ eV. Strikingly, $ASnBr_3$ type perovskites exhibited a clustering around $\alpha' = 920$ eV, resulting in a shift of $\Delta\alpha' = 2$ eV compared to $ASnI_3$ type perovskites. A shift to lower AP of $\Delta\alpha' > 1$ eV was observed when comparing $AX_2$ compounds (i.e. $SnBr_2$ and $SnI_2$) to their respective $ASnX_3$ counterparts. Furthermore, the alloying of the B-site for the $FA(Sn_{1-x}Pb_x)I_3$ type perovskites was investigated. It has to be noted, that a reduction to Sn(0) was observed in these samples, which increased with increasing Pb content (**Figure S 2**). A minor shift of $\Delta\alpha' < 1$ eV compared to $FASnI_3$ was observed (**Figure S 3Error! Reference source not found.**). This is also reflected in the increasing Auger electron kinetic energies with increasing Pb and thus Sn(0) content. The Sn 3d spectra shows two distinct components, so that the respective core level binding energy is not affected. Similar effects were observed for $SnBr_2$ but not for the resulting $ASnBr_3$ perovskites. The contribution of beam damage or exposure to ultra-high vacuum (UHV) to potential degradation was evaluated using repeated measurements and angle-resolved XPS (ARXPS), respectively. While repeated exposure to the X-Ray beam did not appear to highly increase the Sn(0) fraction (**Figure S 4**), ARXPS measurements with a probing depth of approximately 4.6 nm and 6.3 nm revealed the increased presence of Sn(0) on the sample surface (**Figure S 5**). These findings are consistent with the formation of a Sn(0) film on the surface as a result of UHV degradation while the effect of beam damage appeared minuscule.

As empirically and theoretically established, $\alpha'$ depends both on the oxidation state of the core-ionized atom as well as the geometry and electronic interaction with its nearest neighbors.[15] Due to the identical preparation conditions and lack of significant Sn(0) formation for $ASnI_3$ and $ASnBr_3$ type perovskites, the Sn oxidation state for these compounds is expected to be similar.

Based on the perovskite structure, the nearest neighbors of the probed Sn atom are the X-site atoms (**Figure 1a**). The bonding between the B-site (i.e. $Sn^{2+}$) and X-site (i.e. $I^-$ or $Br^-$) can be understood as a mixed set of ionic and covalent interactions.[29] From overlap of the respective atomic orbitals along the bond axis, a $\sigma$-bond is formed with increased B–X electron transfer towards heavier halogenides.[30] Subsequently, the screening of the core-ionized Sn atom is heavily influenced by the type of halogenide (e.g. $Br^-$, $I^-$) surrounding it. With the covalent bonding contribution being more pronounced for Sn-based perovskites than for Pb perovskites,[31] especially large Sn–X electronic interactions result during the photoionization process. In contrast, the A-site mostly interacts with the X-sites of the $SnX_6^{4-}$ octahedra through weak hydrogen bonding.[32]

Consequently, the sensitivity of $\alpha'$ on X-site variations for Sn-based perovskites can be directly linked to their particular bonding mechanism.

As reported for AP studies on different polymorphs of $Al_2O_3$, shifts of $\alpha'$ are highly dependent on the short-range order and largely independent from the long-range order.[33] Similarly, we recently reported on the Zn AP in nitride semiconductors. In this study, the evolution of Zn AP was found to be dominated by the local chemical environment and coordination of the Zn atom, rather than the crystal structure and symmetry of the material.[34] Likewise, the $\alpha'$ shift between $SnI_2$ and $ASnI_3$ compounds can be explained by the shift from a five- to a six-fold coordination environment of the probed Sn atom. In contrast, the performed A-site variations may rather affect the tilting of $BX_6^{4-}$ octahedra on the long-range order and the thereby do not affect $\alpha'$.[35] These dependencies explain the additional sensitivity of $\alpha'$ to the perovskite structure.

For the B-site alloying, less intense effects on electronic interactions can be expected as Sn and Pb atoms would at most interact via the X-site within the solid solution.[36] However, based on the gradual formation of Sn(0) with increased Pb alloying (**Figure S 2**), the shift from $ASnI_3$ perovskites likely resulted from a change in Sn oxidation state. Indeed, the shift towards higher $\alpha'$ values is well in alignment with a reduction of the Sn centers (i.e. a relative increase in the Sn(0) component). Strikingly, the comparably minor shift of $\Delta\alpha' < 1$ eV between $FA(Pb_{1-x}Sn_x)I_3$ and $ASnI_3$ perovskites suggests an increased sensitivity of $\alpha'$ on the Sn–X interactions compared to the Sn oxidation state.

Based on these findings, the AP of Sn-based perovskites exhibits a high sensitivity on the Sn–halide bonding with lesser dependencies on the Sn coordination and oxidation state. More recently, surface-sensitive solid state nuclear magnetic ressonance (ssNMR) spectroscopy studies on perovskites have been reported.[37] Besides the herein identified factors, the $^{119}$Sn chemical shift was identified to be also dependent on A-site variations amongst other factors.[38] As a result, the exceptionally high sensitivity of $\alpha'$ on the Sn–X interactions alone makes it well-suited for selective studies on X-site variations even when multiple sites are being varied.

Furthermore, these features enable to observe the surface degradation of perovskite samples based on the change of the Sn chemical state. The herein presented references may facilitate such studies.

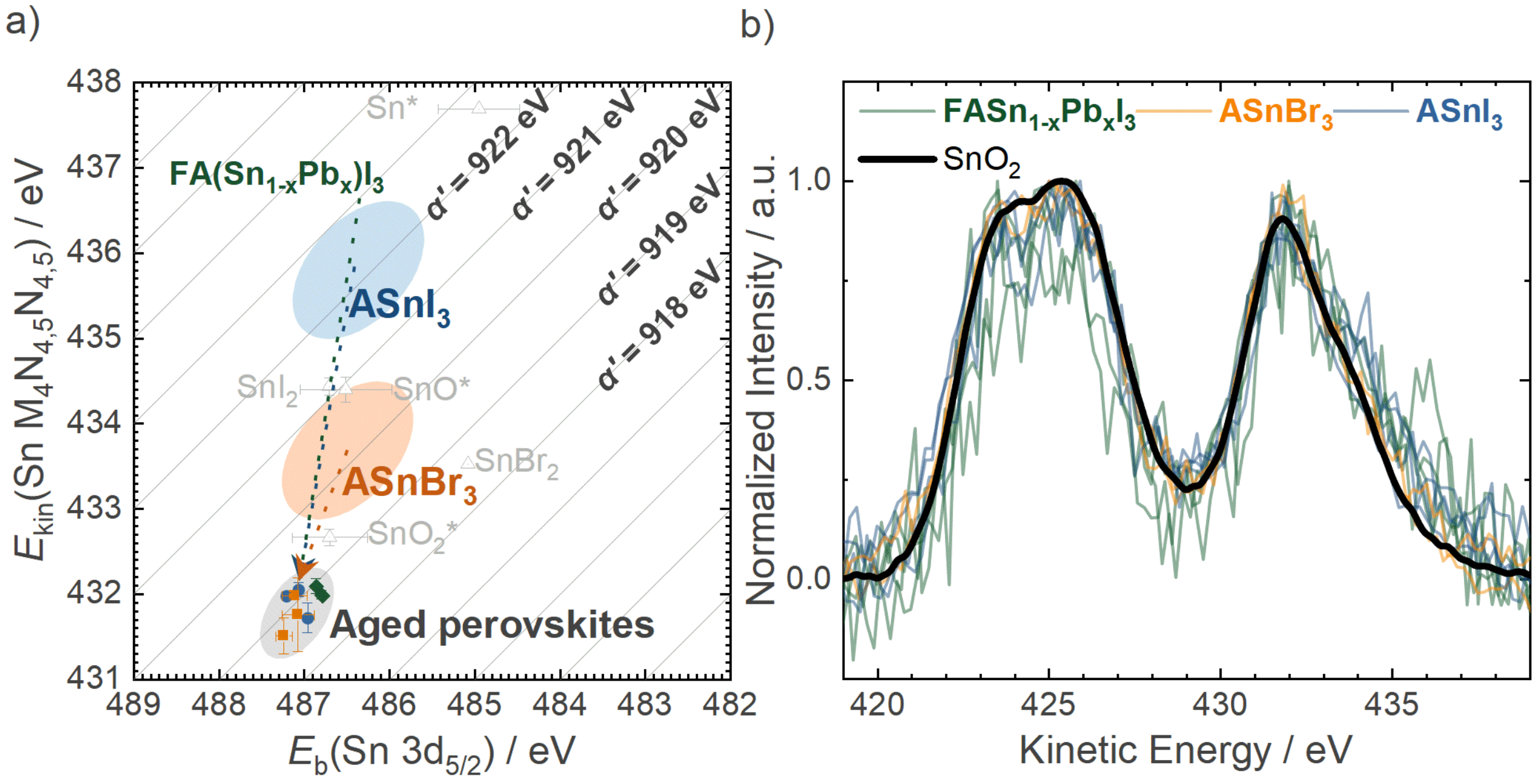

**Figure 5. a)** Wagner plot depicting $\alpha'$ for each aged reference sample. In contrast to the pristine samples previously transferred under inert-gas, the values converge to $\alpha' = 919$ eV for all samples regardless of the perovskite chemistry. **b)** Sn MNN region of all aged perovskite samples compared to the $SnO_2$ reference following charge correction. The similar peak shape further suggests the formation of $SnO_2$ on the surface regardless of the employed perovskite chemistry.

Following aging of the pristine perovskite samples under ambient air, the AP was determined again for each sample. Remarkable, regardless of the perovskite chemistry, all aged samples exhibit an AP at $\alpha' = 919$ eV, similar to the AP of $SnO_2$ according to the NIST database (**Figure**

**5a**). Furthermore, the feature-richness of the Auger emission allows it to be utilized as a chemical fingerprint without the need for peak fitting. This eliminates the need for potential assumptions on the underlying processes occurring during the Auger emission. By normalizing all spectra to their maxima, the similar shape of the Auger emission for all samples becomes clear (**Figure 5b**). Additionally, this peak shape is highly similar to the one observed for $SnO_2$. Indeed, the decomposition of Sn-based perovskites in air has been reported to proceed through the formation of both tin(IV) oxide (i.e. $SnO_2$) and their corresponding double perovskite (i.e. $A_2SnX_6$).[6] Thus, based on the Sn chemical state alone, tin(IV) oxide as the sole degradation product of Sn on the surface for all perovskite samples could be identified, regardless of the herein employed A- B- or X-site chemistry.

These findings clearly highlight how observation of the easily accessible Sn MNN feature can augment studies based on the Sn 3d feature alone. Combining both results in AP studies allows to eliminate issues resulting from charge referencing which are especially profound for hybrid semiconductors. Resulting shifts of $\alpha'$ are largely driven by a shift in the peak of the Sn MNN feature which highlights the higher sensitivity of the Auger emission to the chemical environment and Sn–X interactions for perovskites specifically.

## 3. Conclusions and Outlook

In this article, we established how AP studies can be used to perform more robust surface analysis of Sn-based perovskite surfaces using XPS. By performing comprehensive XPS analysis on high-quality reference materials, we identified the high sensitivity of $\alpha'$ on Sn–X interactions in perovskites. Additional dependencies on the Sn oxidation state were identified, albeit being less pronounced. Due to the aforementioned vast chemical space of Sn based semiconductors, similar studies, which include additional reference materials, may aid in the more robust characterization of these compounds. Likewise, the importance of B–X interactions unraveled for Sn-based perovskites are likely to be similar for other types of perovskites.

Since no additional equipment is required for AP studies, this method can be widely adopted for the surface and interface analysis of Sn semiconductors with minimal additional effort. The high tolerance of the AP studies towards erroneous energy calibration and artifacts resulting from surface band bending makes them an ideal tool to study semiconductor surfaces and compare data sets between laboratories. Furthermore, the high sensitivity of $\alpha'$ on Sn–halide interactions makes it especially interesting to study the effects on additives affecting the X-site at perovskite interfaces. Lastly, the facile data analysis makes this method also ideal for high-throughput studies which are increasingly being adopted in the development of new semiconducting materials.

## 4. Experimental section

### 4.1 Materials

25 mm×25 mm soda lime glasses were purchased from Advanced Election Technology Co., Ltd. Cesium iodide (CsI, 99%), methylammonium bromide (MABr, ≥98%), formamidinium bromide (FABr, ≥99%) were purchased from Tokyo Chemical Industry Co., Ltd. Methylammonium iodide (MAI, 98%), formamidinium iodide (FAI, ≥99.99%) were purchased from Greatcell Solar Ltd. Cesium bromide (CsBr, 99.999%), lead(II) iodide ($PbI_2$, 99.999%) and tin(IV) oxide ($SnO_2$, 15% $H_2O$ in colloidal dispersion) were purchased from Alfa-Aesar. Dimethyl sulfoxide (DMSO, anhydrous, 99.9%, anhydrous), dimethylformamide (DMF, 99.8% anhydrous), lead(II) thiocyanate ($Pb(SCN)_2$, (99.5%), tin(II) iodide ($SnI_2$, 99,99%), Ethyl acetate (EtOAc, 99.8%, anhydrous), tin(II) bromide ($SnBr_2$) were purchased from Sigma-Aldrich Pty Ltd. All the materials were used as received.

## 4.2 Sample preparation

**$ASnI_3$ perovskites, $ASnBr_3$ perovskites and $SnX_2$ references**

Stoichiometric amounts of A cation source chemicals (CsI, MAI, FAI, CsBr, MABr, or FABr) and corresponding B cation metal salts ($SnI_2$ or $SnBr_2$) were weighed in a $N_2$-filled glove box ($H_2O$<0.5 ppm. $O_2$<0.5 ppm) and dissolved in DMSO to yield 0.3 M $CsSnI_3$, $MASnI_3$, $FASnI_3$, $CsSnBr_3$, $MASnBr_3$, $FASnBr_3$ precursors, respectively. $SnI_2$ and $SnBr_2$ were also weighed separately to get 0.3 M $SnI_2$ and $SnBr_2$ solutions in DMSO, respectively. The precursors were kept on a 75 °C hotplate during the spin-coating. The hot precursors were spin-coated onto pre-cleaned soda lime glasses at 1500 rpm with a ramp rate of 1500 rpm for 30 s, followed by a 100 °C annealing for 10 min.

$SnI_2$ and $SnBr_2$ films were prepared by only spin-coating the $SnX_2$ solutions in DMSO. Similarly, $SnO_2$ films were directly prepared from the aqueous dispersion. The annealing step was likewise performed as specified for the perovskite films.

**$FA(Sn_{1-x}Pb_x)I_3$ perovskites**

The $FASnI_3$ precursor solution was prepared by dissolving 372 mg of $SnI_2$ and 172 mg of FAI in a mix of DMF (424 µL) and DMSO (212 µL) solvents. The $FAPbI_3$ precursor solution was prepared by dissolving 461 mg $PbI_2$ and 159 mg MAI with 3.5 mol% $Pb(SCN)_2$ in 630 µL DMF and 70 µL DMSO. The two solutions were heated and dissolved at 65 °C for 2 h. Then, stoichiometric amounts of $FASnI_3$ and $FAPbI_3$ perovskite precursors were mixed to obtain 1.5M $FA(Sn_{1-x}Pb_x)I_3$ perovskites (x = 0.25, 0.5, and 0.75) precursor solutions.

The perovskite films were deposited using a two-step spin coating procedure: (1) 1000 rpm for 10 s with an acceleration of 1000 rpm $s^{-1}$ and (2) 5000 rpm for 50 s with a ramp-up of 10000 rpm $s^{-1}$. EtOAc (150 µL) was dropped onto the spinning substrate during the second spin-coating step at 20 s before the end of the procedure. The substrates were then transferred on a hotplate and annealed at 65 °C for 3 min and then 105 °C for 7 min.

### 4.3 Sample characterization

X-ray photoelectron spectroscopy was performed in a PHI Quantum system to which samples were transferred using a custom-made inert-gas transfer vessel in Ar atmosphere. XPS measurements were performed at a pressure of $10^{-9} - 10^{-8}$ Torr. The monochromatic Al Kα radiation was generated from an electron beam at a power of 12.6 W and a voltage of 15 kV. To minimize beam damage during measurements, the beam spot with a diameter of 50 µm was continuously scanned over an area of 500 × 1000 $\mu m^2$. Charge neutralization was performed using a low-energy electron source. Short-term measurements (<3 min) of the Sn 3d core level before and after each presented measurement were conducted to rule out changes in the chemical state due to X-ray induced beam damage. The binding energy scale was reference to the main component of adventitious carbon at 284.8 eV, resulting in a typical inaccuracy of ±0.2 eV. Peak fitting of photoelectron features was performed in Casa XPS using Voigt profiles with GL ratios of 60 following Shirley-background subtraction. For Auger electron features, the area around the peak was fit using a cubic function. This was performed twice while shifting the area within approximately 80% of the peaks' full width at half maximum. The resulting maximum was then calculated based on the average of the maxima of the fitted functions while the error margin was calculated based on the deviation between these values. Atomic ratios were calculated using the instrument specific relative sensitivity factors. To estimate the information depth depending on the observed feature, the inelastic mean free path (IMFP) was calculated from the kinetic energy of the detected electrons based on the Tanuma, Powell, Penn formula.[39] A detailed descripted of the XPS characterization workflow for the herein investigated Sn-based perovskites can be found in the Supporting Information (SI).

XRD analysis was performed in a Bruker D8 Discovery diffractometer using Cu kα radiation in Bragg-Brentano geometry. The full conversion of each Sn-based perovskite sample to the perovskite phase was confirmed by the absence of signals related to $SnI_2$ or $SnBr_2$.

ASSOCIATED CONTENT

**Supporting Information**.

Additional XPS spectra, detailed procedures and measurements taken to prevent sample degradation during measurements, full XPS and XRD characterization of each reference sample (PDF)

AUTHOR INFORMATION

Corresponding Author

Correspondence regarding the synthesis of the reference Sn-based perovskites should be directed at fan.fu@empa.ch. Correspondence regarding XPS/Auger parameter analysis should be directed at sebastian.siol@empa.ch.

Author Contributions

Conceptualization and design of the study was performed by A.W., F.F. and S.S. Synthesis of the reference materials was performed by H.L. and J.P. under supervision of F.F. XPS analysis, as well as XRD analysis of the reference materials was performed by A.W. under supervision of S.S. Analysis and interpretation of the XPS results was performed by A.W. and S.S. A.W. wrote the initial draft of the manuscript with contributions from S.S. All authors edited and contributed to the final version of the manuscript. All authors have given approval to the final version of the manuscript.

Funding Sources

Financial support from the Strategic Focus Area – Advanced Manufacturing (SFA-AM) through the project Advancing manufacturability of hybrid organic-inorganic semiconductors for large area optoelectronics (AMYS) is acknowledged by all authors.

## ACKNOWLEDGMENT

All author acknowledge funding from the the Strategic Focus Area – Advanced Manufacturing (SFA-AM) through the project Advancing manufacturability of hybrid organic-inorganic semiconductors for large area optoelectronics (AMYS). A.W. acknowledges Jyotish Patidar for his support with the XRD measurement setup. H. L. thanks the funding of China Scholarship Council (CSC) from the Ministry of Education of P. R. China.

## ABBREVIATIONS

$MA^+$, methylammonium; $FA^+$, formamidinium; XPS, X-Ray photoelectron spectroscopy; XRD, X-Ray diffraction; NMR, nuclear magnetic resonance spectroscopy; $E_b$, binding energy; $E_{kin}$, kinetic energy, $\alpha'$ modified Auger parameter; UHV, ultra-high vacuum; ARXPS, angle-resolved X-Ray photoelectron spectroscopy.

**For Table of Contents Only**

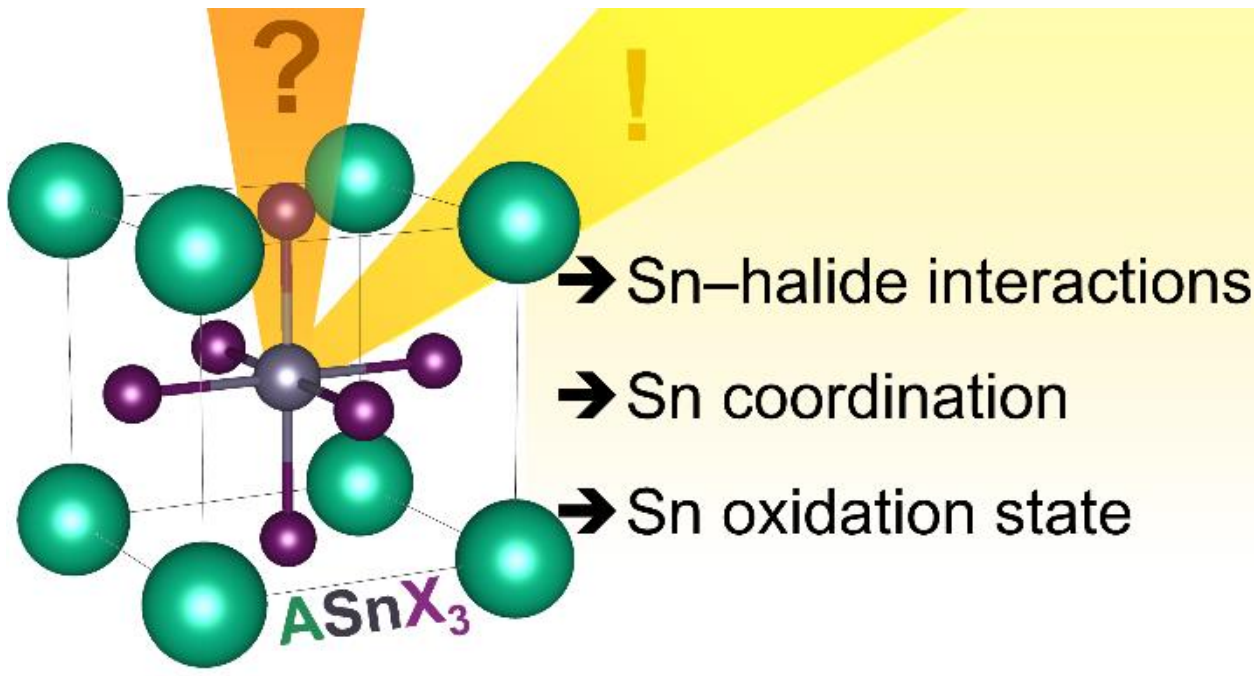

Supporting information of

# Resolving oxidation states and Sn–halide interactions of perovskites through Auger parameter analysis in XPS

*Alexander Wieczorek[1], Huagui Lai[2], Johnpaul Pious[2], Fan Fu*[2], Sebastian Siol*[1]*

[1]Laboratory for Surface Science and Coating Technologies, Empa − Swiss Federal Laboratories for Materials Science and Technology, Switzerland.

[2]Laboratory for Thin Films and Photovoltaics, Empa − Swiss Federal Laboratories for Materials Science and Technology, Switzerland.

Corresponding authors: fan.fu@empa.ch, sebastian.siol@empa.ch

**Table S 1.** Values for the Sn $M_4N_{4,5}N_{4,5}$ feature maximum $E_{kin}$(Sn $M_4N_{4,5}N_{4,5}$) and binding energy of the Sn $3d_{5/2}$ feature maximum $E_b$(Sn $3d_{5/2}$) as well as the modified Auger parameter $\alpha'$ for each measured reference compound.

| **Compound** | $E_b$**(Sn 3d$_{5/2}$) / eV** | $E_{kin}$**(Sn $M_4N_{4,5}N_{4,5}$) /eV** | **α' / eV** |
|---|---|---|---|
| **$SnI_2$** | 486.7 ± 0.0 | 434.4 ± 0.1 | 921.1 ± 0.1 |
| **$SnBr_2$** | 485.1 ± 0.0 | 433.5 ± 0.2 | 918.6 ± 0.2 |
| **$CsSnI_3$** | 485.9 ± 0.1 | 436.0 ± 0.2 | 922.0 ± 0.3 |
| **$MASnI_3$** | 486.5 ± 0.3 | 435.6 ± 0.3 | 922.1 ± 0.4 |
| **$FASnI_3$** | 486.3 ± 0.7 | 436.0 ± 0.4 | 922.3 ± 0.8 |
| **$CsSnBr_3$** | 486.1 ± 0.4 | 434.2 ± 0.3 | 920.3 ± 0.5 |
| **$MASnBr_3$** | 486.6 ± 0.6 | 433.5 ± 0.2 | 920.1 ± 0.7 |
| **$FASnBr_3$** | 486.8 ± 0.4 | 433.1 ± 0.2 | 919.9 ± 0.5 |
| **$FASn_{0.25}Pb_{0.75}I_3$** | 486.5 ± 0.2 | 436.6 ± 0.2 | 923.1 ± 0.3 |
| **$FASn_{0.5}Pb_{0.5}I_3$** | 486.2 ± 0.0 | 436.9 ± 0.3 | 923.1 ± 0.3 |
| **$FASn_{0.75}Pb_{0.25}I_3$** | 486.1 ± 0.1 | 437.0 ± 0.2 | 923.1 ± 0.3 |

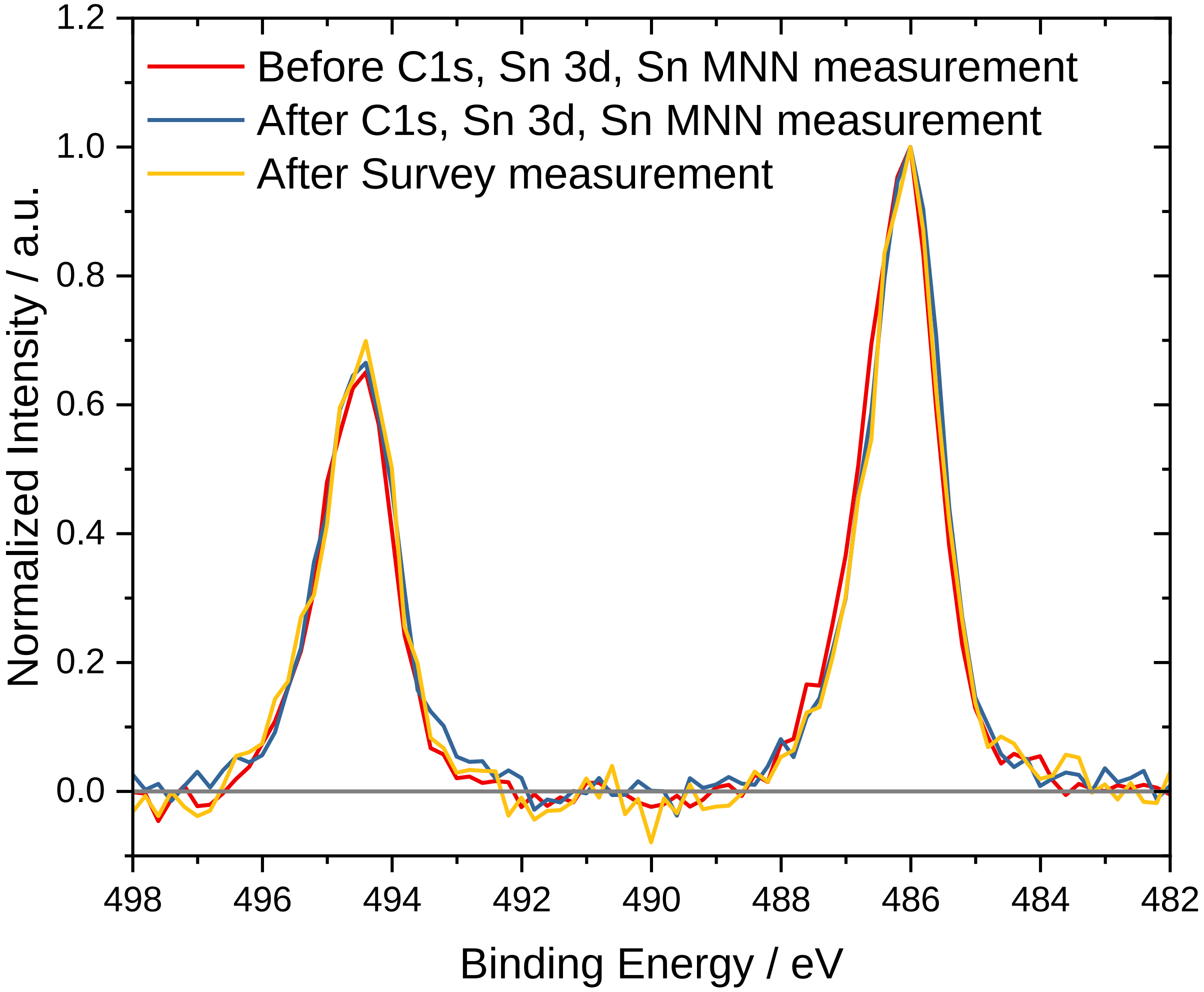

**Figure S 1.** Exemplary short-term measurements performed of the Sn 3d region on $CsSnI_3$ before and after each long-term measurement. Most importantly, due to the effectively identical peak shape during each measurement, no significant beam damage during the long-term XPS measurements of the C 1s, Sn 3d and Sn MNN regions occurred. Similar to reports on Pb perovskites,[40] the clear emergence of a separate Sn(0) feature would be expected as a result of X-Ray induced photolysis. For the herein presented spectra, charge referencing based on the C 1s feature determined during the long-term measurement was performed. Further detailed information on the measurement workflow and detection of possible surface oxides can be found in the chapter "Detailed measurement workflow for the determination of the Auger parameter".

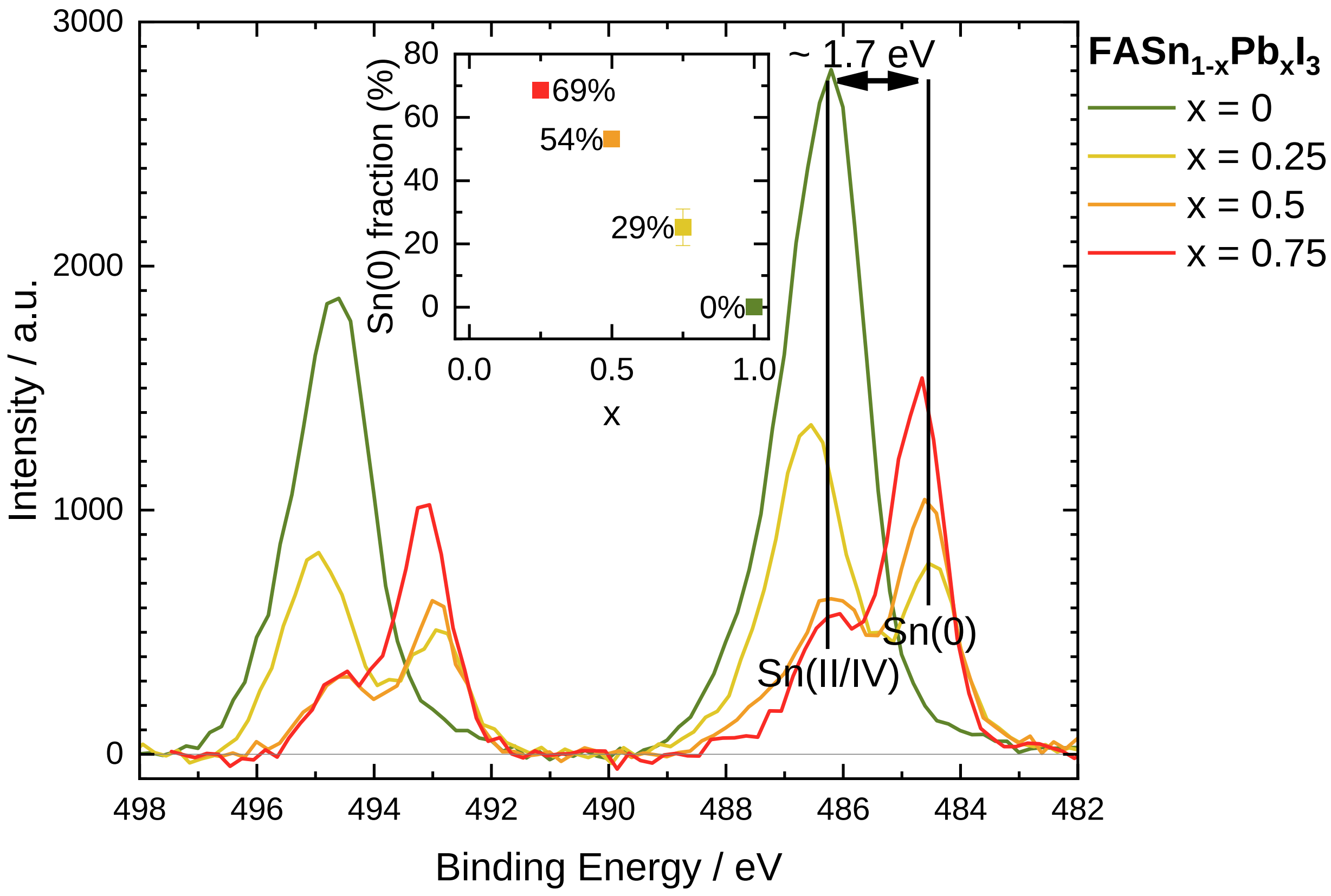

**Figure S 2.** Sn 3d regions of FA($Sn_{1-x}Pb_x$)$I_3$ type perovskites for x = 0, 0.25, 0.5, 0.75. The inset depicts the fraction of Sn(0) from Sn(total) found in these compounds based on the fitting of the Sn(II/IV) 3d and Sn(0) 3d features. While XPS spectra indicating the presence of Sn(0) on Sn-Pb perovskites have been reported before, the signals were wrongfully attributed to the Sn(II) and Sn(IV) species. As presented here, the peak separation of approx. 1.7 eV clearly indicates the presence of Sn(0) and Sn(II/IV) in alignment with reported NIST values.[20]

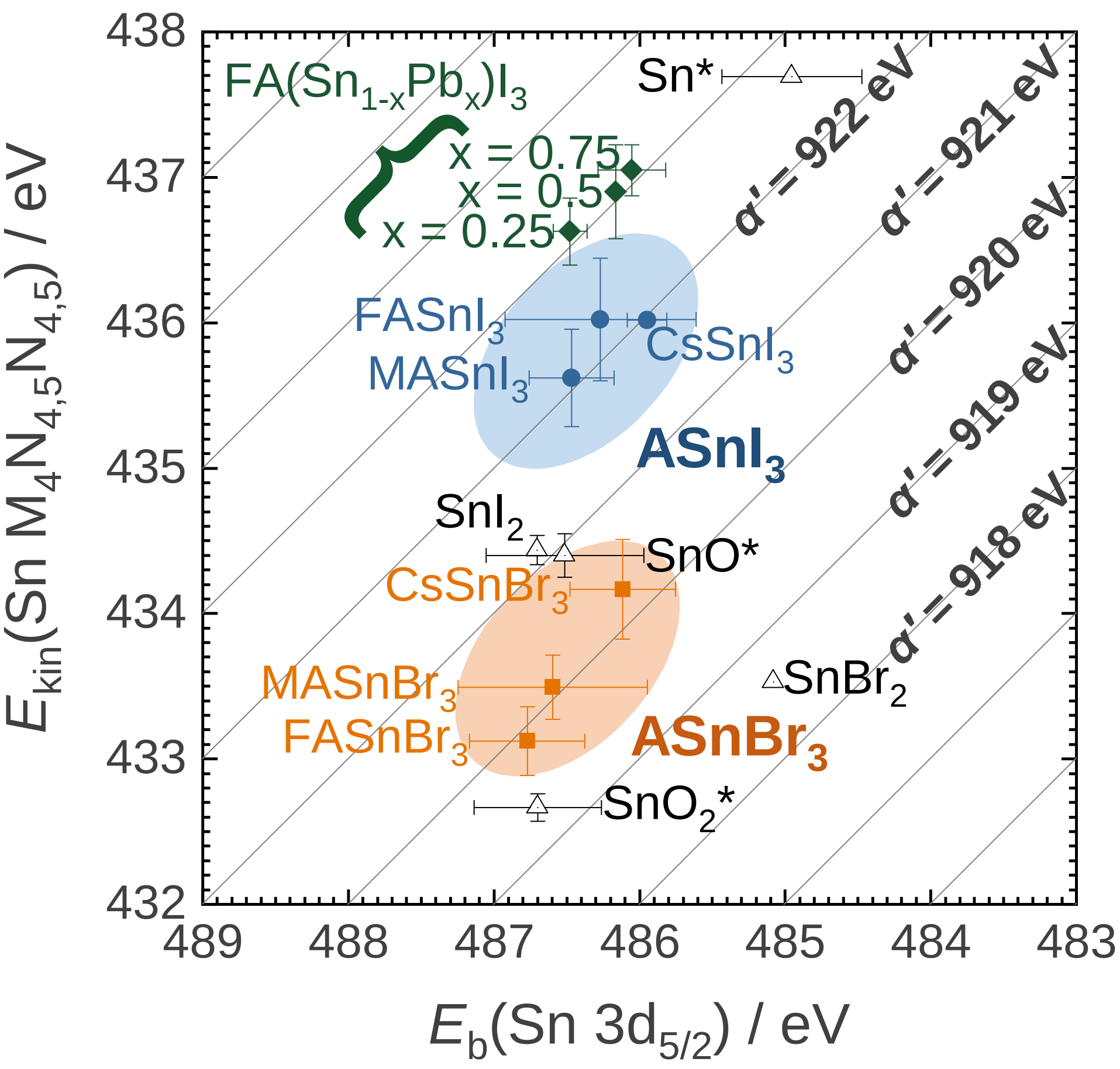

**Figure S 3.** Wagner plot depicting all reference Sn-based perovskites of this study. Most strikingly, clear regions for $ASnI_3$ and $ASnBr_3$ perovskites could be identified on the Wagner plot. The difference of the AP by $\Delta\alpha' \approx 2$ eV highlights its sensitivity for Sn–X interactions. The $FA(Sn_{1-x}Pb_x)I_3$ were found to exhibit varying degrees of Sn(0), which is also reflected in the increasing Auger electron kinetic energy with increasing alloying concentration. $Pb(SCN)_2$ was used in trace amounts as an additive for the preparation of $FA(Sn_{1-x}Pb_x)I_3$ perovskites which might result in marginally altered X-site interactions. $E_{kin}$ error bars were derived from the estimated error for each peak fit as well as the standard deviation of maxima across measurements. $E_b$ error bars were derived from the standard deviation of the maxima across measurements alone. Reference values obtained from the NIST database are marked with an asterisk and related error bars are calculated based on the standard deviation between these values.

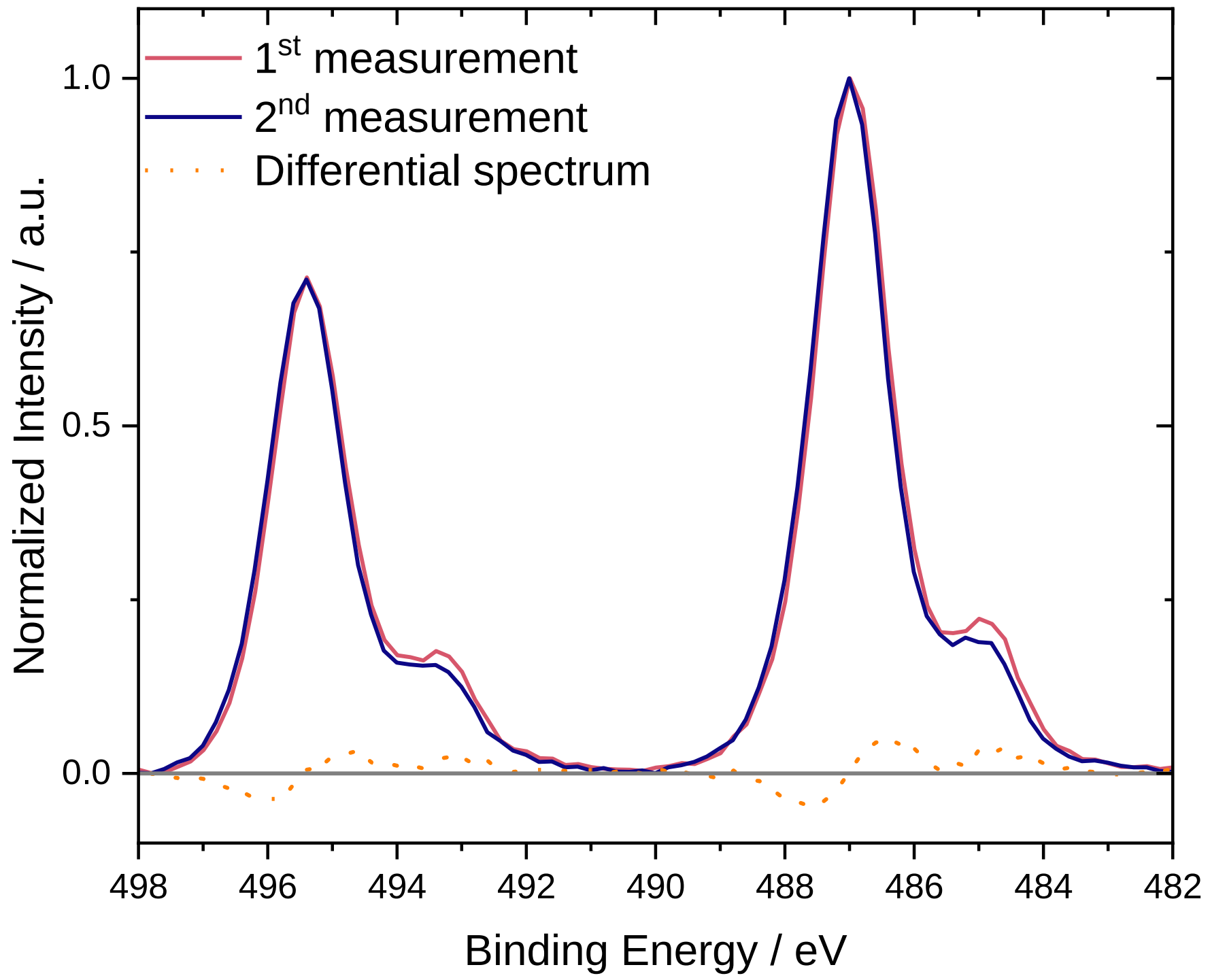

**Figure S 4.** Repeated long-term measurements of the Sn 3d region measured at an emission angle of 15 °. As indicated by the differential spectrum, a minor increase of the Sn(0) was found, in alignment with minor $SnBr_2$ photolysis due to X-Ray beam damage over these extended irradiation times on the same sample surface area.

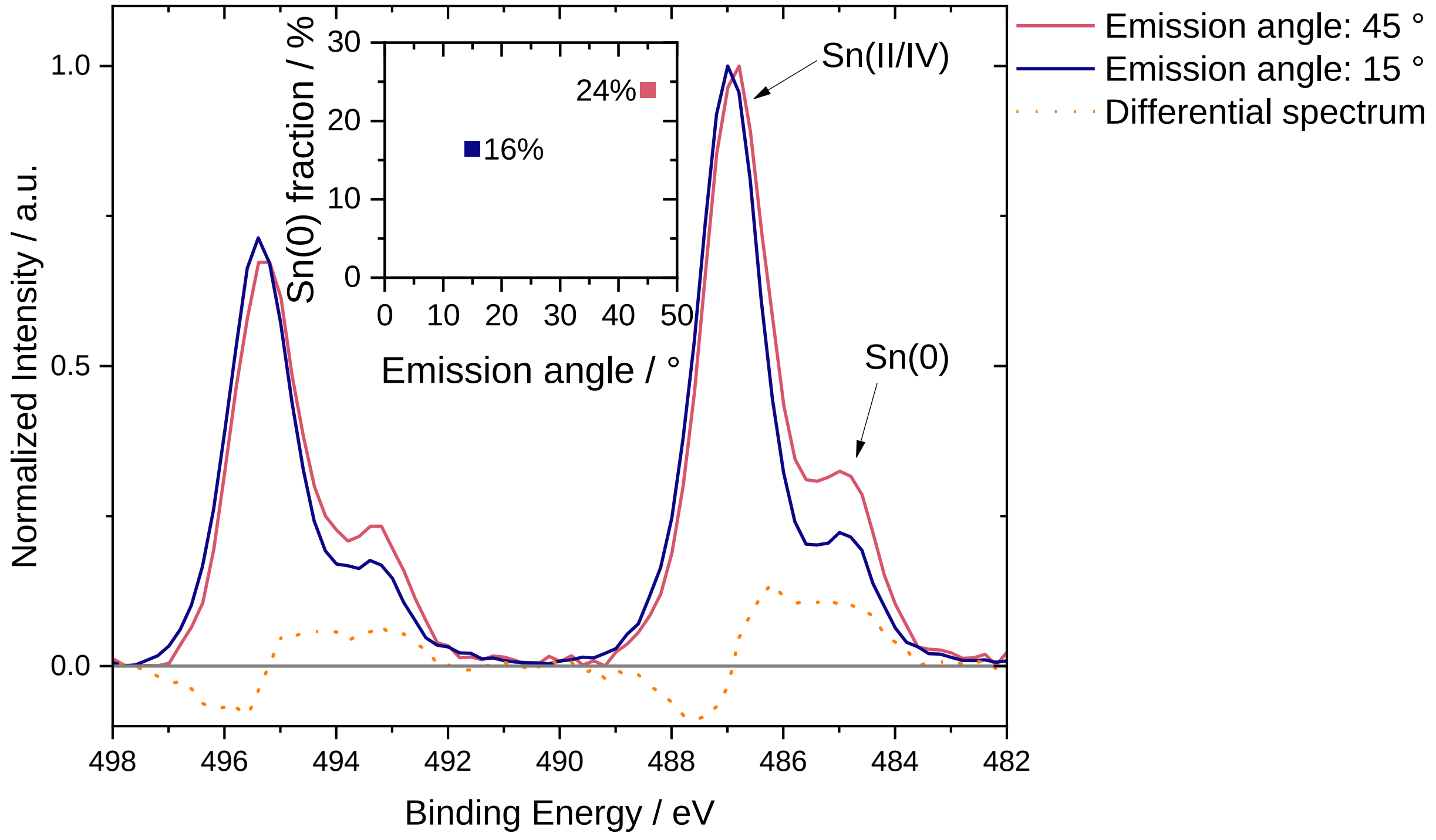

**Figure S 5.** Normalized spectra from the ARXPS measurements of the Sn 3d region on $SnBr_2$. The emission angle of 45 ° was used for the other measurements of this study and corresponds to a probing depth of 4.6 nm. In contrast, the emission angle of 15 ° corresponds to a deeper probing depth of 6.3 nm. The depicted differential spectrum indicates a larger difference between directly followed measurements at varied angles than those subsequently performed at the same emission angle (**Figure S 4**).

**Detailed measurement workflow for the determination of the Auger parameter**

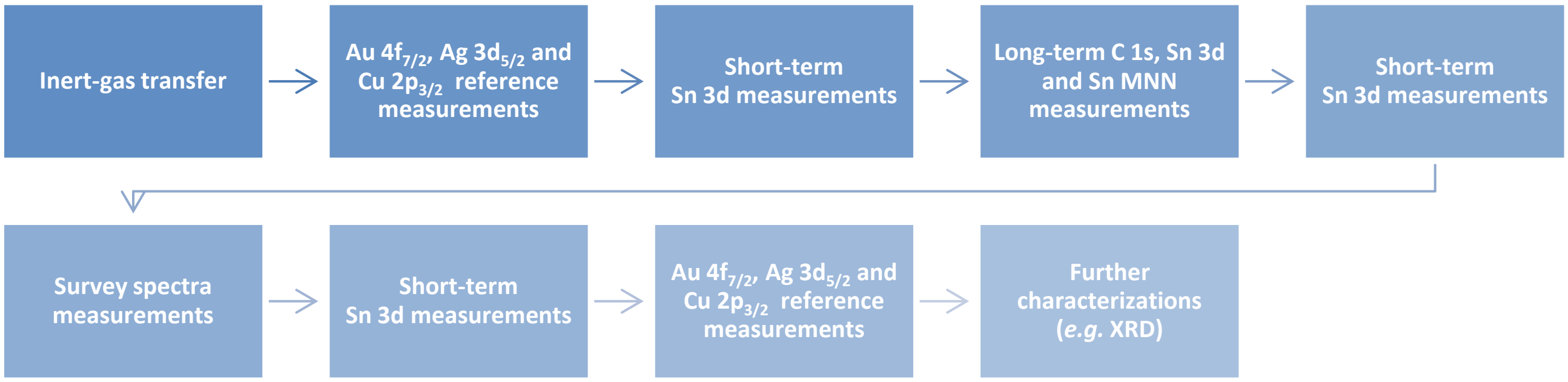

**Figure S 6.** Flow-chart depicting the major measurement steps for the herein reported modified Auger parameter study.

To ensure the reproducibility of results as well as relevance of the measured spectra for pristine samples, measures were taken to minimize materials degradation during the measurement process and reduce artifacts due to degradation (**Figure S 6**). First, each sample was directly transferred from the glovebox to the XPS introduction chamber using a custom-made inert-gas vessel in Ar atmosphere. This served to prevent the rapid formation of surface oxides during sample transfer.

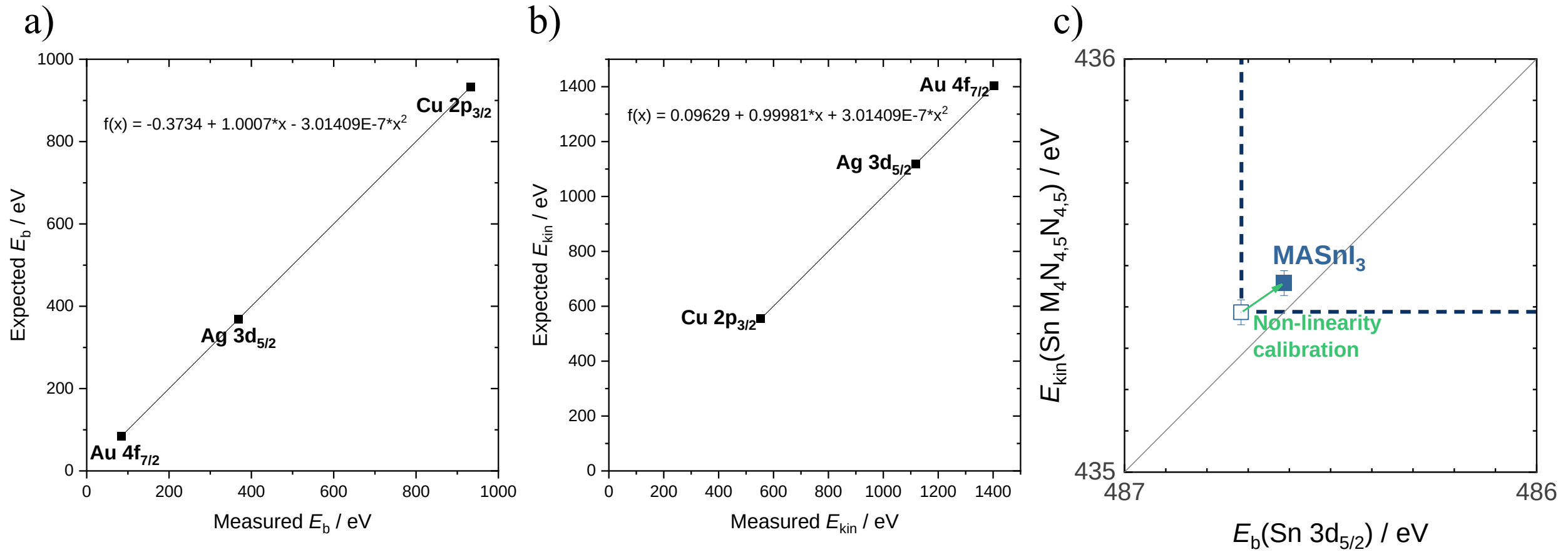

**Figure S 7.** Correlation of measured and expected values of Au $4f_{7/2}$, Ag $3d_{5/2}$ and Cu $2p_{3/2}$ maxima on the a) $E_b$ and b) $E_{kin}$ scale. Based on this relationship, a quadratic function was calculated to linearize each energy scale and minimize instrument-related errors during the modified Auger parameter study. c) Data point for single measurement of $MASnI_3$ on the Wagner plot before (□) and after (■) linearization of the energy scales. Following the introduction of samples into the XPS main chamber, Au $4f_{7/2}$, Ag $3d_{5/2}$ and Cu $2p_{3/2}$ standards were measured. Their expected $E_b$ and $E_{kin}$ values for XPS measurements with Al kα radiation are well defined according to ISO #15472 as reported elsewhere.[17] The resulting calibration curve was modelled using a quadratic function (**Figure S 7.b+c**) and further used to correct the later recorded $E_b$ and $E_{kin}$ positions of maxima as shown on the Wagner plot (**Figure S 7.c**). For each perovskite sample measured, the Sn 3d core-level were quickly measured before the analysis within a measurement time of < 1 min.

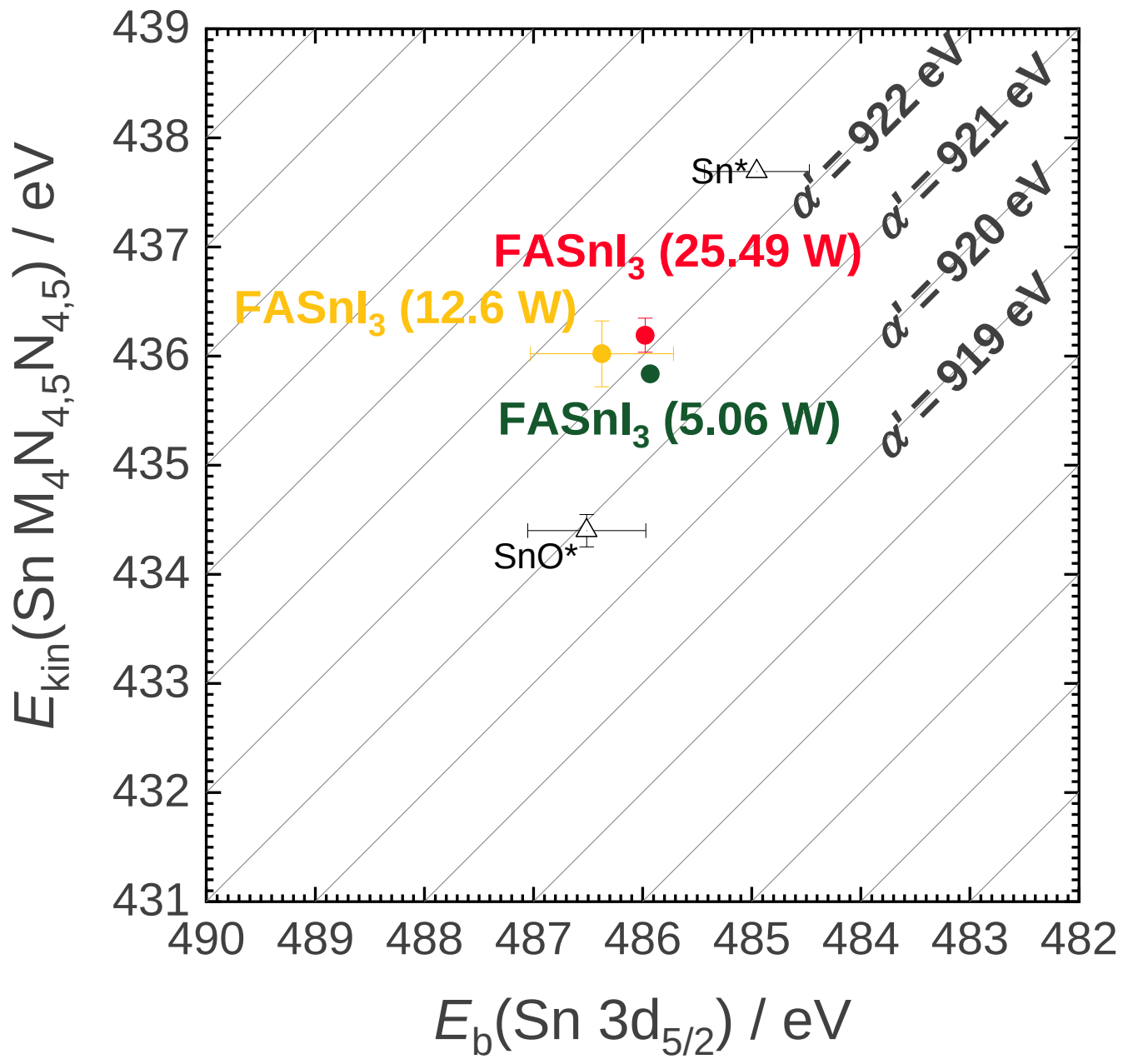

**Figure S8.** Wagner plot depicting Auger parameters of $FASnI_3$ determined using varied X-Ray powers. The data point at 12.6 W represents the conditions used for all other depicted measurements of this study and includes error estimation based on multiple measurements. The results of single measurements at lower (5.06 W) and higher (25.49 W) X-Ray powers are well within these error bounds. The clustering around $\alpha' = 922$ eV is in alignment with the modified Auger parameter regime identified for $ASnI_3$ type perovskites in this study.

Then, using a reduced X-Ray beam power of 12.6 W and a voltage of 15 kV, longer measurements of the C 1s, Sn 3d and Sn MNN regions were performed within approx. 30 min. The scanning of the 50 µm beam over an area of 500 × 1000 $\mu m^2$ served to reduce the overall radiation exposure of the sample during this process. As demonstrated for UPS measurements on perovskites, surface photo voltage generation may induce probing depth dependent shifts.[11] To verify the absence of similar effects of this study, the X-Ray flux was varied, with marginal effects on the resulting values below the herein presented error bounds (**Figure S8**).

Following this long-term measurement, the short-term measurement of the Sn 3d core-level was repeated. By comparing the spectra before and after the long-term measurement, the stability of the organic fraction as well as the Sn fraction was ensured. Since radiation damage of perovskites is well reported to result in the formation of metallic B-sites,[40] the comparison of the Sn 3d core-level alone is well-suited to detect the change of the chemical state during measurements.

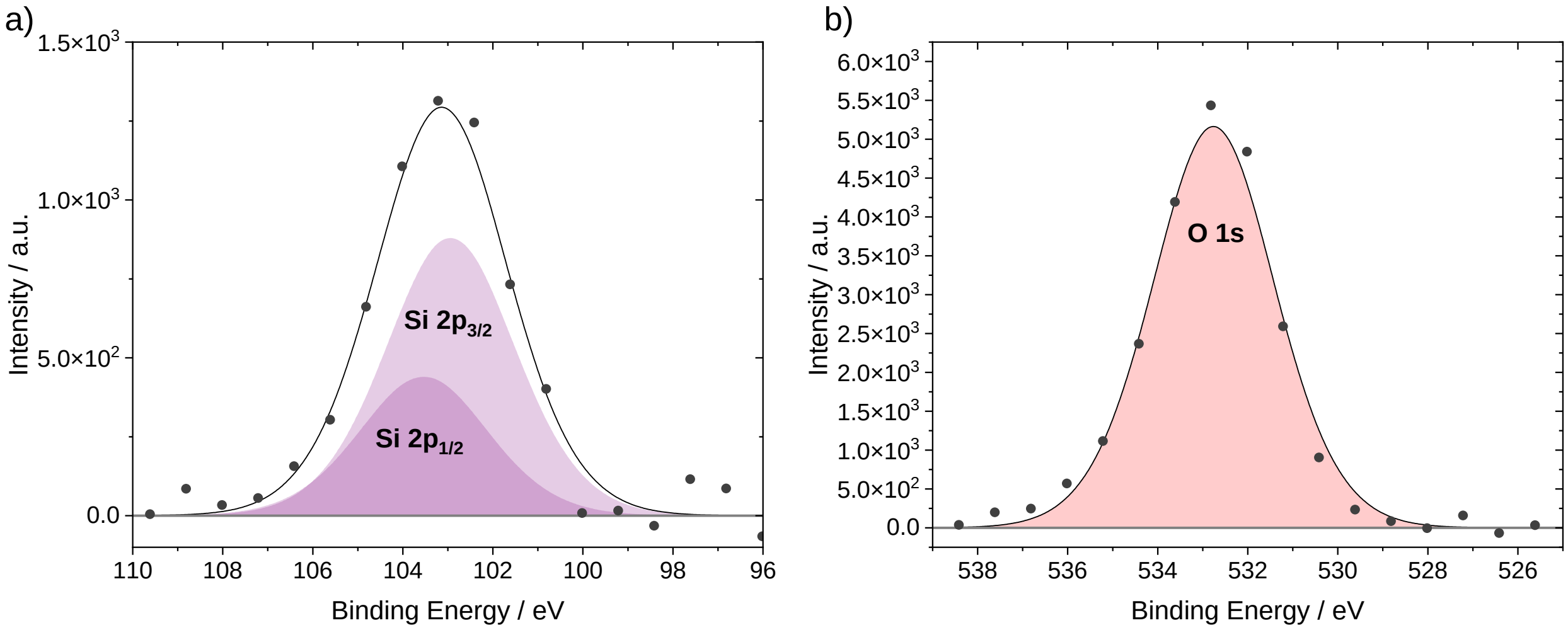

**Figure S 9. a)** Si 2p and **b)** O 1s regions of a Survey measurement performed on a $CsSnI_3$ sample. Based on the fitted Si 2p region (purple areas), the Signal area of the O 1s region was constructed (red area) using the device-specific relative sensitivity factors.

Survey-spectra ranging from -5 to 1200 eV were then recorded to obtain information about the chemical species found on the perovskite surface. In multiple instances, signals related to the used soda-lime glass substrate (e.g. Si 2p, O 1s) were identified. These are likely to result from pin-holes in the perovskite film over the scanned surface area. To ensure the detected oxygen related to this substrate-signal, the relative sensitivity factors were used to calculate the area of the O 1s signal based on the Si 2p signal (**Figure S 9**). In all instances, > 99% of the observed O 1s signal could be accounted for this way, strongly suggesting no significant formation of surface oxides.

This procedure should be used to validate alternative custom inert-gas transfers of perovskite samples from the glovebox to the XPS introduction chamber in other laboratories.

Following this survey measurement, another short-term measurement of the C 1s and Sn 3d core-level was performed. Since the chemical state is typically not determined for Survey measurements, the stability of the overall signal intensities before and after the measurement is more important than the emergence of new signals. To test the stability of results determined by the electron energy analyzer during each measurement session, repeated measurements of the Au $4f_{7/2}$, Ag $3d_{5/2}$ and Cu $2p_{3/2}$ standards may be performed.

Lastly, following these XPS measurements, additional characterizations may be performed to validate the crystal structure or composition of the measured samples. For this investigation, the formation of the perovskite phase was verified using X-Ray Diffraction.

## Characterization of the reference samples

$SnI_2$

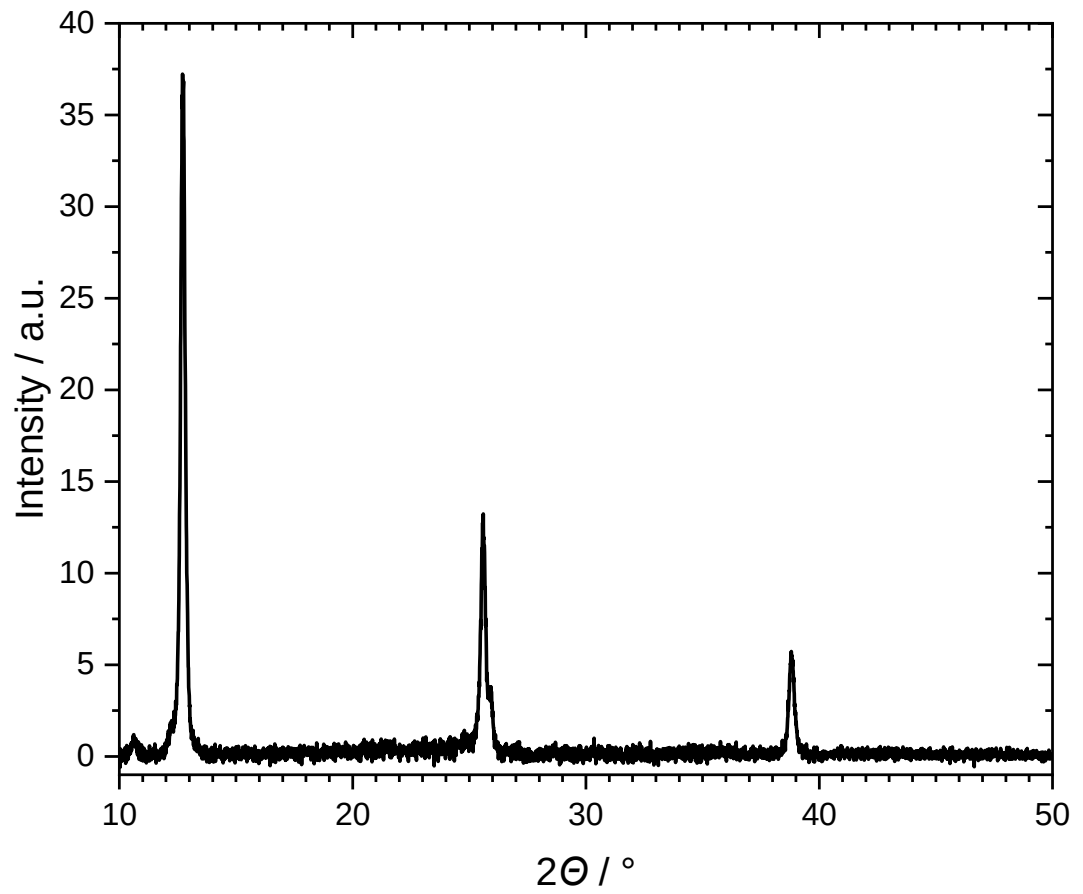

XRD diffractogram of $SnI_2$.

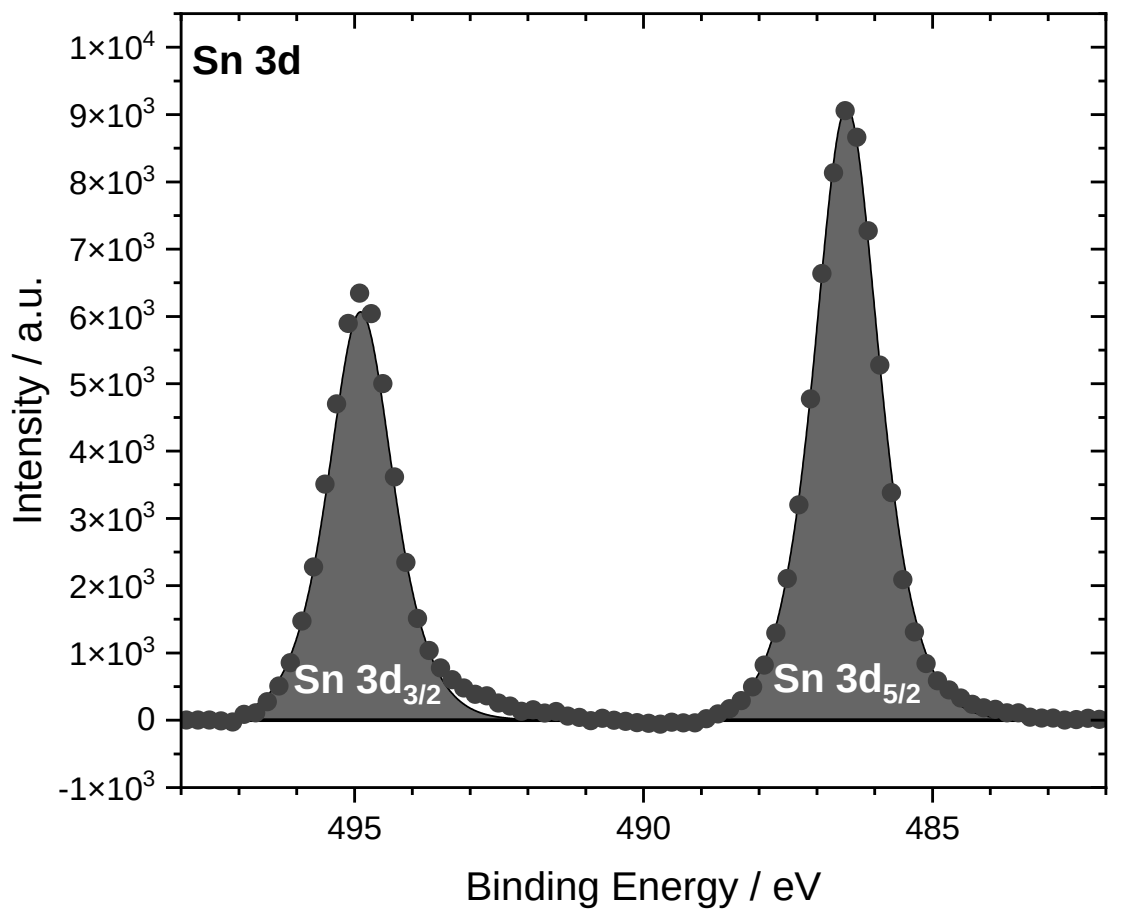

Sn 3d region of $SnI_2$.

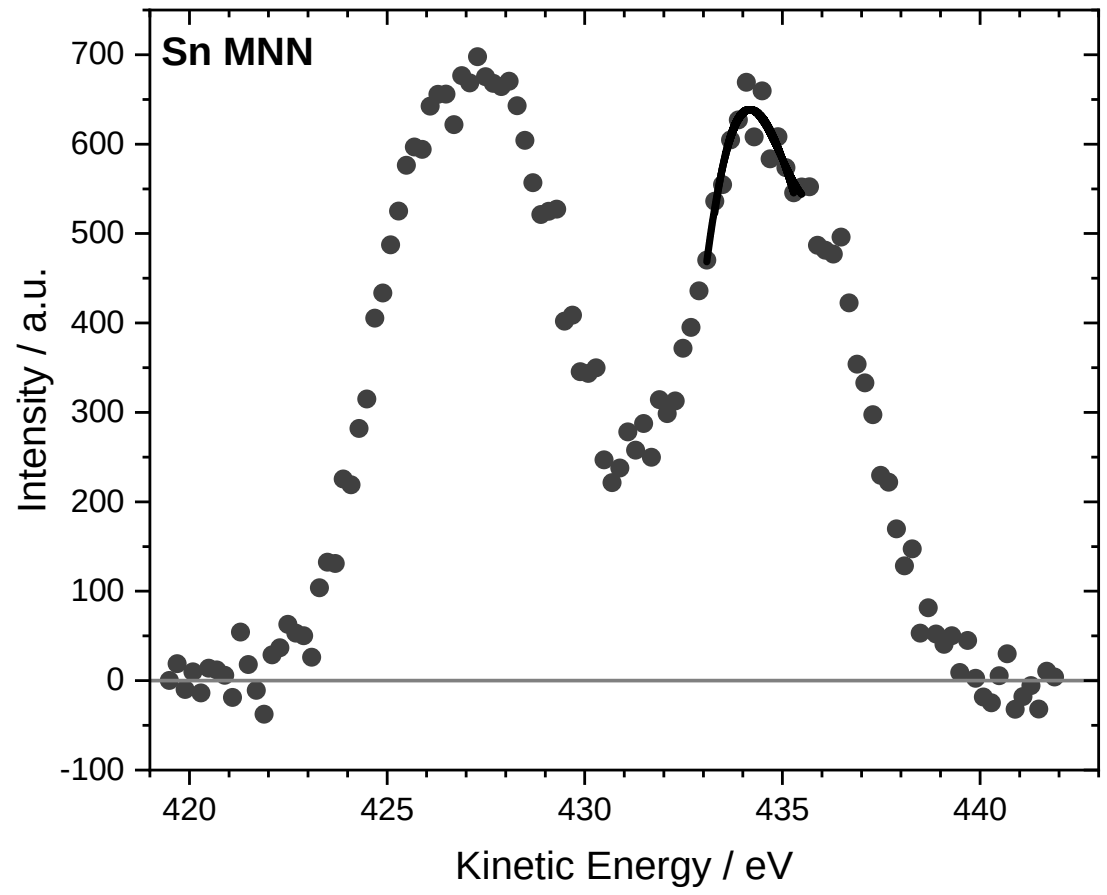

Sn MNN region of $SnI_2$.

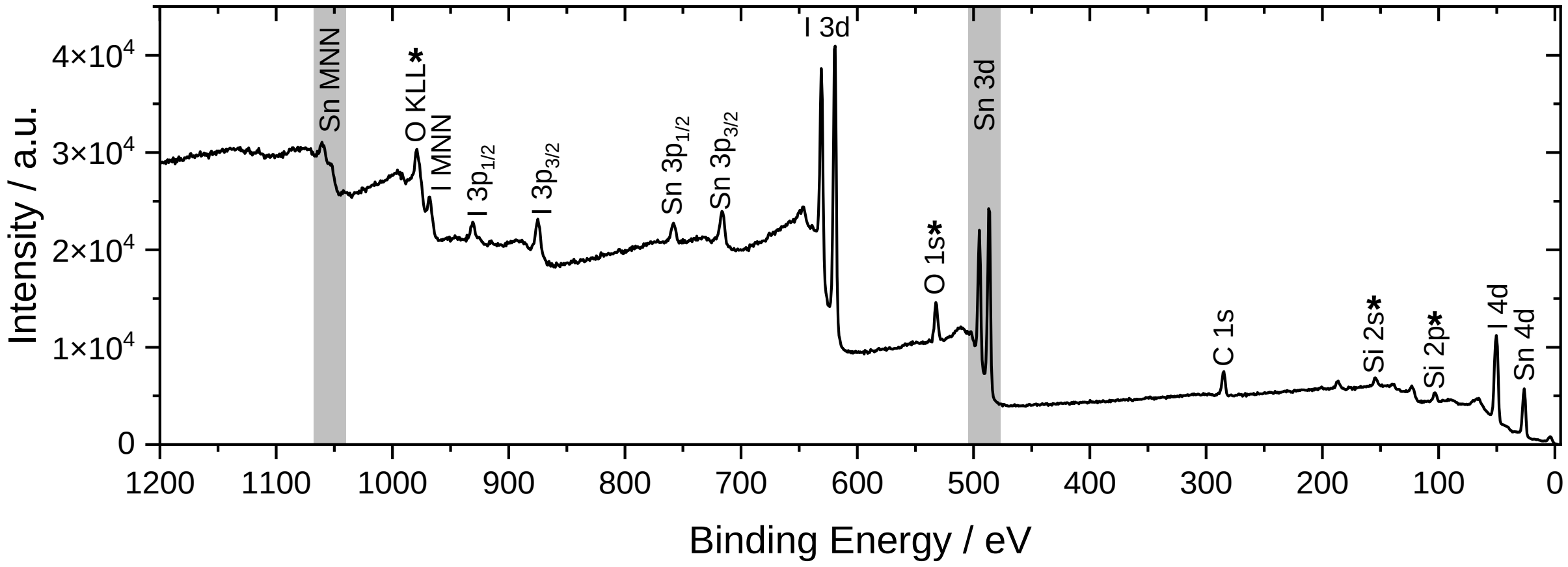

Survey spectrum of $SnI_2$.

Peaks assigned to the probing of the substrate through pin-holes are marked with an asterisk.

SnBr2

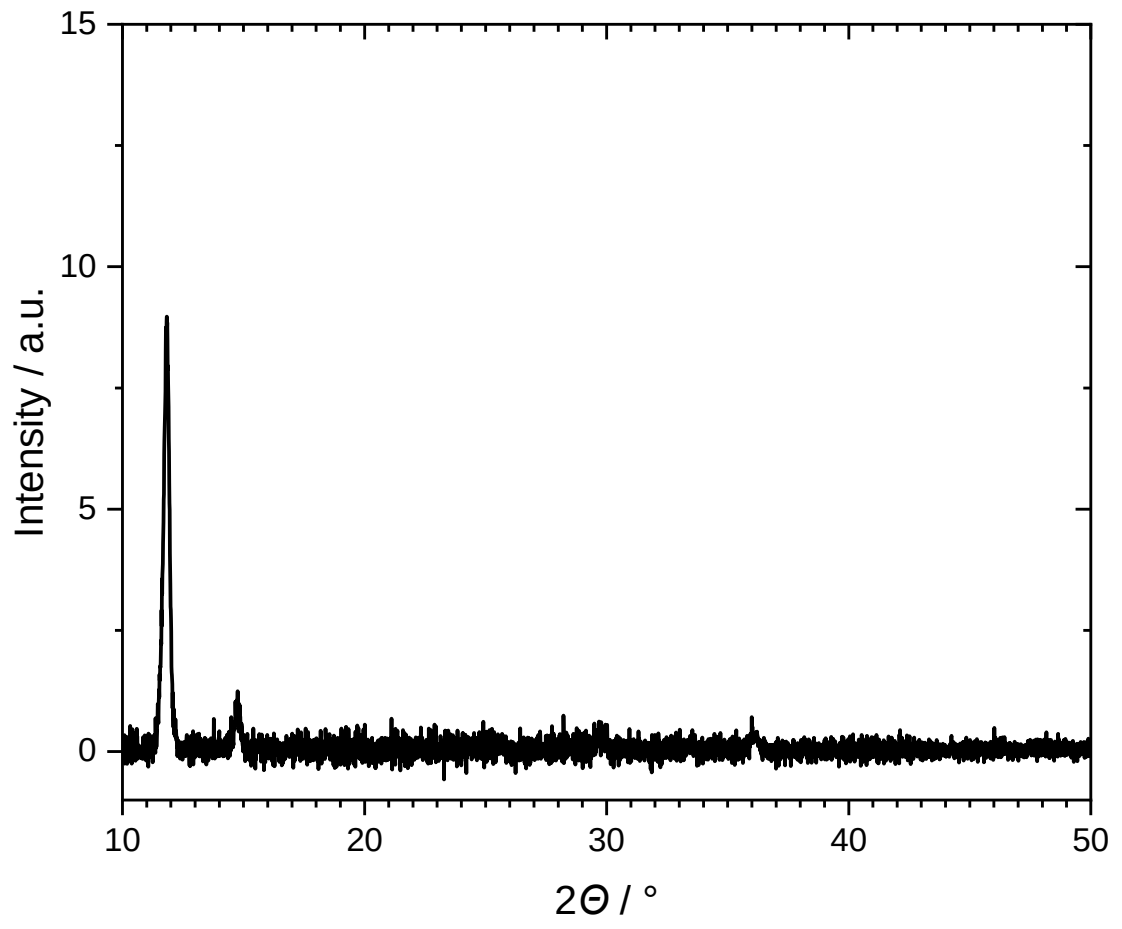

XRD diffractogram of $SnBr_2$.

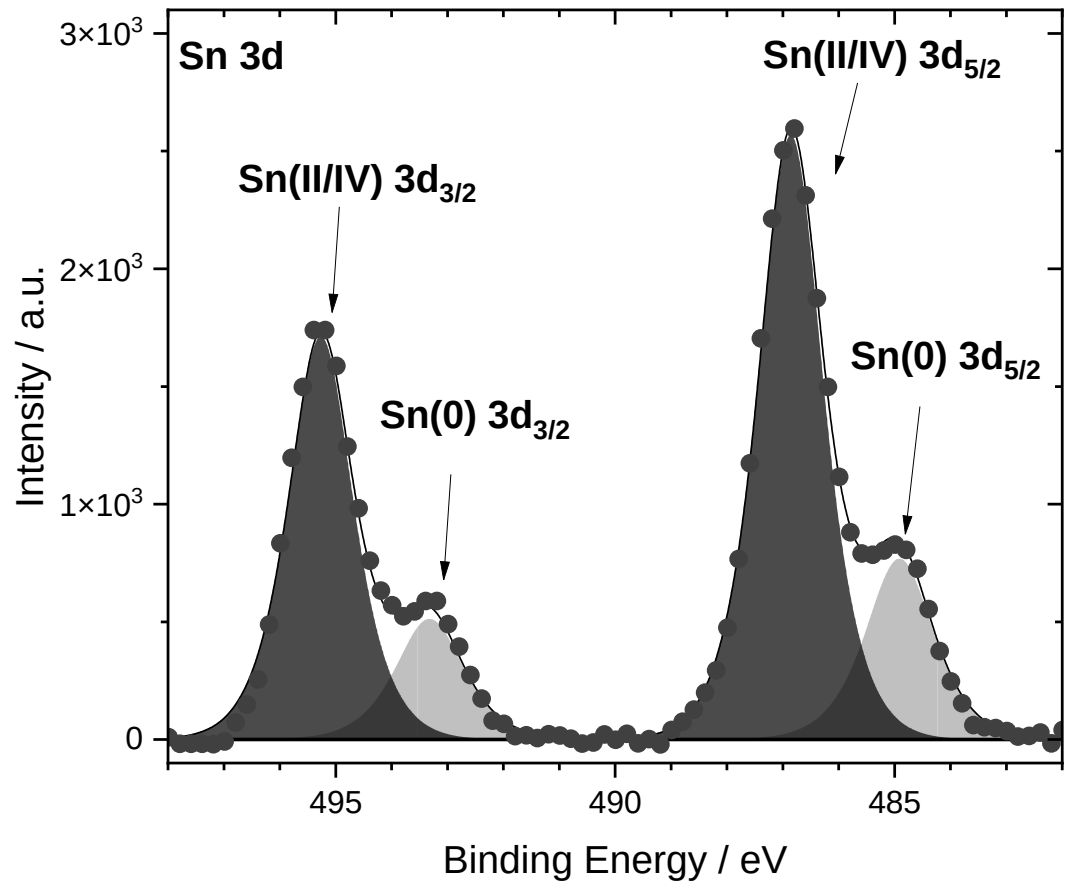

Sn 3d region of $SnBr_2$.

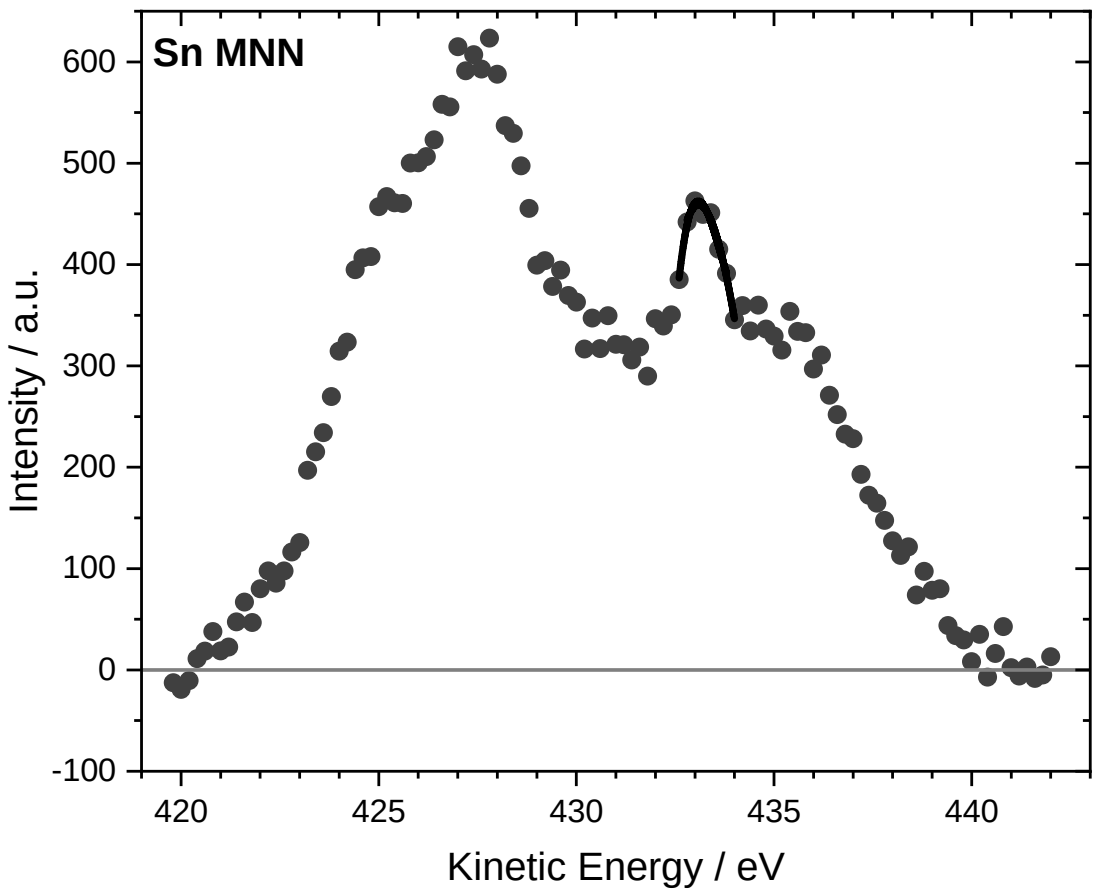

Sn MNN region of $SnBr_2$.

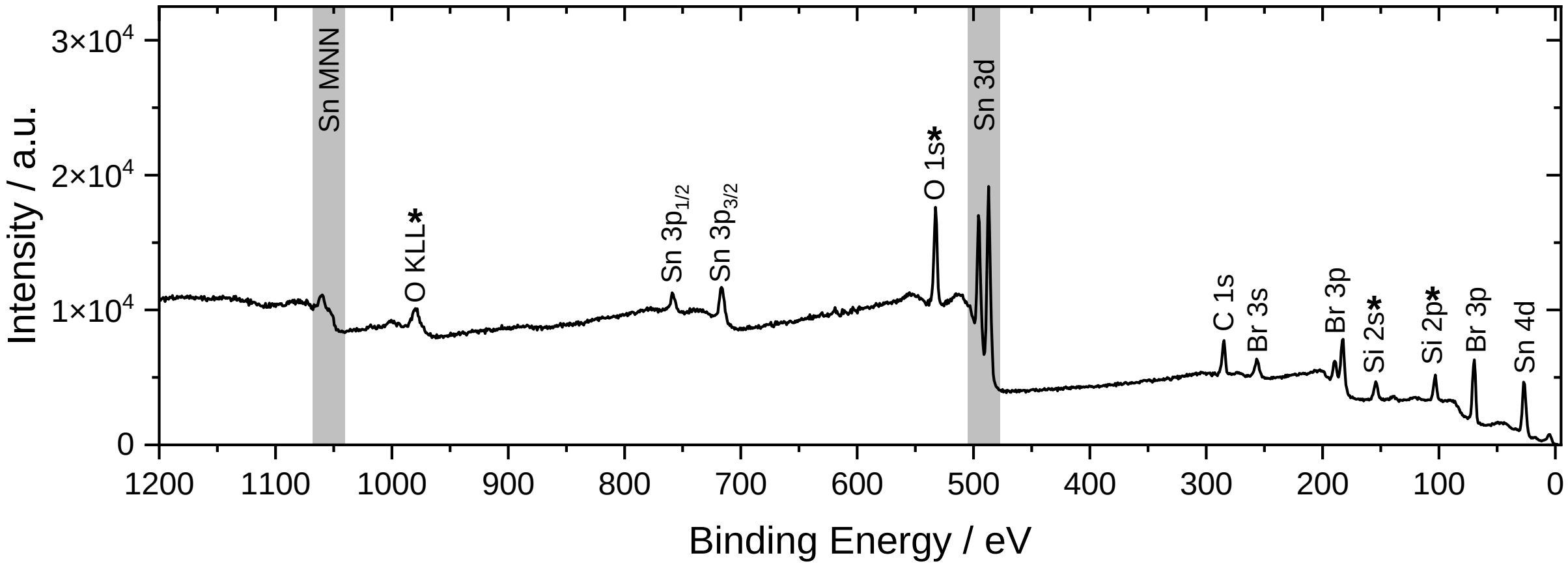

Survey spectrum of $SnBr_2$.

Peaks assigned to the probing of the substrate through pin-holes are marked with an asterisk.

CsSnI3

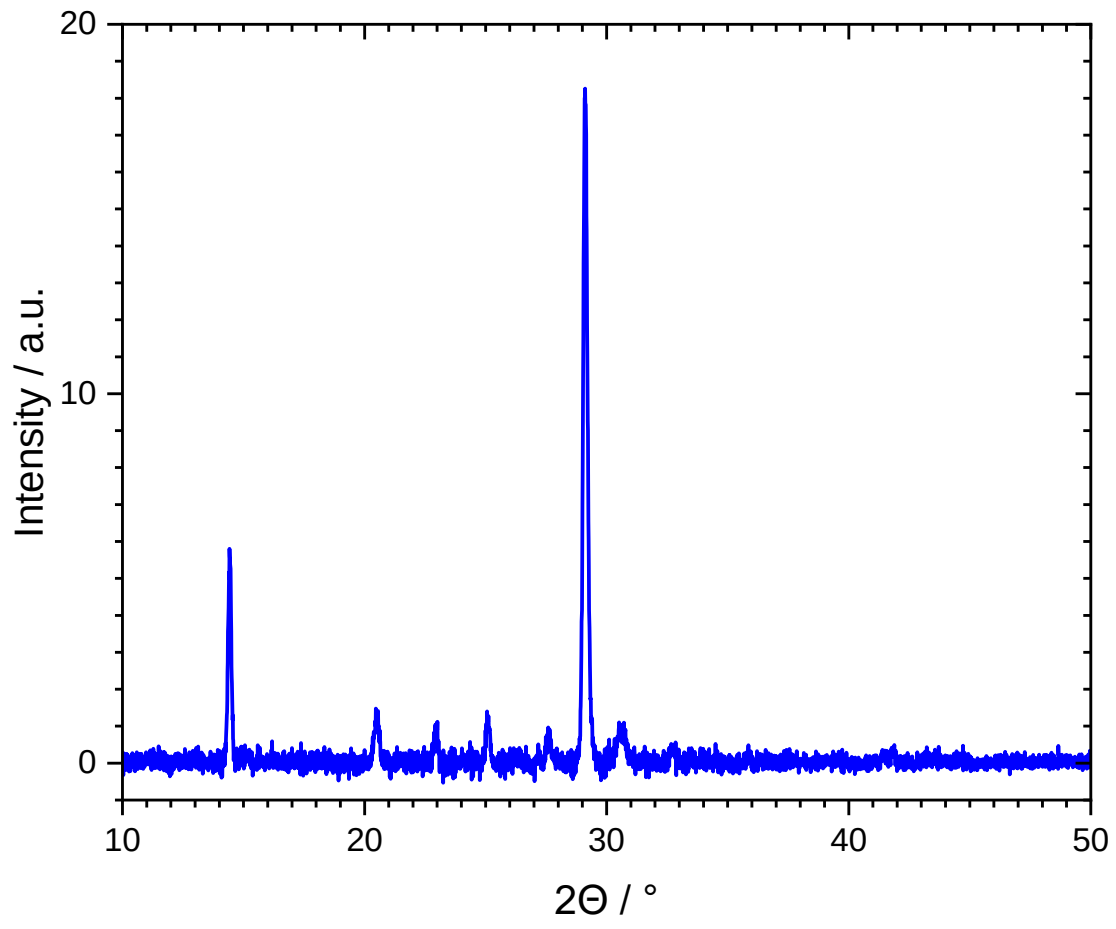

XRD diffractogram of $CsSnI_3$.

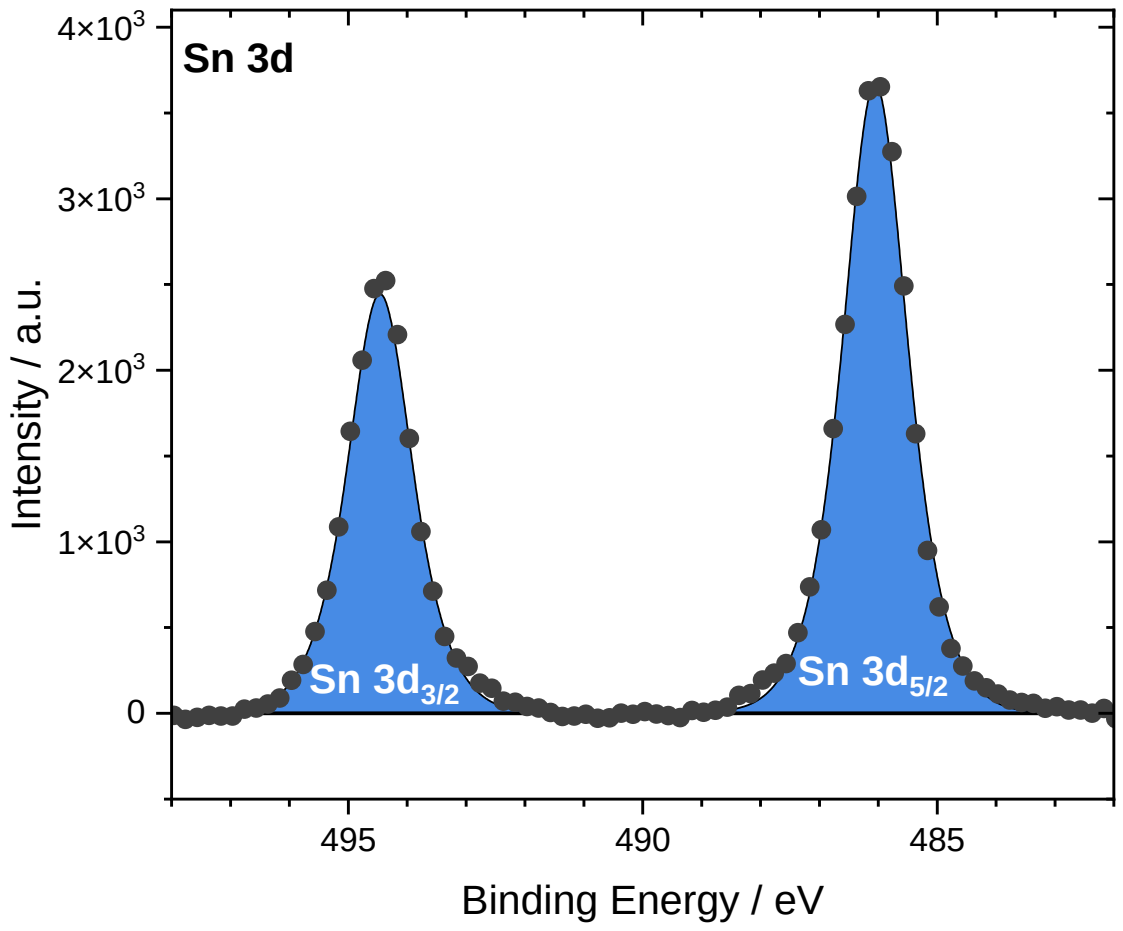

Sn 3d region of $CsSnI_3$.

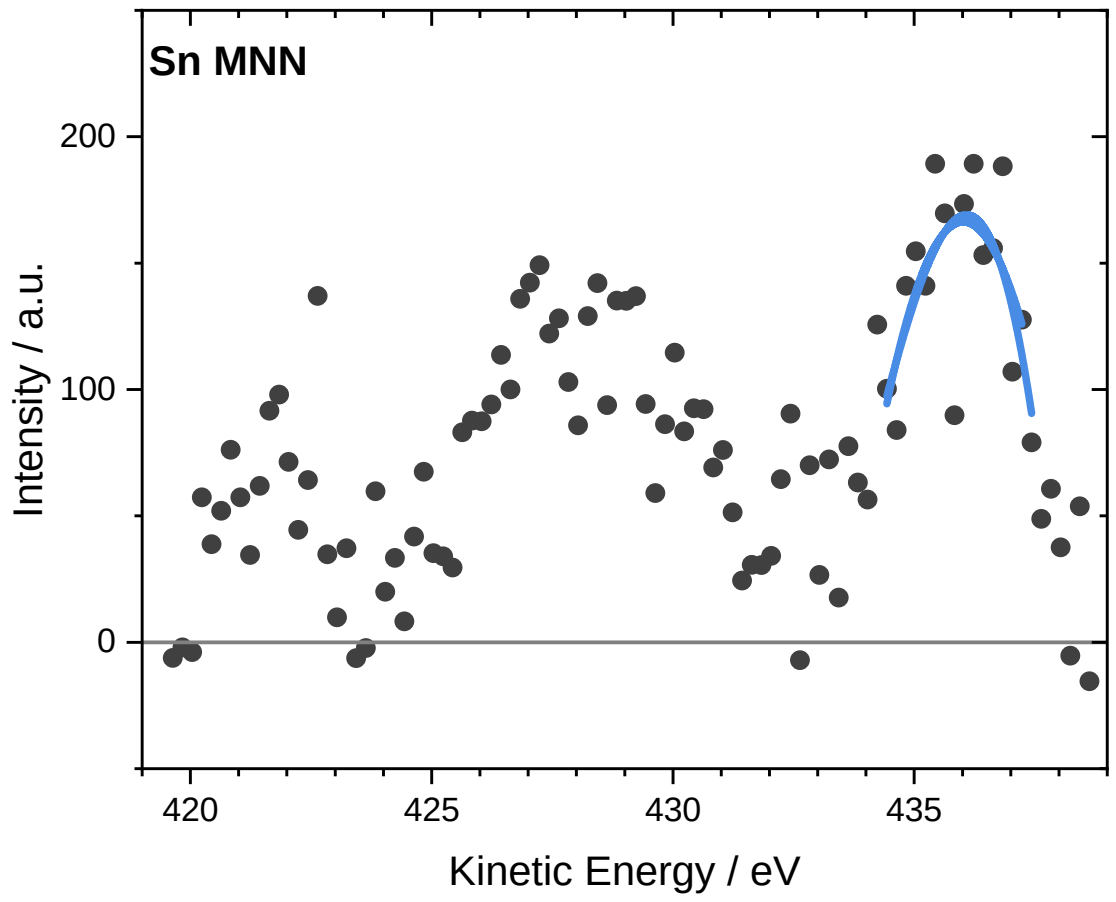

Sn MNN region of $CsSnI_3$.

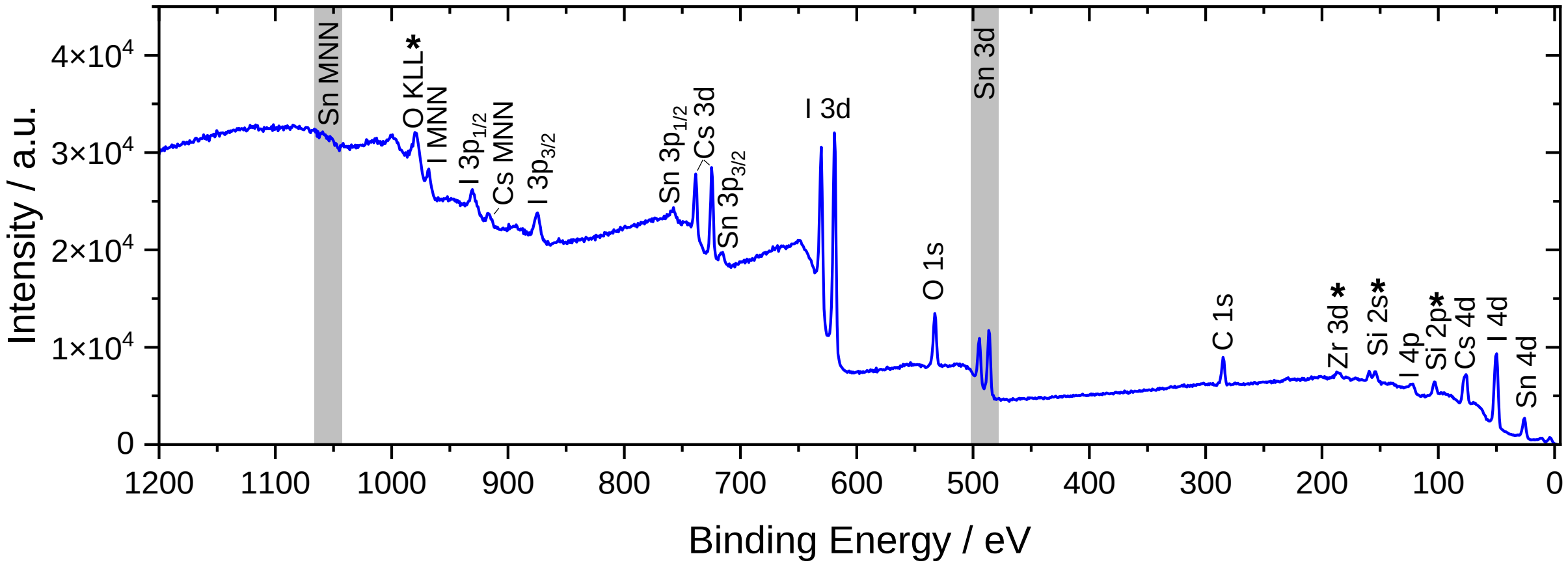

Survey spectrum of $CsSnI_3$.

Peaks assigned to the probing of the substrate through pin-holes are marked with an asterisk.

MASnI₃

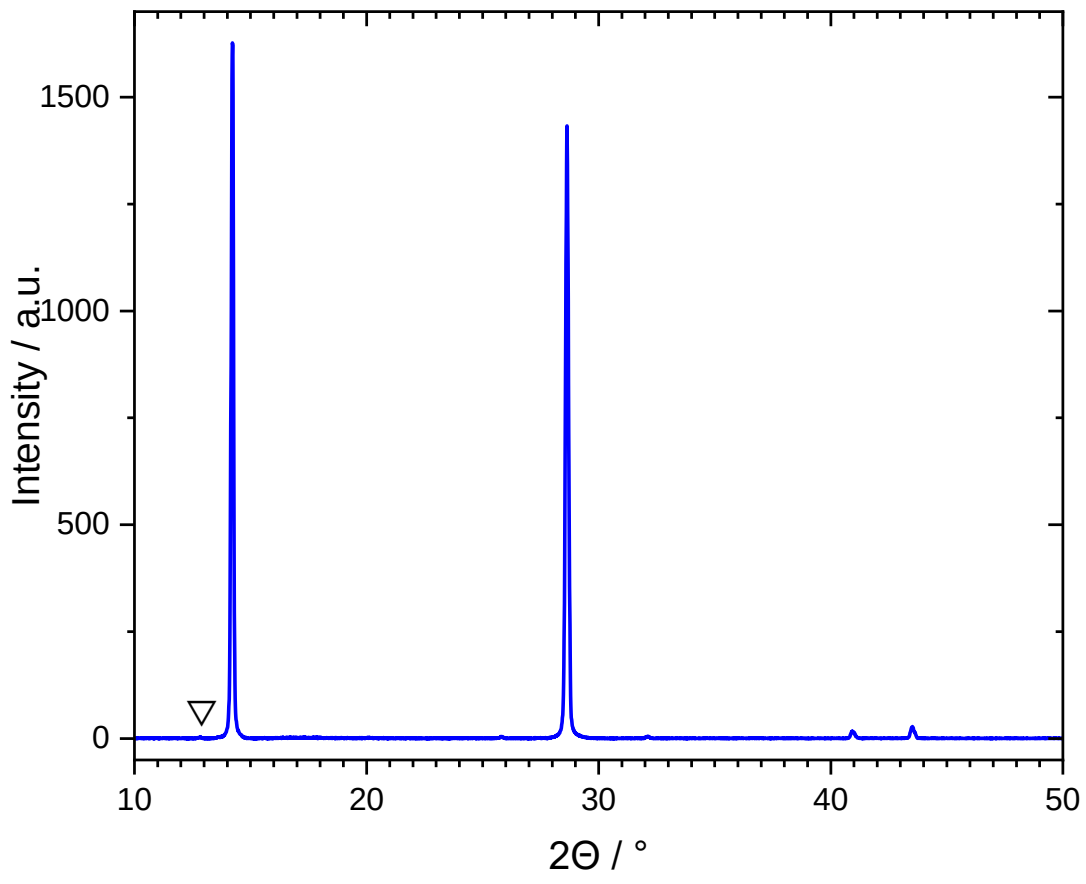

XRD diffractogram of $MASnI_3$.

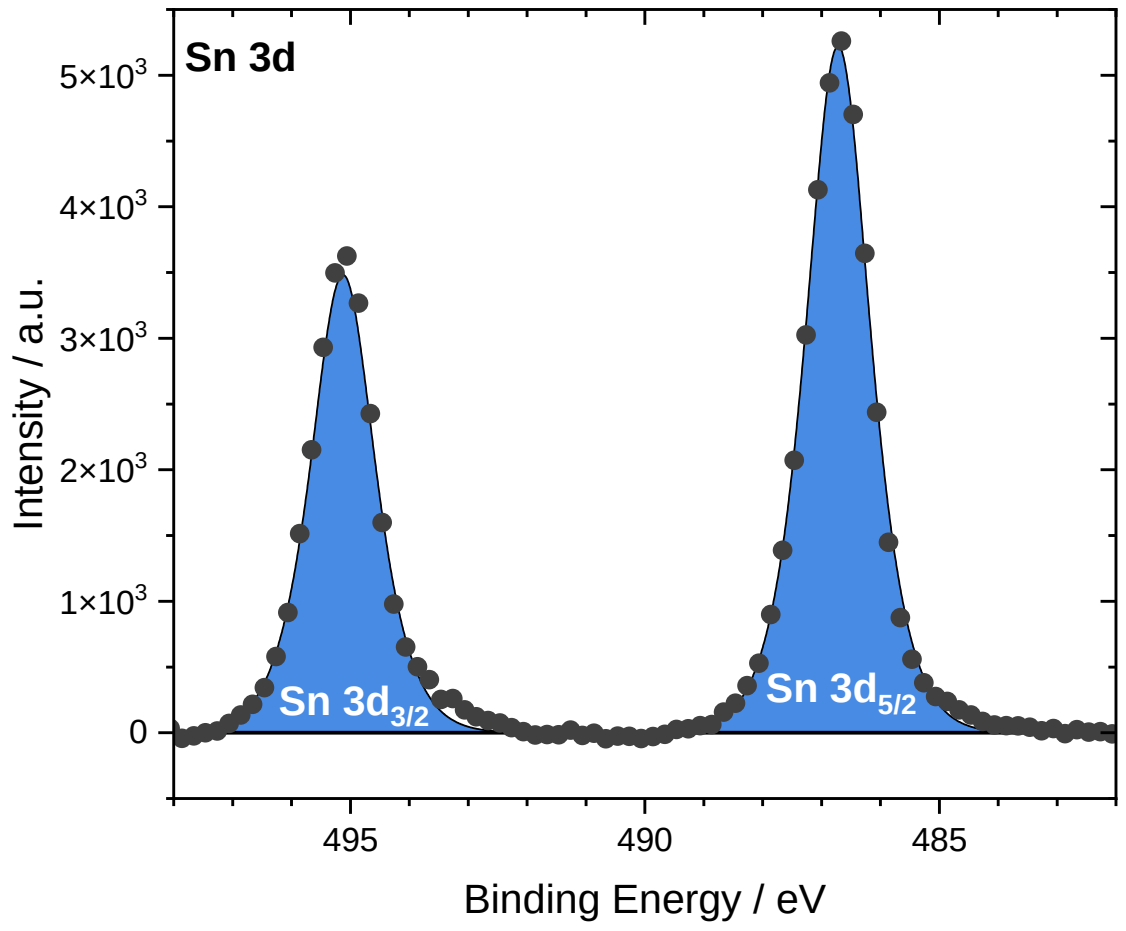

Sn 3d region of $MASnI_3$.

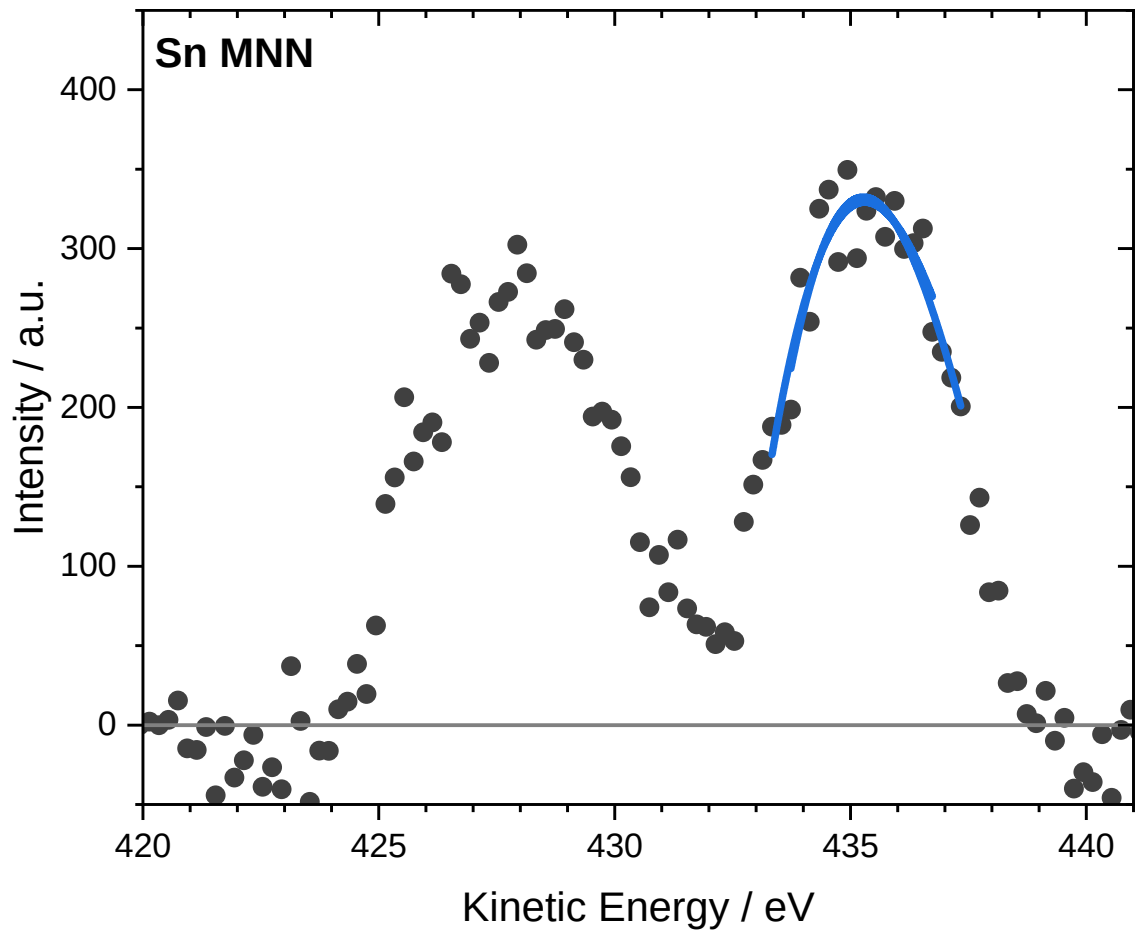

Sn MNN region of $MASnI_3$.

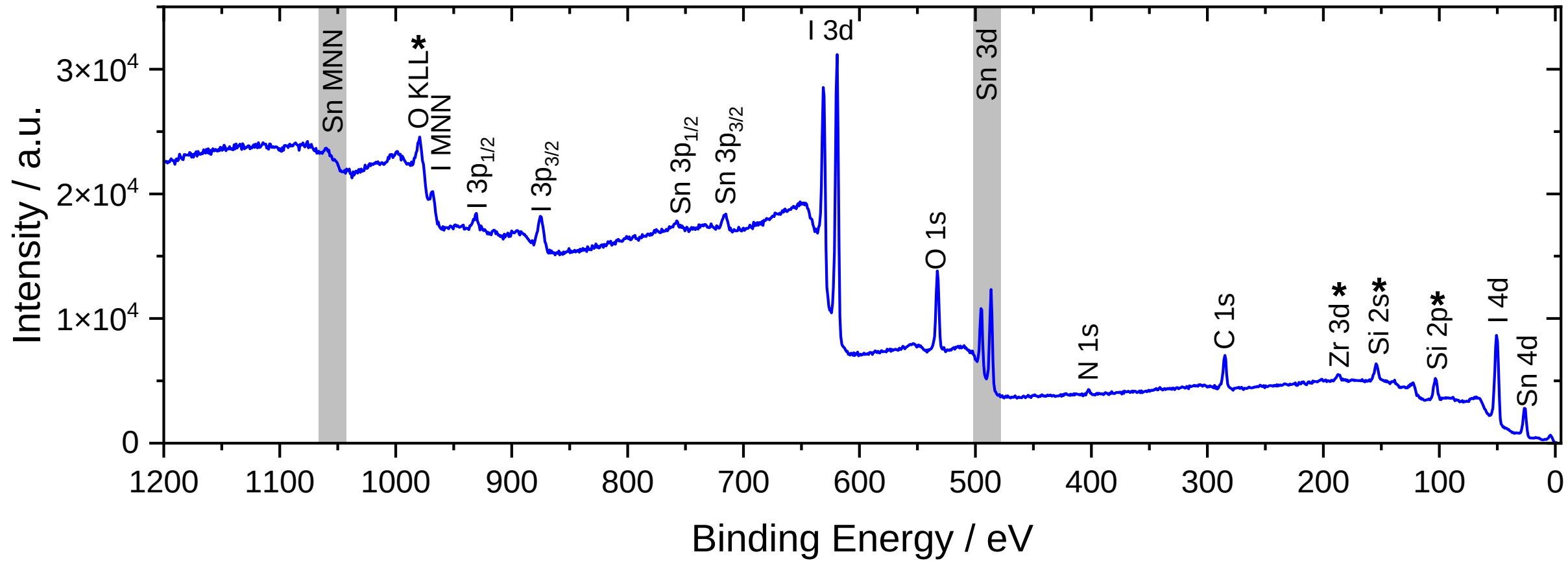

Survey spectrum of $MASnI_3$.

Peaks assigned to the probing of the substrate through pin-holes are marked with an asterisk. Trace amounts of $SnI_2$ could be identified in the XRD pattern which are marked with triangle.

FASnI3

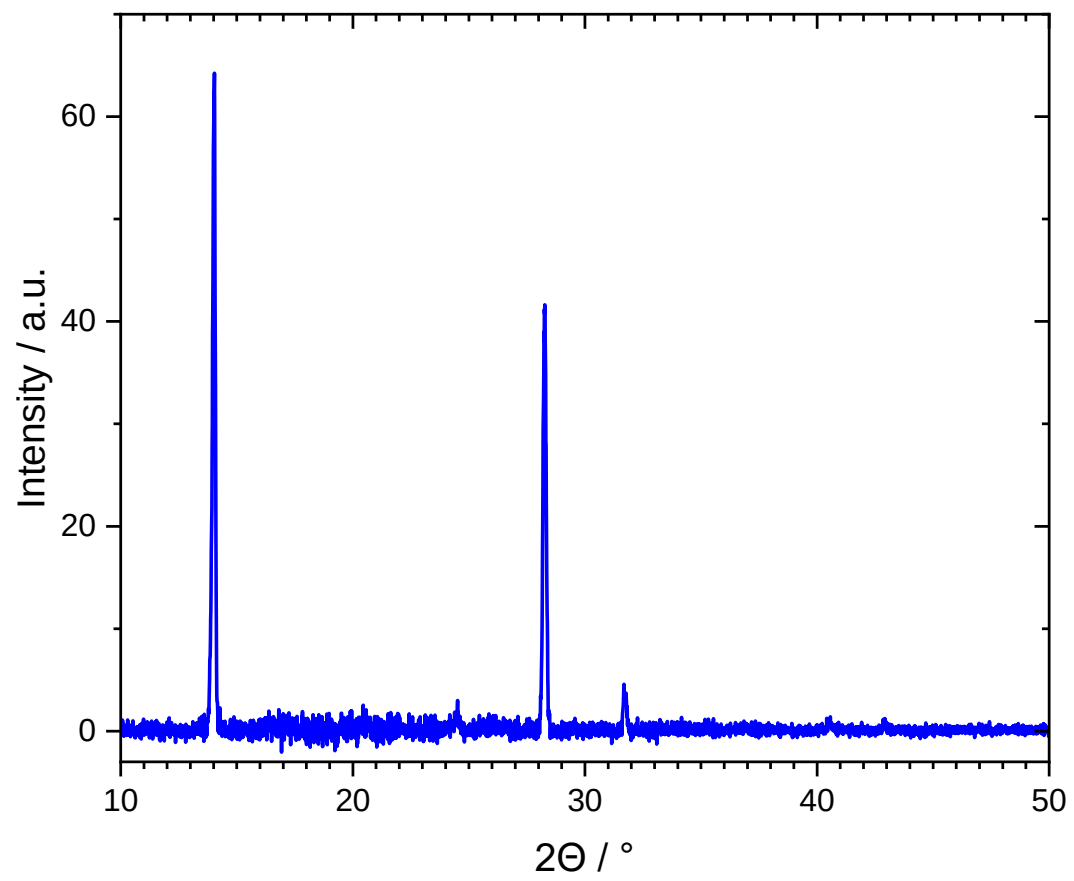

XRD diffractogram of $FASnI_3$.

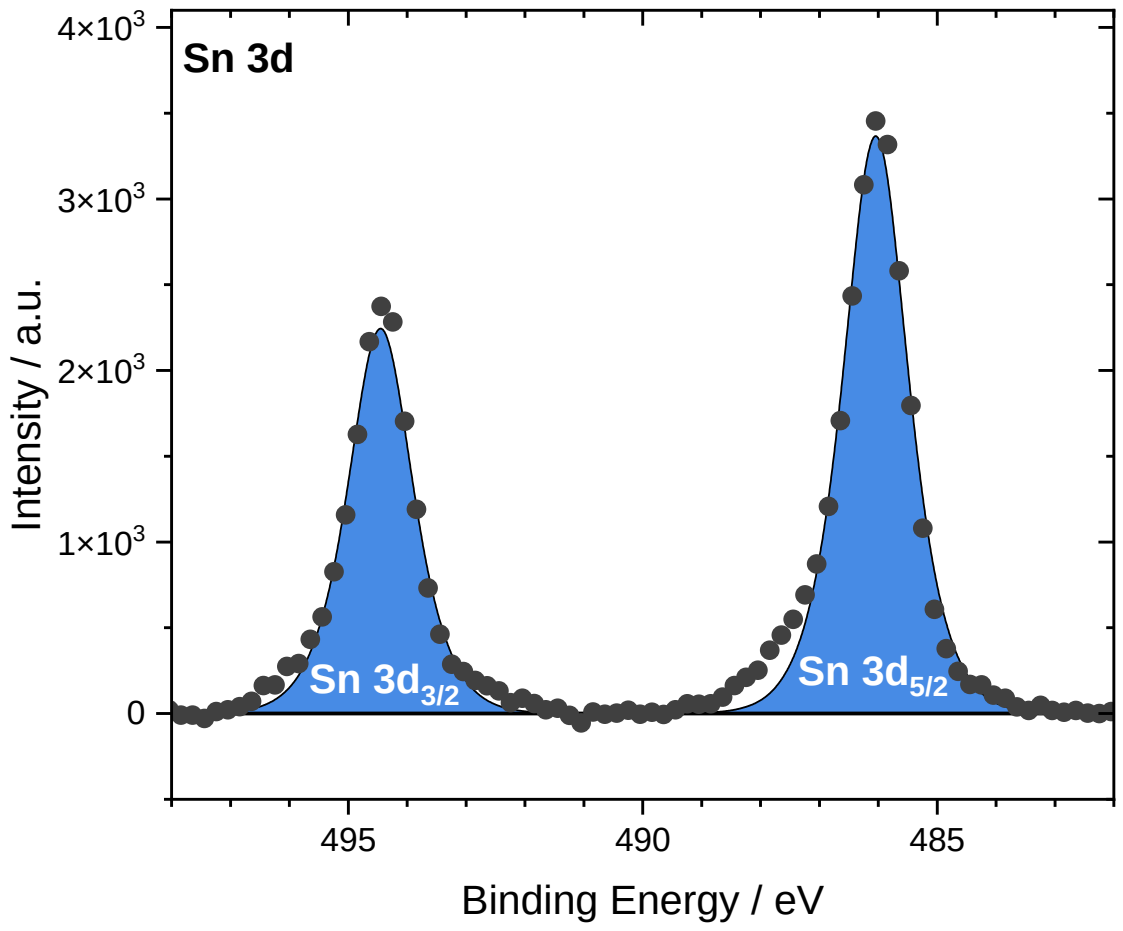

Sn 3d region of $FASnI_3$.

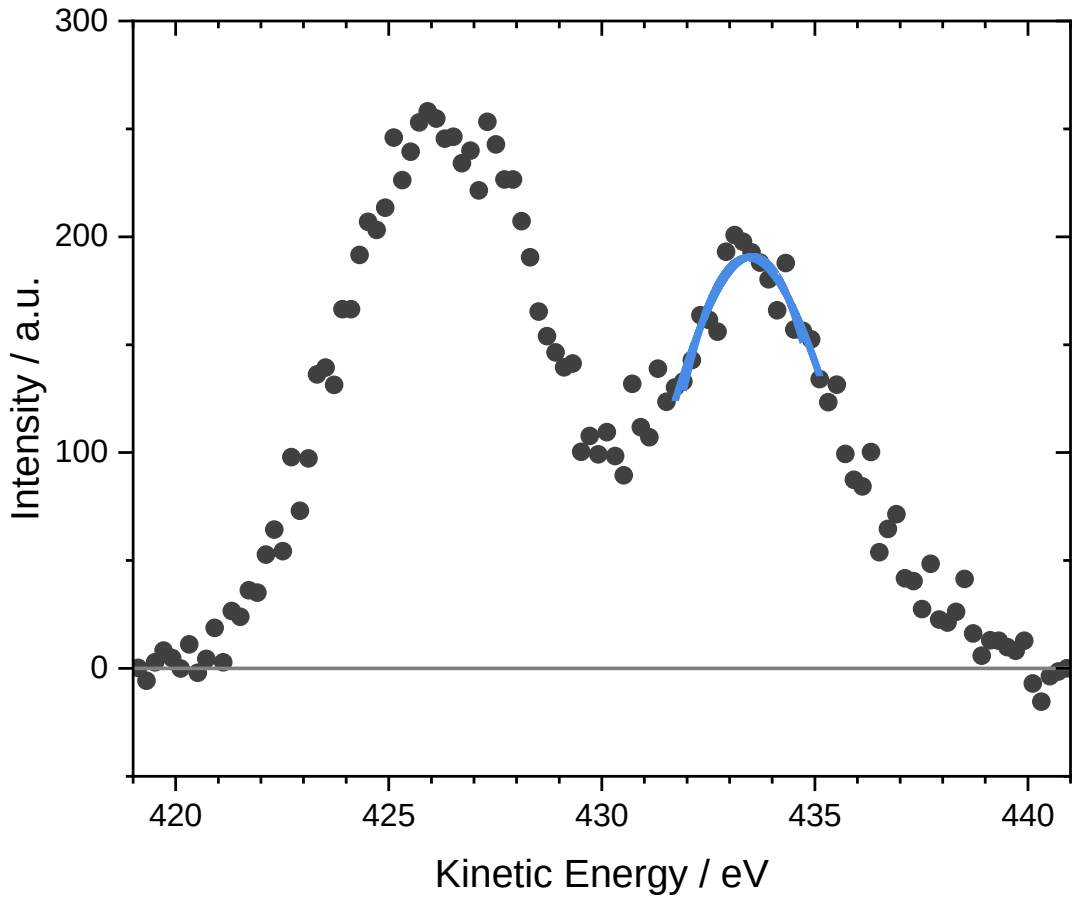

Sn MNN region of $FASnI_3$.

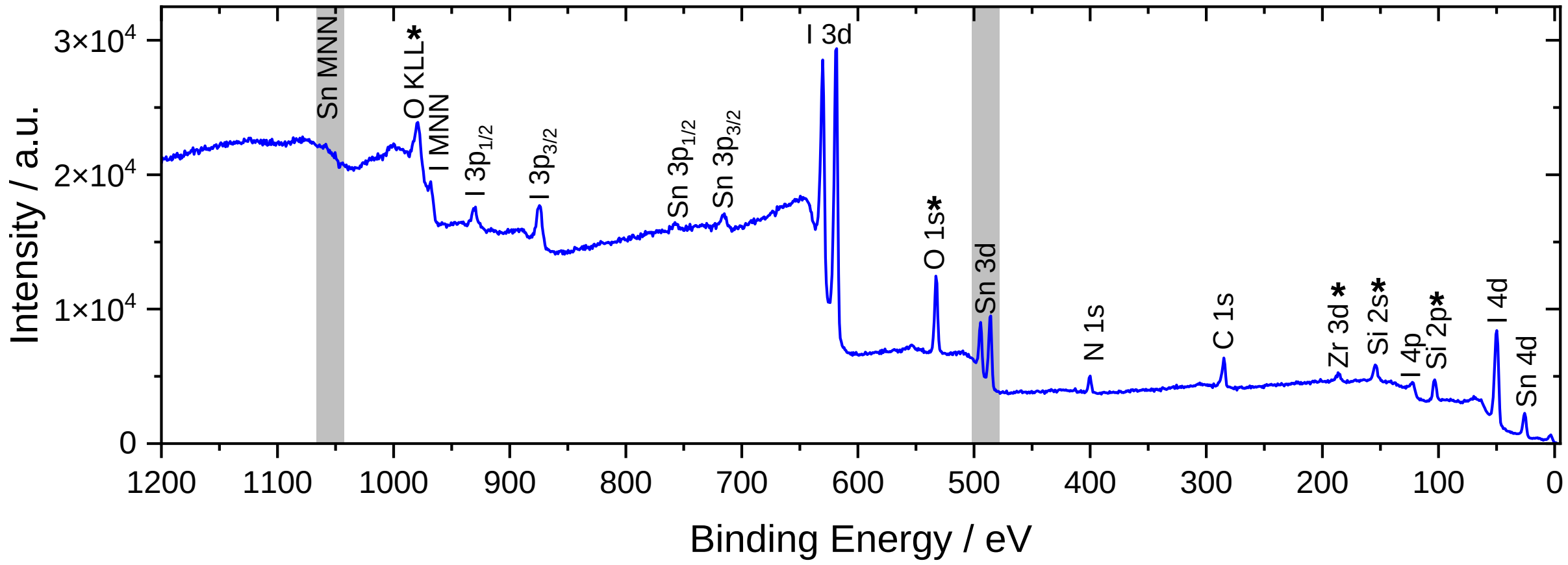

Survey spectrum of $FASnI_3$.

Peaks assigned to the probing of the substrate through pin-holes are marked with an asterisk.

CsSnBr3

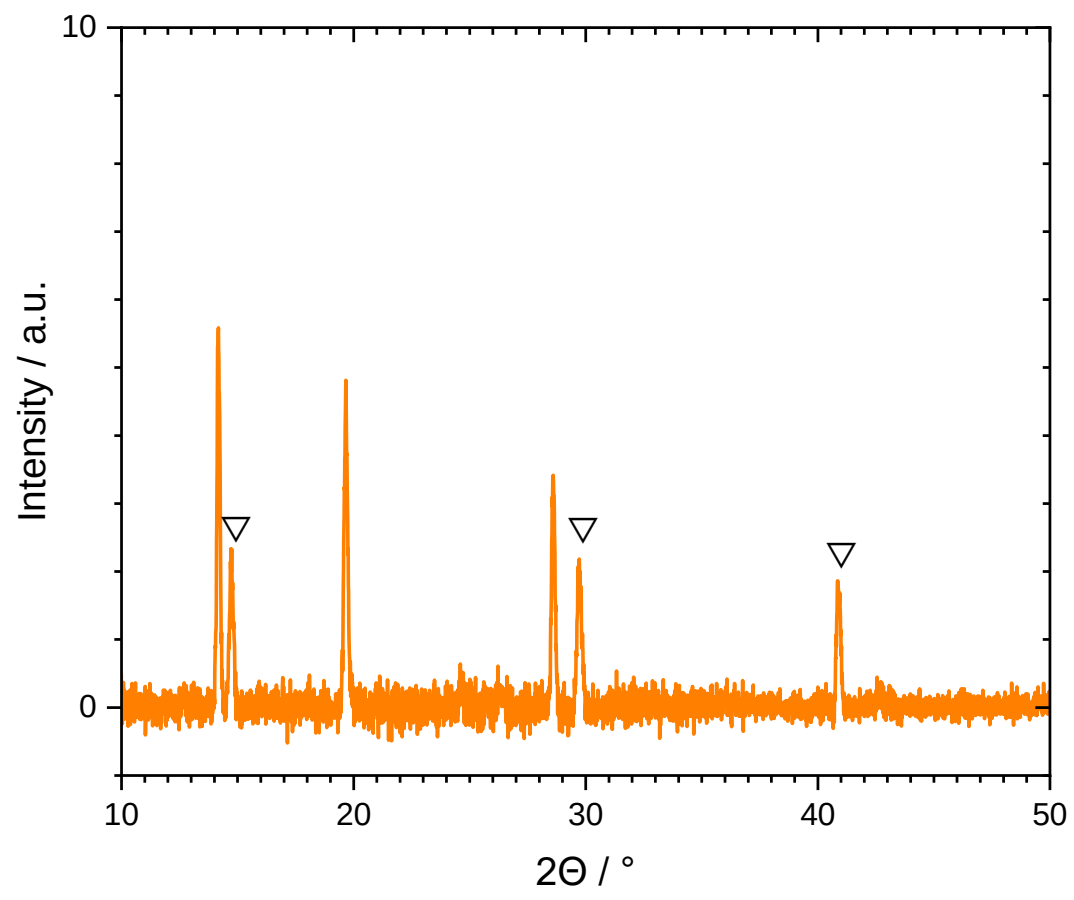

XRD diffractogram of $CsSnBr_3$.

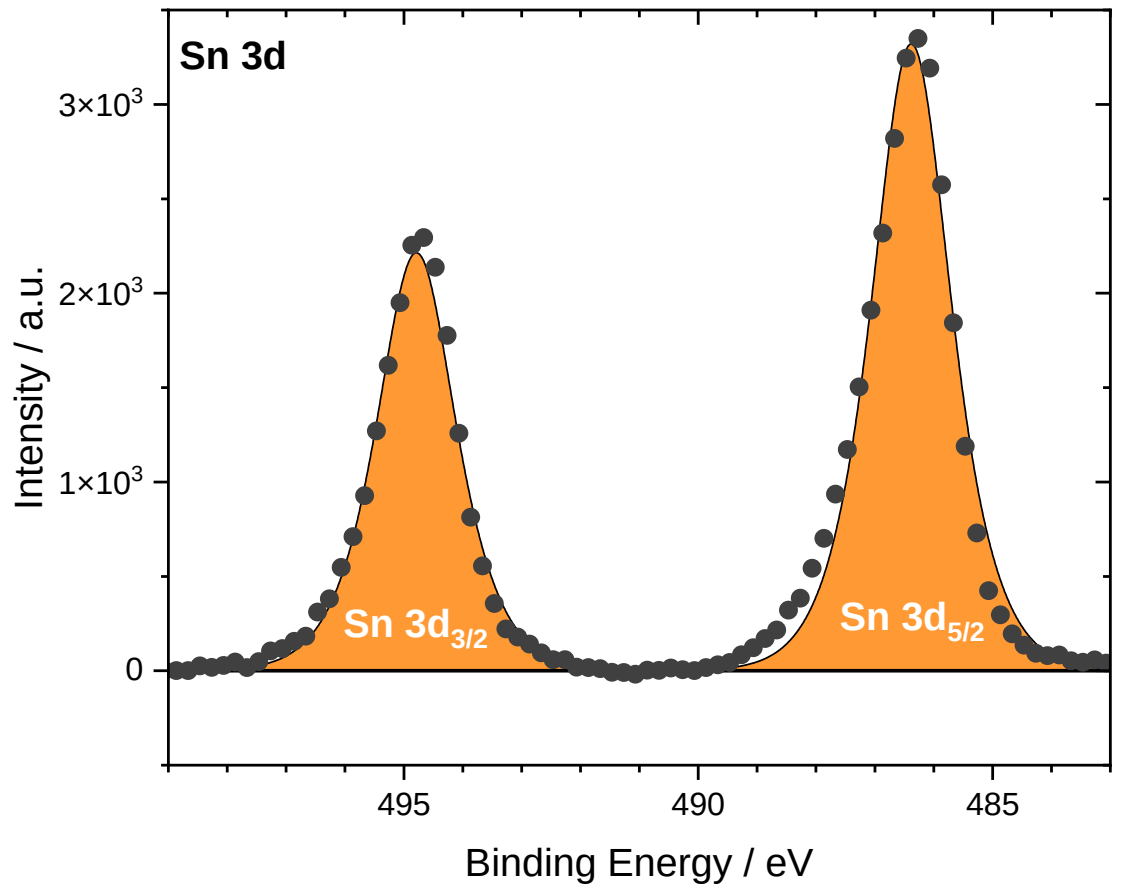

Sn 3d region of $CsSnBr_3$.

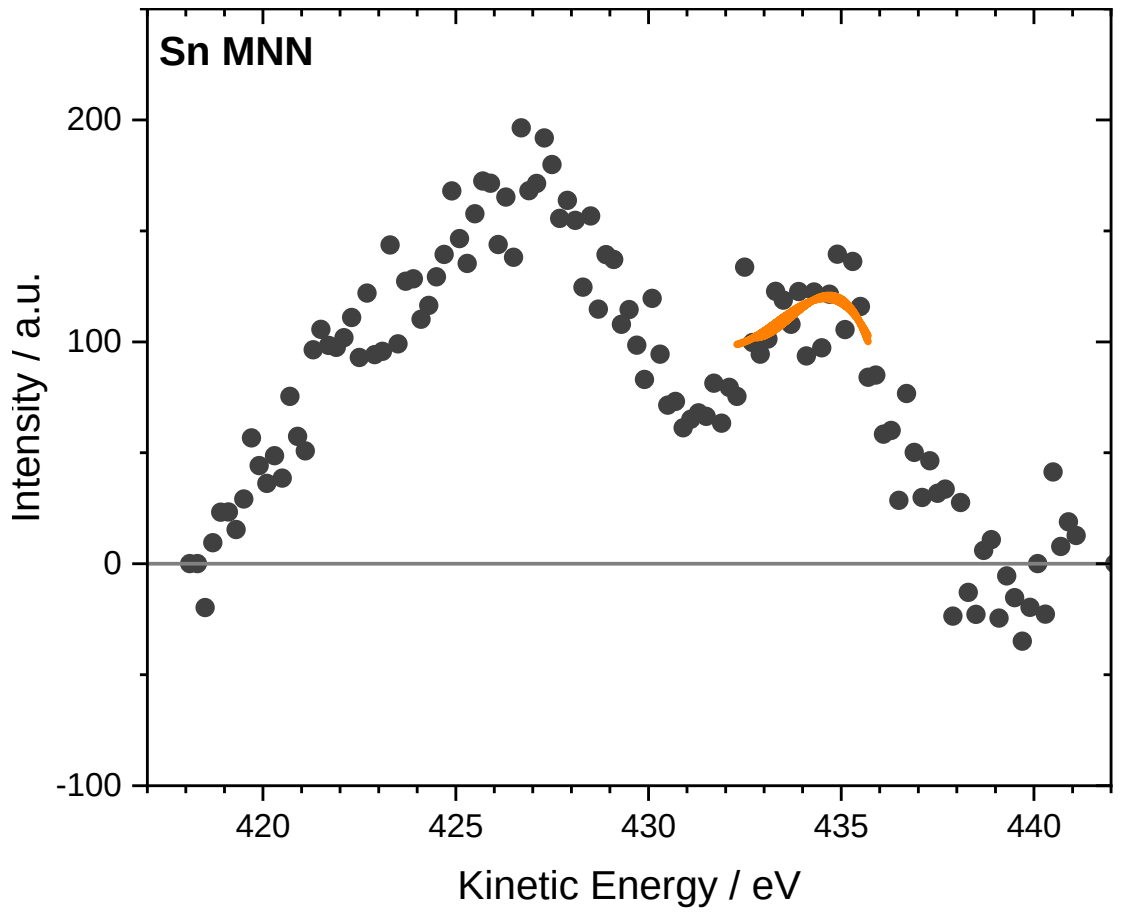

Sn MNN region of $CsSnBr_3$.

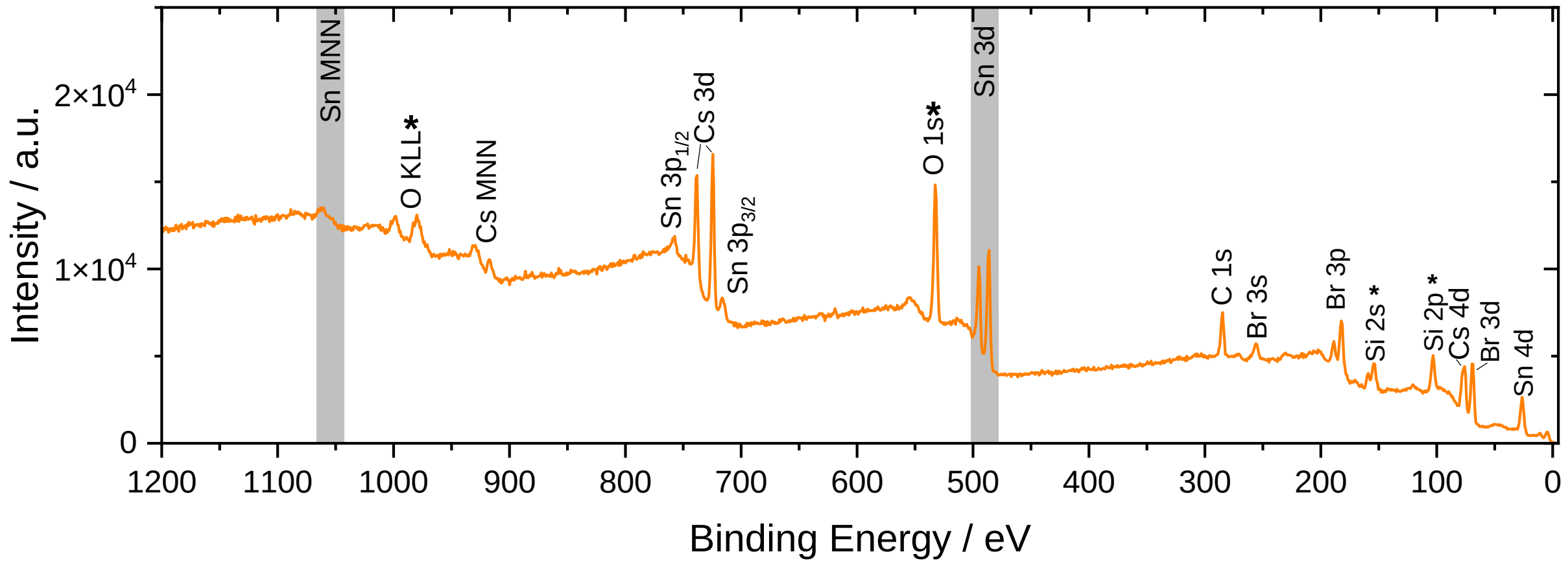

Survey spectrum of $CsSnBr_3$.

Peaks assigned to the probing of the substrate through pin-holes are marked with an asterisk. Amounts of CsI could be identified in the XRD pattern which are marked with triangle.

MASnBr3

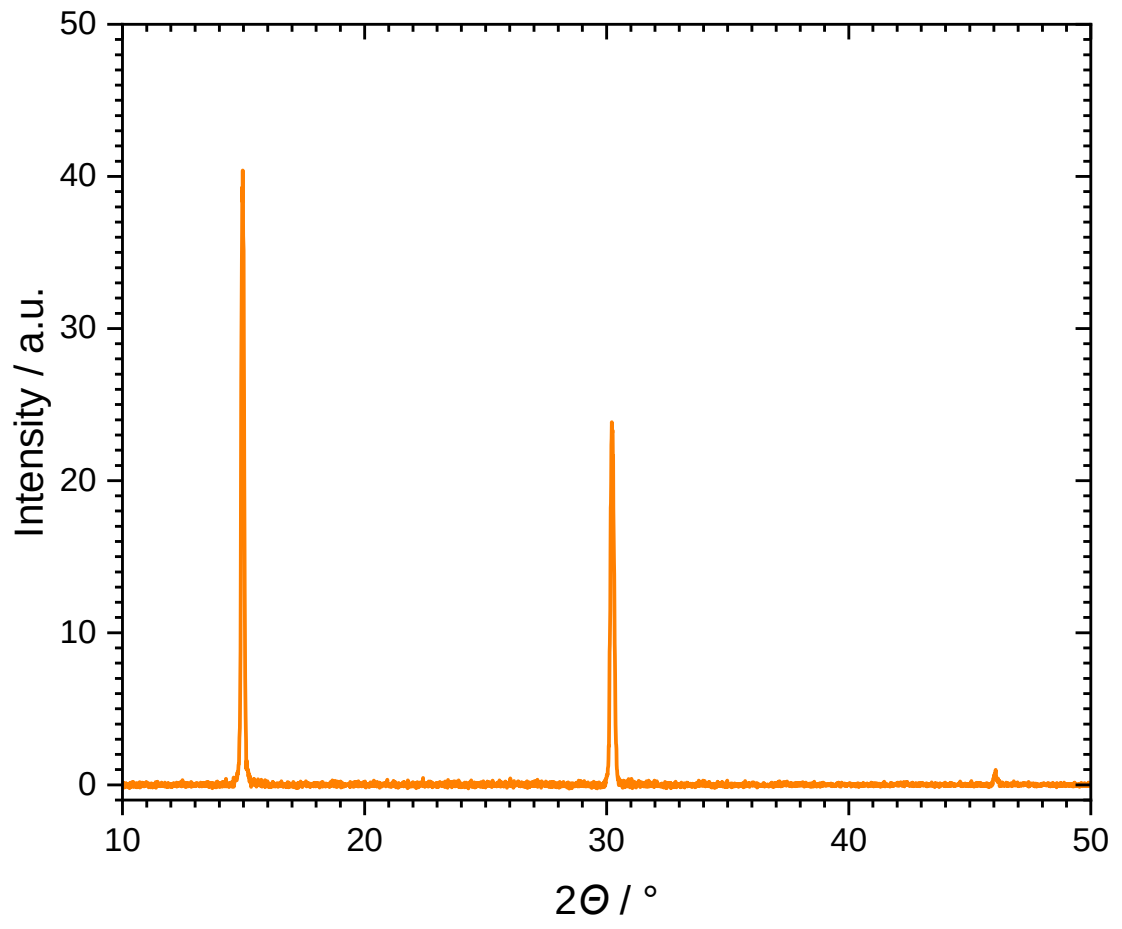

XRD diffractogram of $MASnBr_3$.

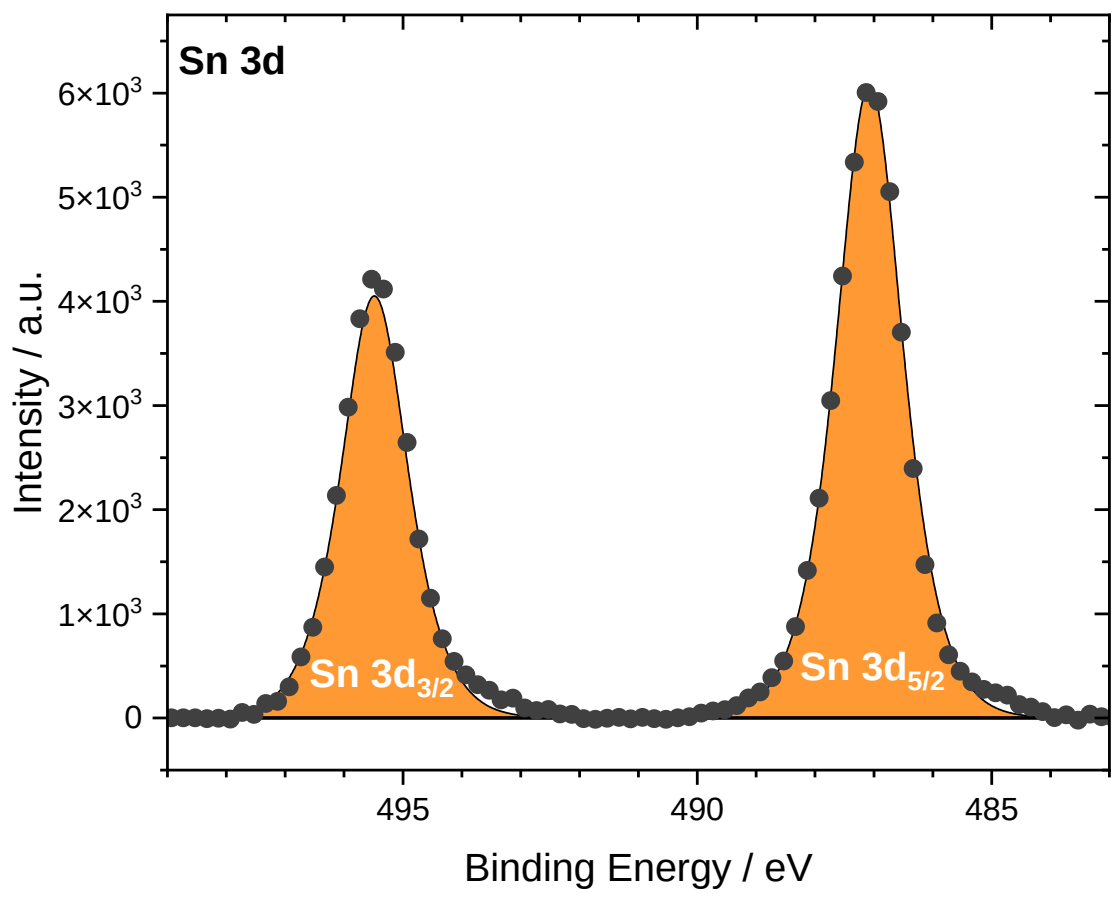

Sn 3d region of $MASnBr_3$.

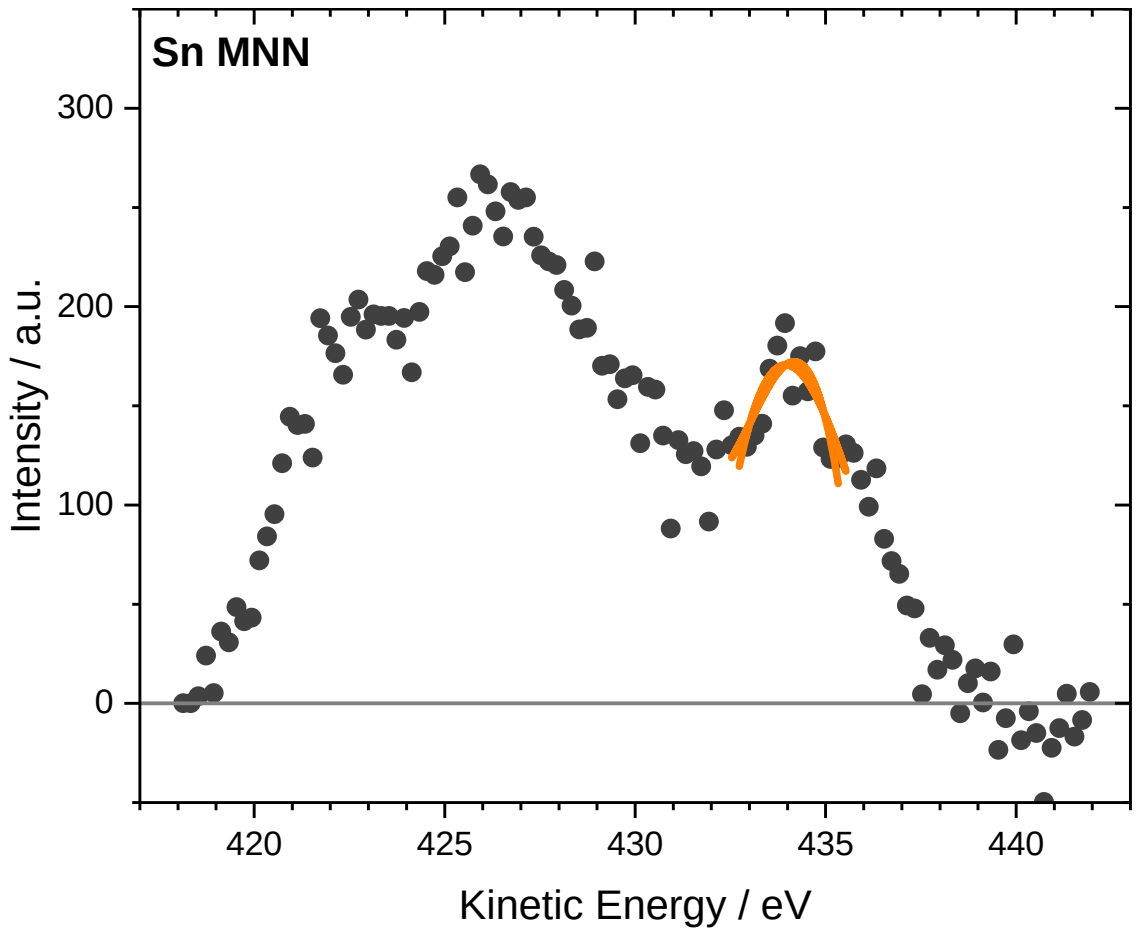

Sn MNN region of $MASnBr_3$.

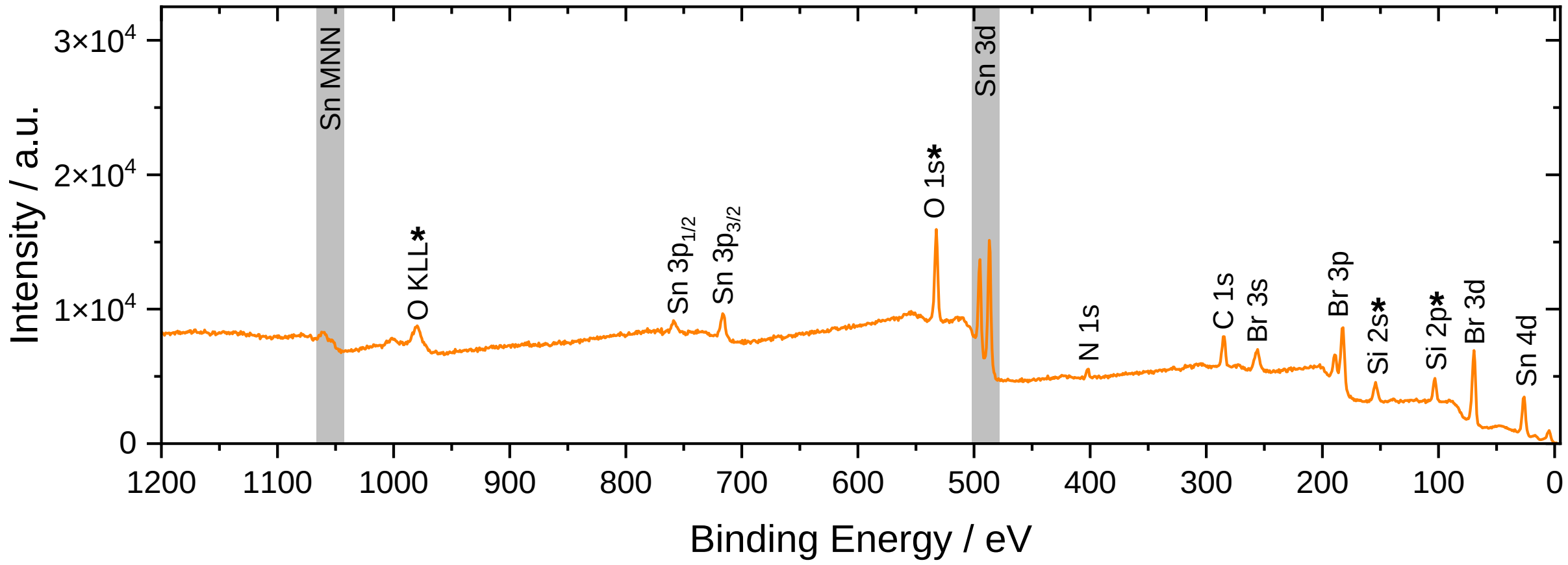

Survey spectrum of $MASnBr_3$.

Peaks assigned to the probing of the substrate through pin-holes are marked with an asterisk.

FASnBr3

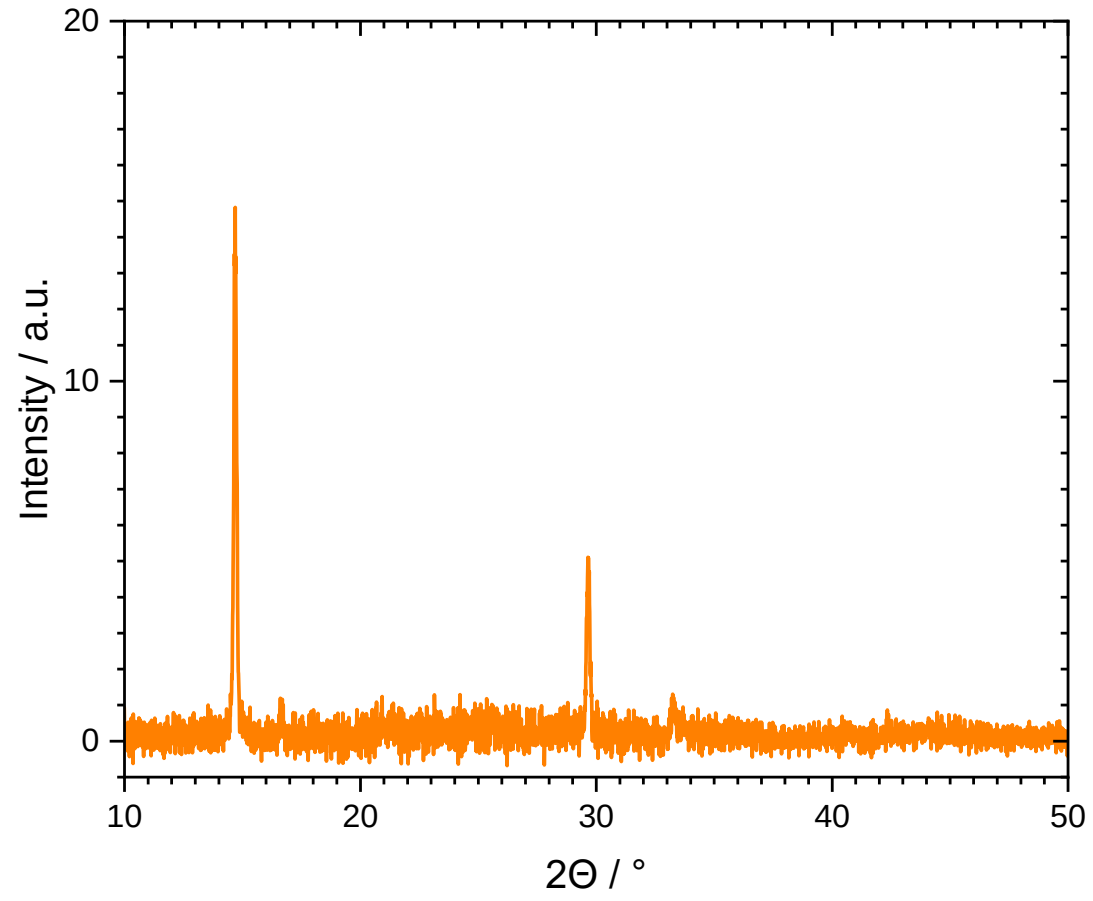

XRD diffractogram of $FASnBr_3$.

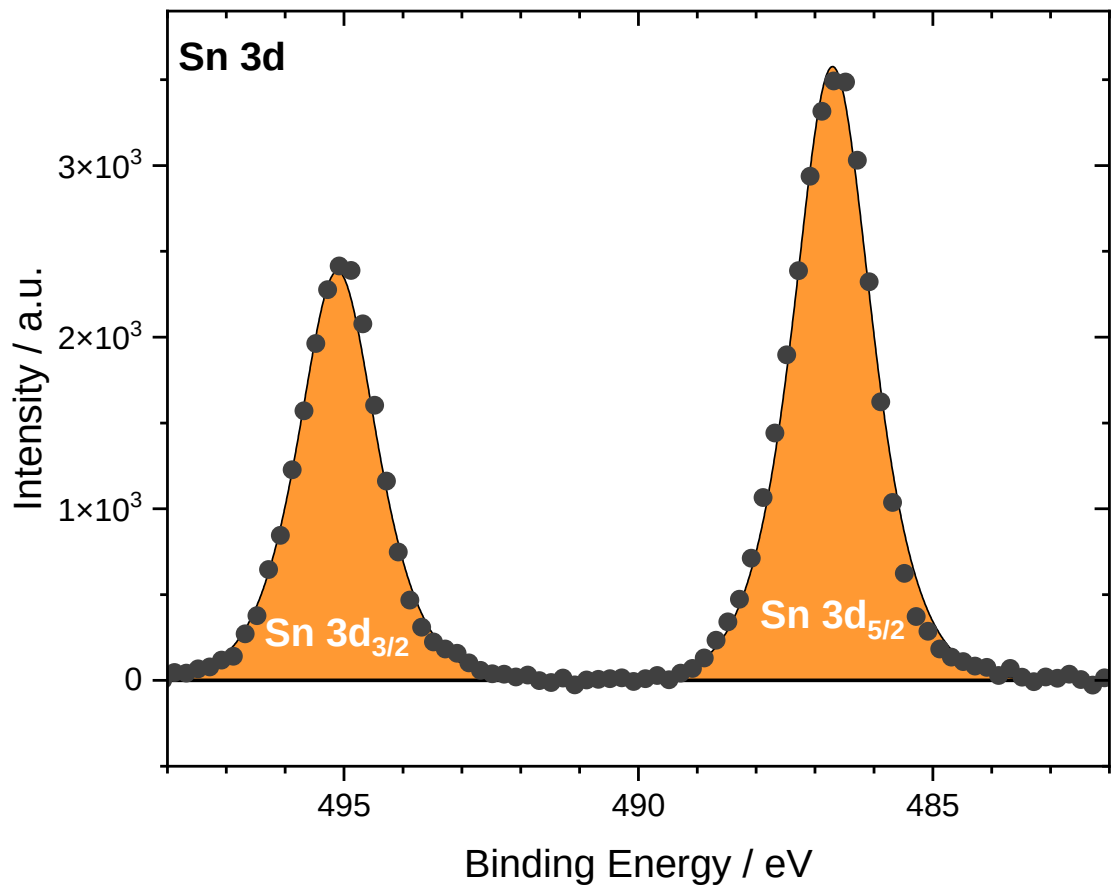

Sn 3d region of $FASnBr_3$.

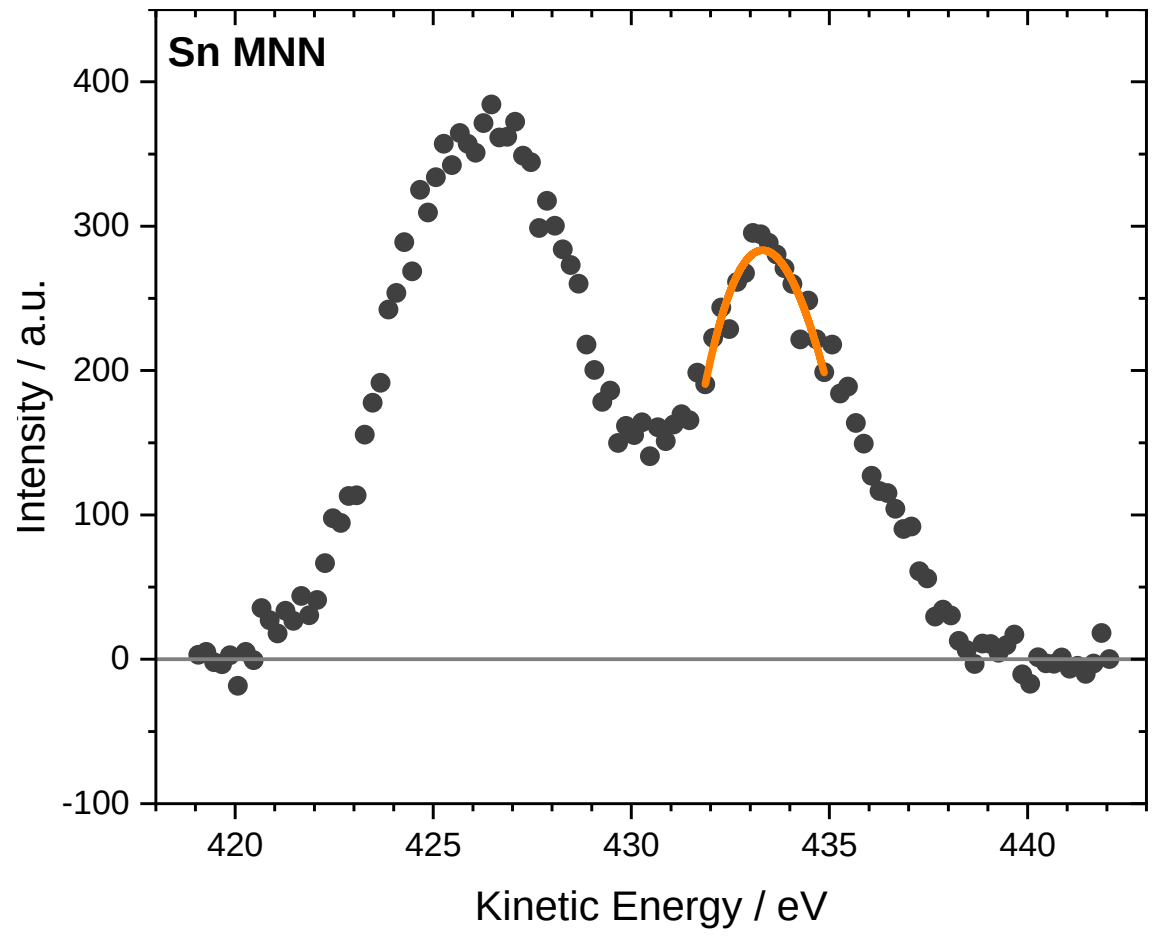

Sn MNN region of $FASnBr_3$.

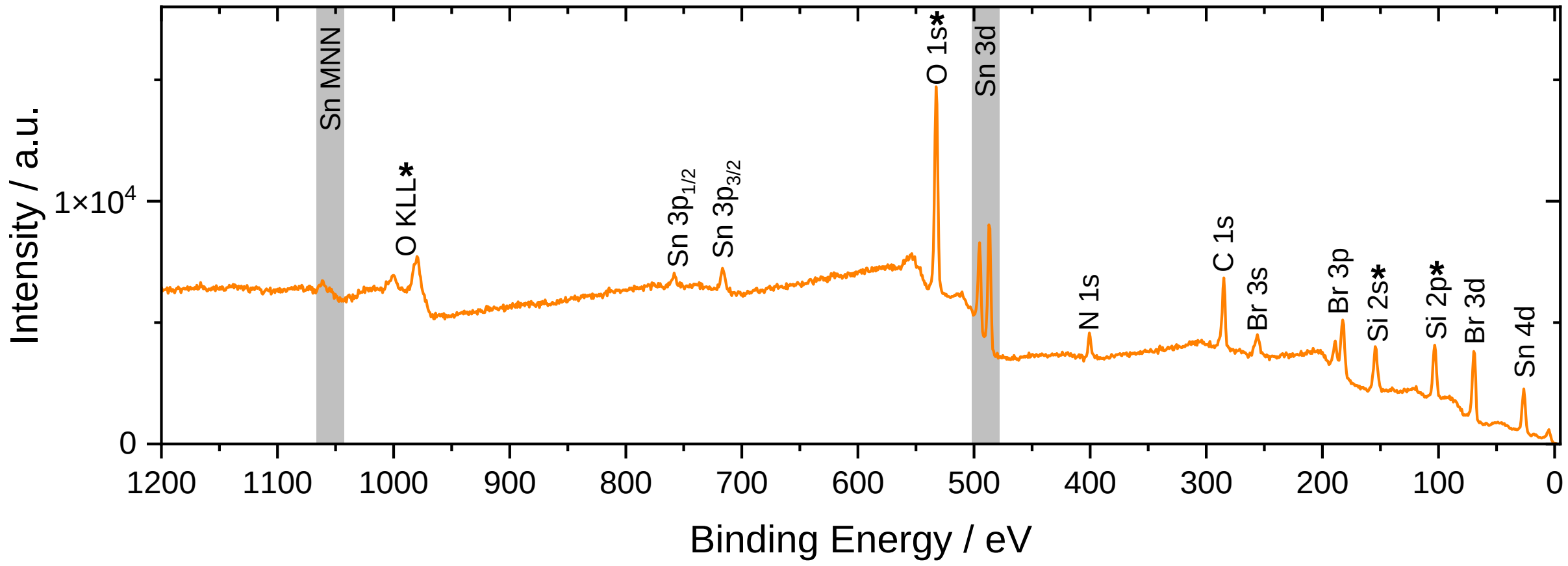

Survey spectrum of $FASnBr_3$.
Peaks assigned to the probing of the substrate through pin-holes are marked with an asterisk.

$FA(Sn_{0.75}Pb_{0.25})I_3$

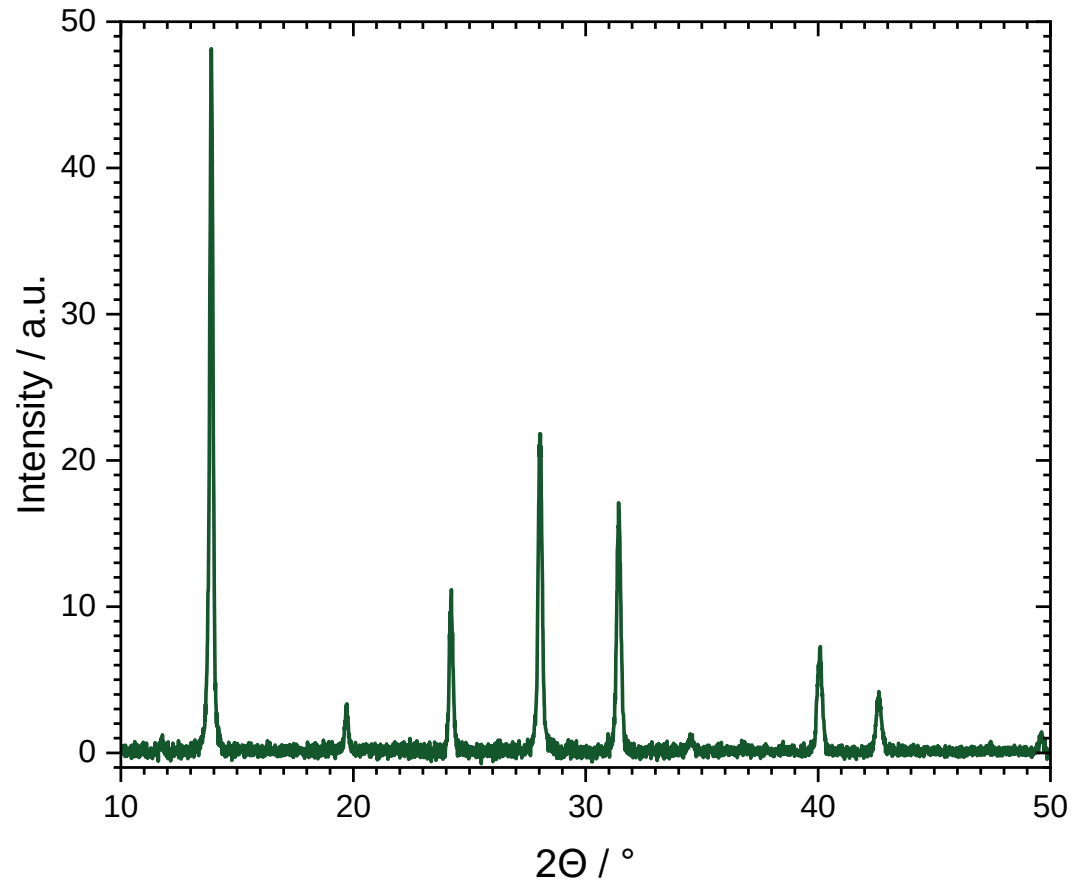

XRD diffractogram of $FA(Sn_{0.75}Pb_{0.25})I_3$.

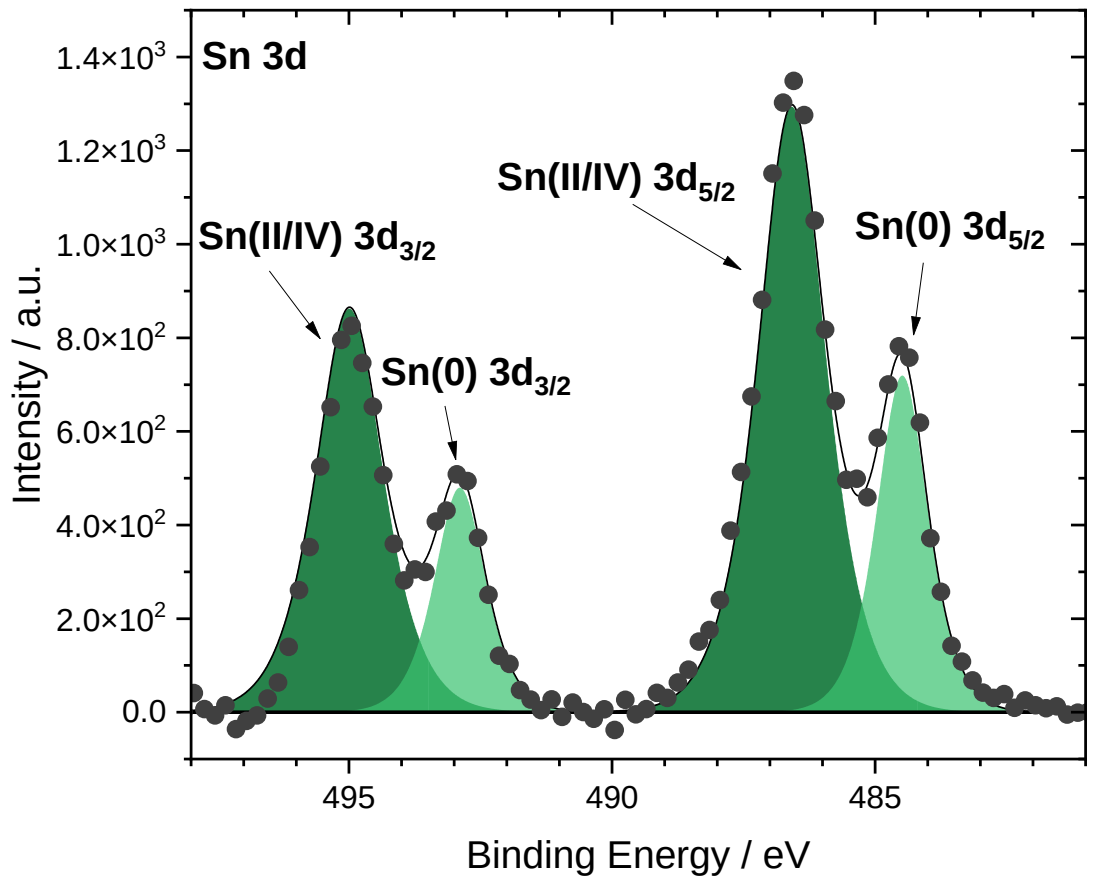

Sn 3d region of $FA(Sn_{0.75}Pb_{0.25})I_3$.

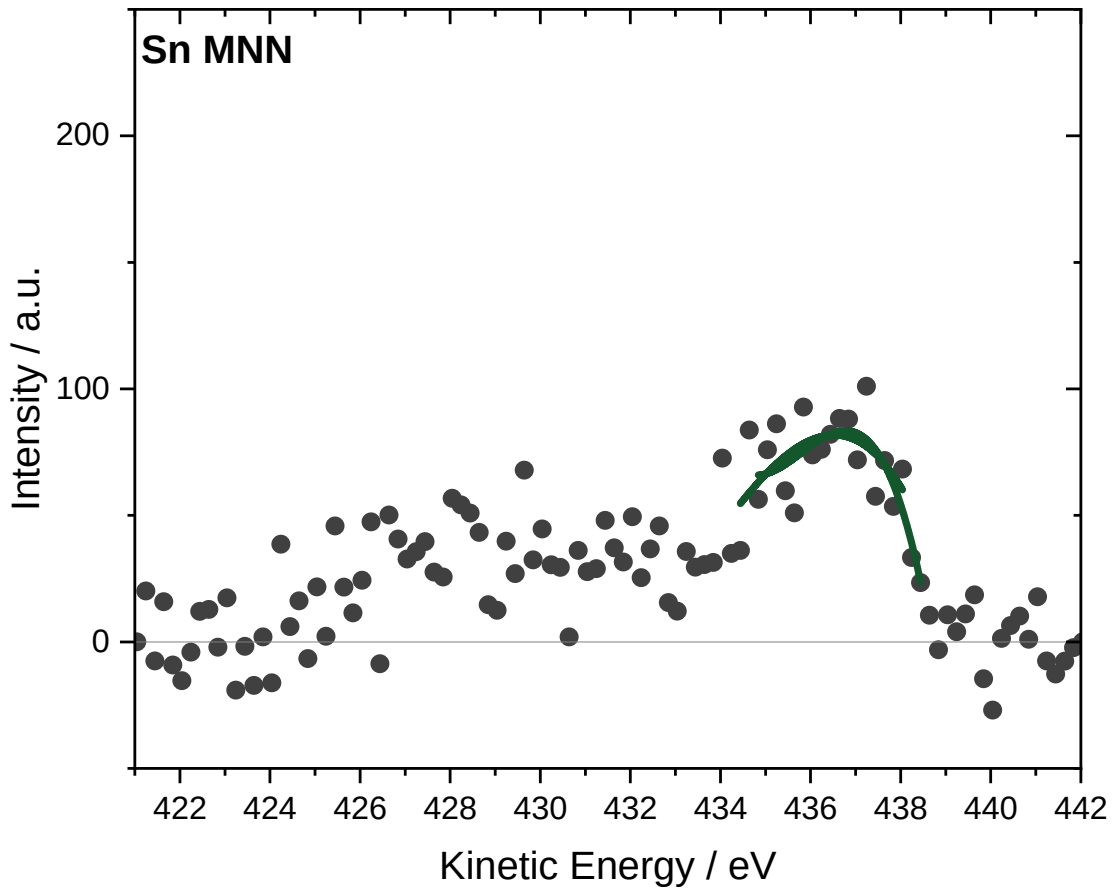

Sn MNN region of $FA(Sn_{0.75}Pb_{0.25})I_3$.

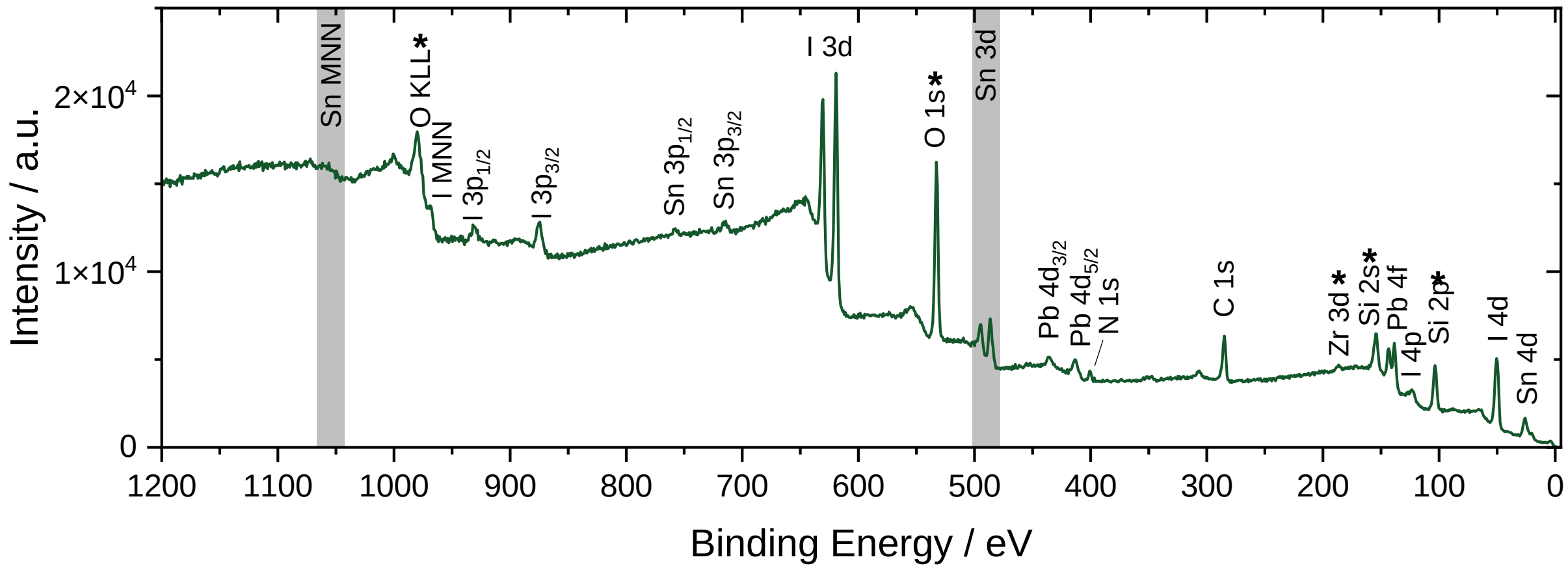

Survey spectrum of $FA(Sn_{0.75}Pb_{0.25})I_3$.
Peaks assigned to the probing of the substrate through pin-holes are marked with an asterisk.

$FA(Sn_{0.5}Pb_{0.5})I_3$

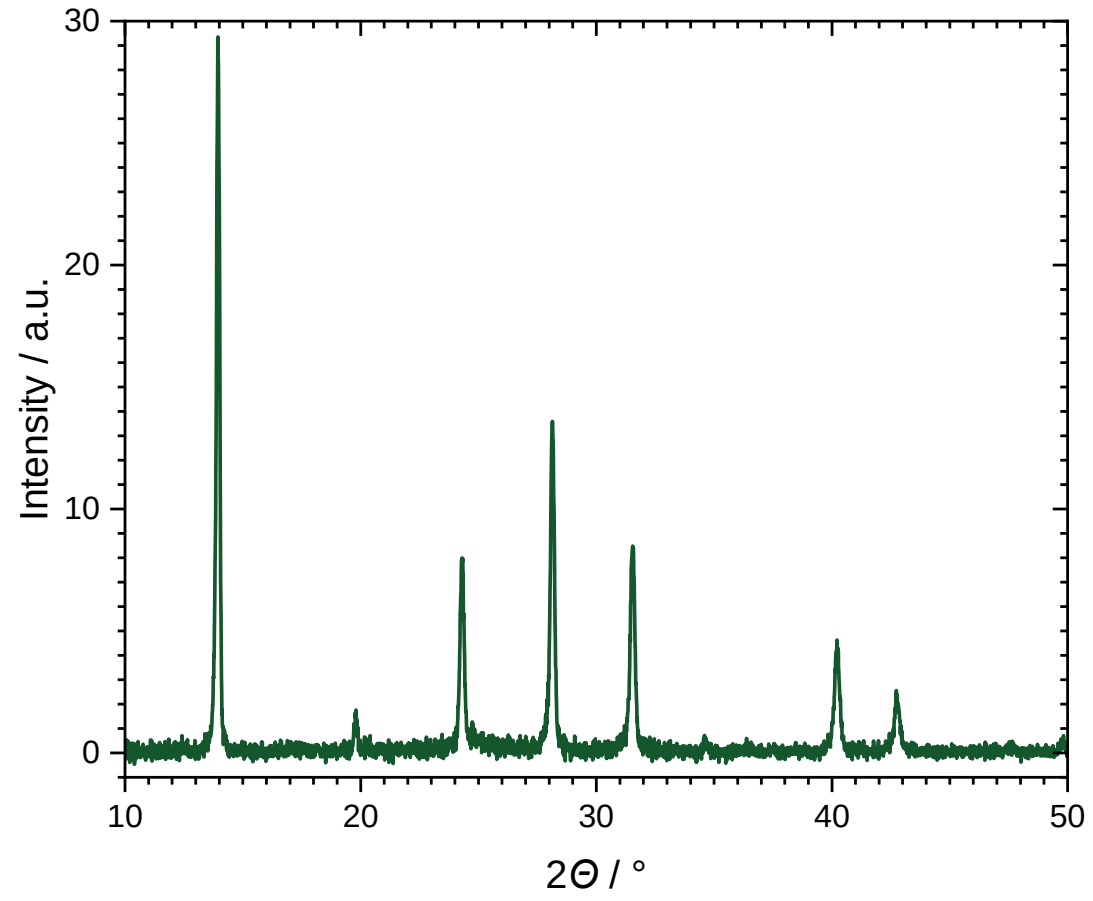

XRD diffractogram of $FA(Sn_{0.5}Pb_{0.5})I_3$.

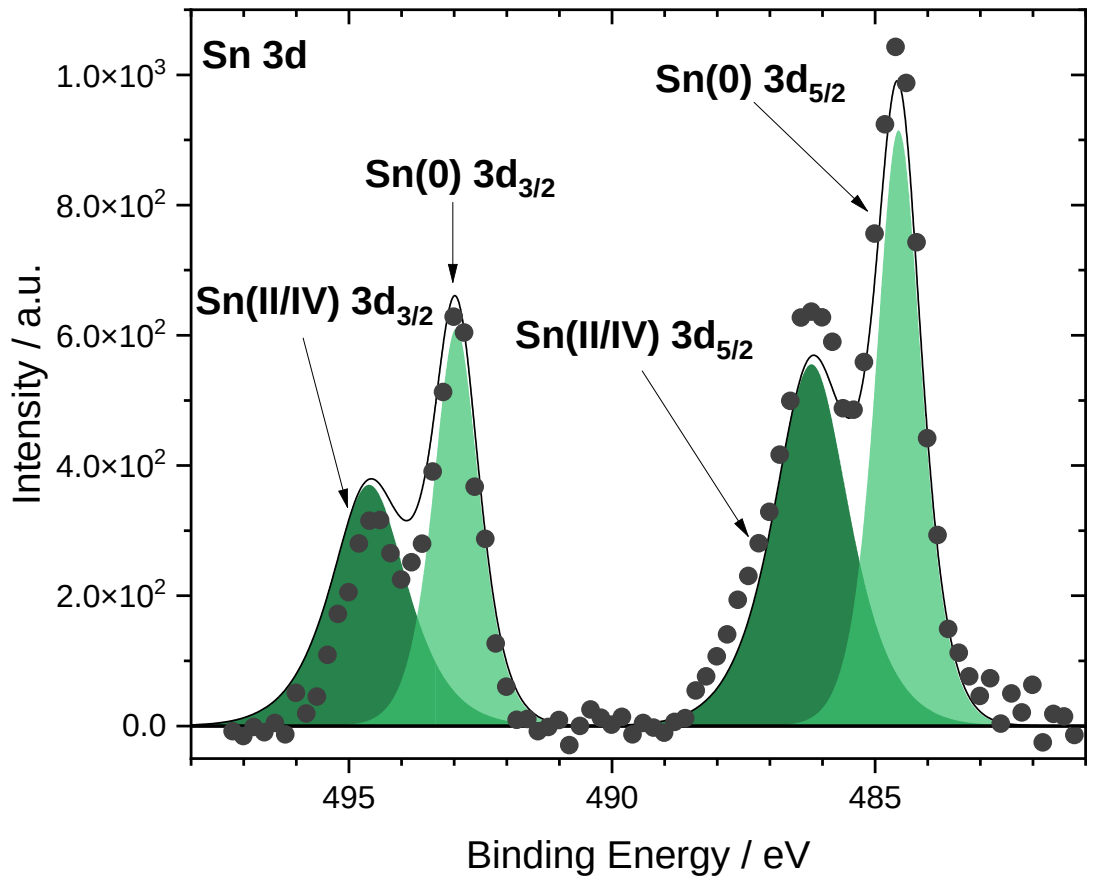

Sn 3d region of $FA(Sn_{0.5}Pb_{0.5})I_3$.

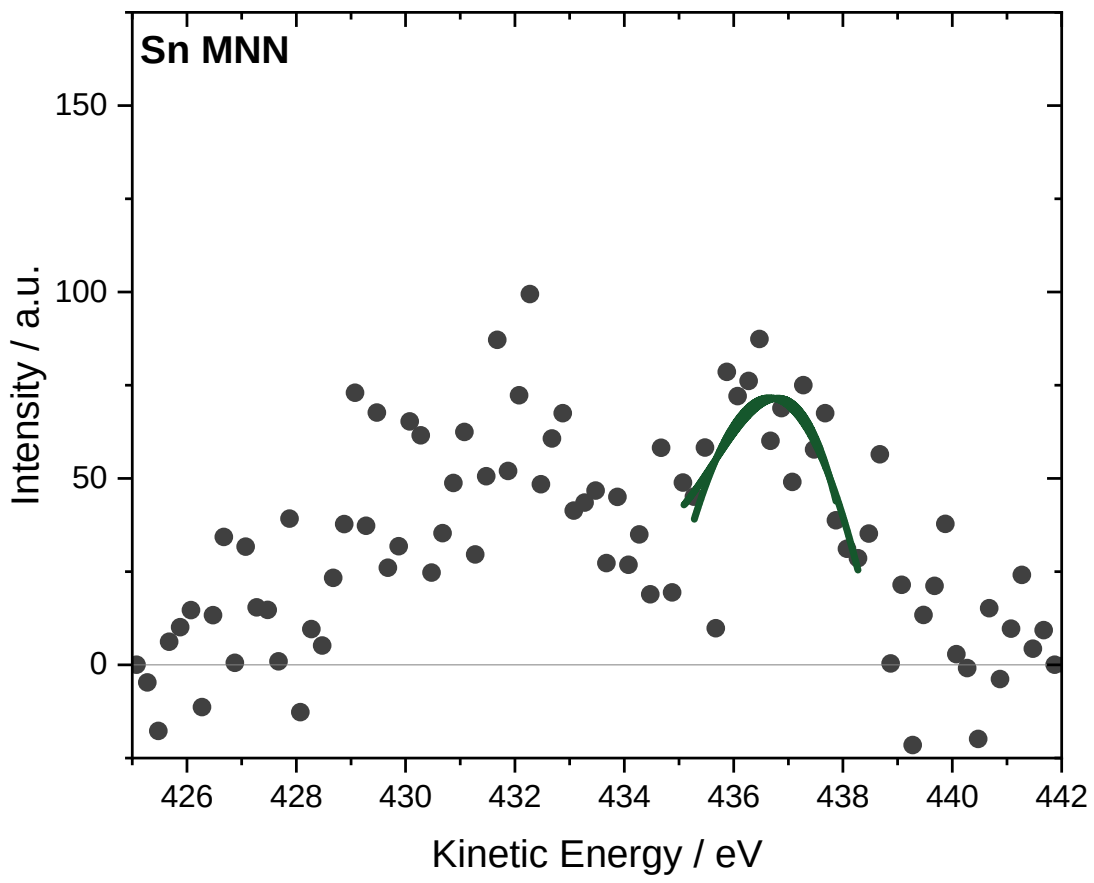

Sn MNN region of $FA(Sn_{0.5}Pb_{0.5})I_3$.

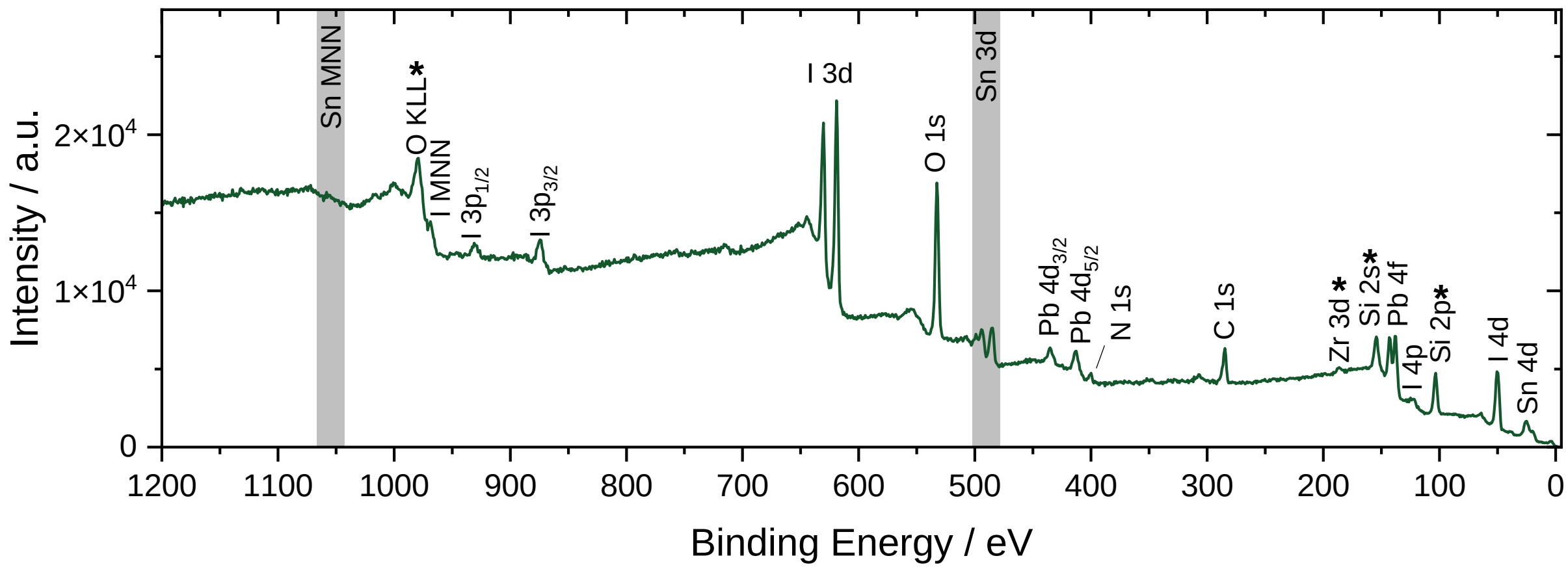

Survey spectrum of $FA(Sn_{0.5}Pb_{0.5})I_3$.
Peaks assigned to the probing of the substrate through pin-holes are marked with an asterisk.

FA($Sn_{0.25}Pb_{0.75}$)$I_3$

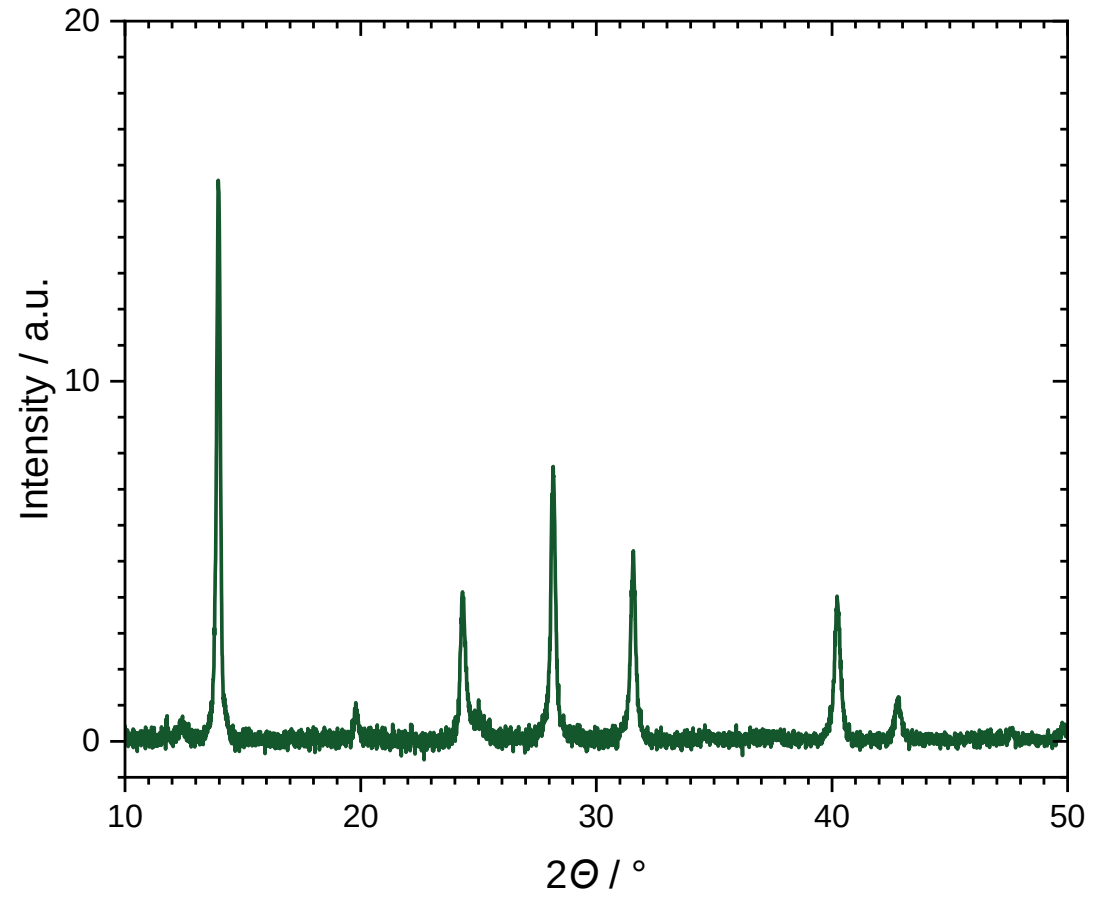

XRD diffractogram of FA($Sn_{0.25}Pb_{0.75}$)$I_3$.

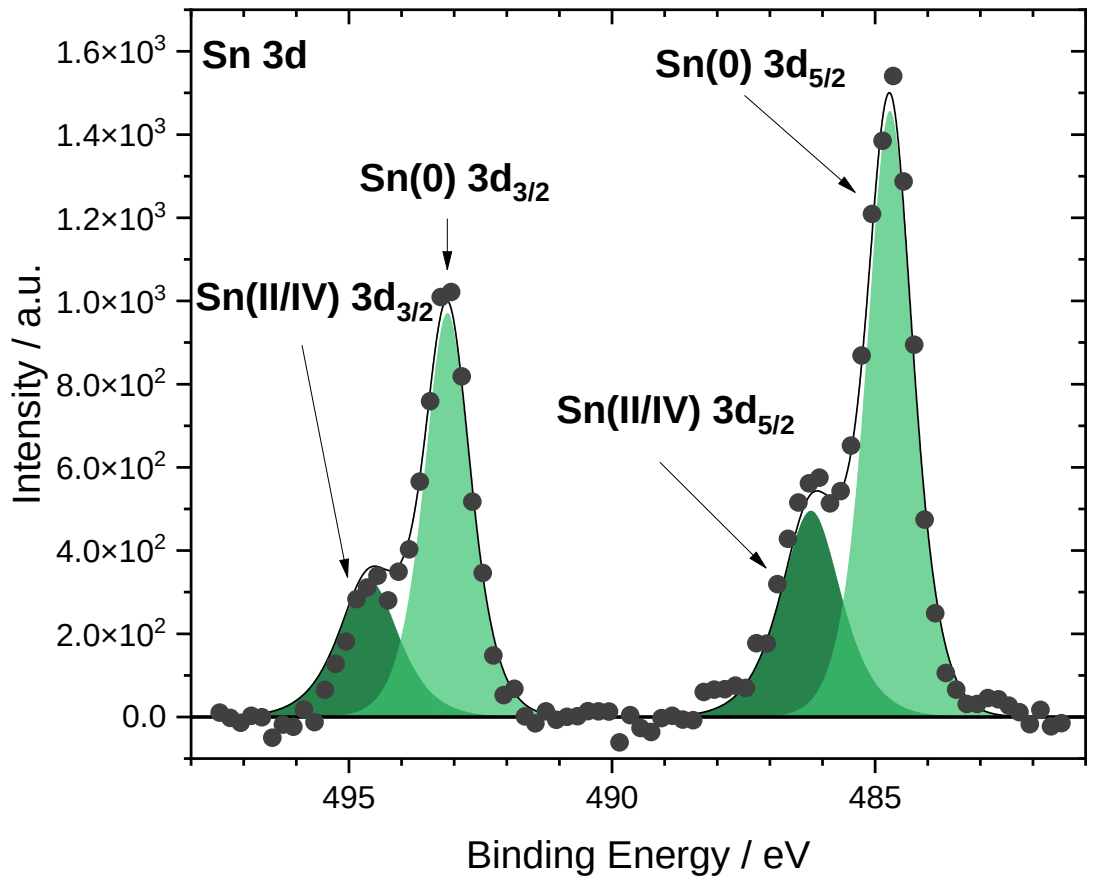

Sn 3d region of FA($Sn_{0.25}Pb_{0.75}$)$I_3$.

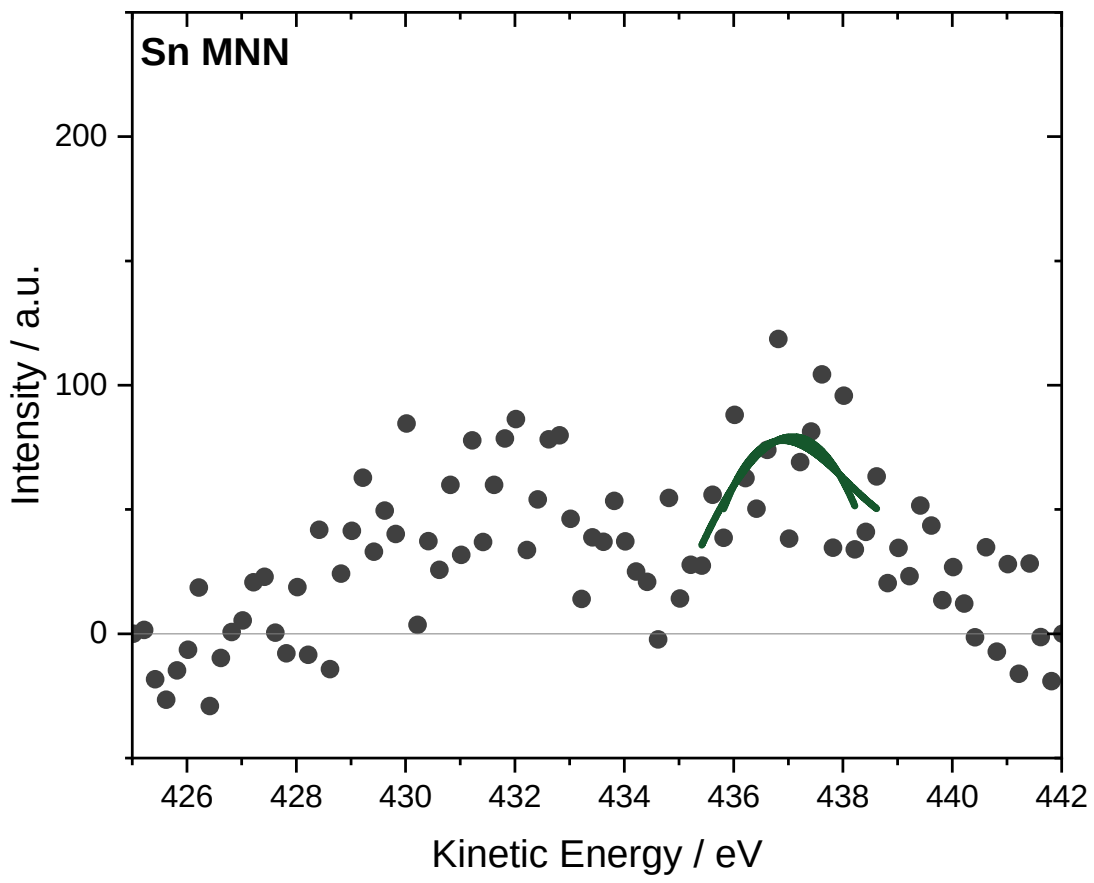

Sn MNN region of FA($Sn_{0.25}Pb_{0.75}$)$I_3$.

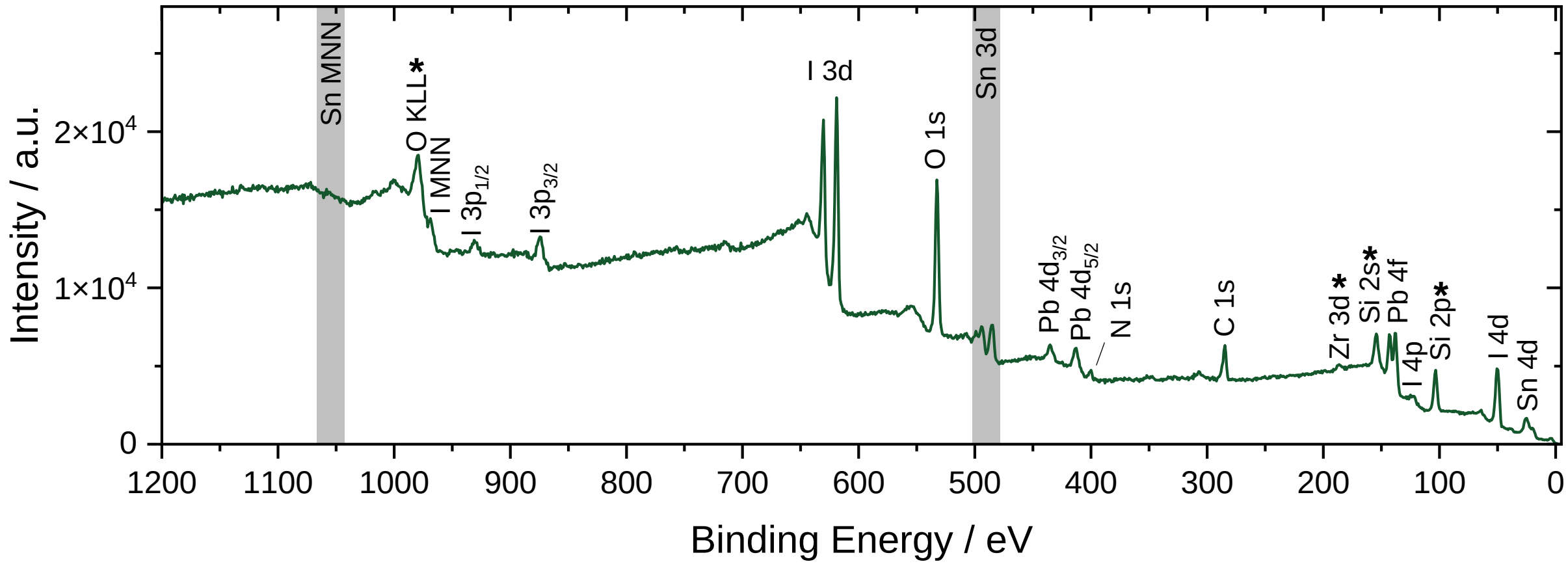

Survey spectrum of FA($Sn_{0.25}Pb_{0.75}$)$I_3$.
Peaks assigned to the probing of the substrate through pin-holes are marked with an asterisk.